 \documentclass[11pt]{article}
\usepackage{amsmath,amsfonts,amssymb}

 \usepackage[numbers,sort&compress]{natbib}
 \usepackage{mathrsfs}
\usepackage{graphicx}

\usepackage{adjustbox}

\usepackage{hyperref}
\hypersetup{
   colorlinks   =  true,
    linkcolor    = blue,
    citecolor    = red,
     urlcolor	=blue,
	urlbordercolor = {1 1 0}
}

\newcommand\be{\begin{equation}}
\newcommand\ee{\end{equation}}
\newcommand\bea{\begin{eqnarray}}
\newcommand\eea{\end{eqnarray}}
\usepackage{lmodern}
\usepackage[english]{babel}
\usepackage[utf8]{inputenc}
\usepackage[T1]{fontenc}
\usepackage{doi}
\renewcommand{\doitext}{\noexpand\textsc{doi:}}
\usepackage{pgf,tikz,pgfplots}
\usepackage{tikz-cd}
\usetikzlibrary{babel}
\usepackage{adjustbox}
\usepackage{subcaption,booktabs}

\usepackage{amsthm}
\usepackage{physics}
\usepackage{amssymb}
\usepackage{bm}

\newcommand{\set}[1]{\left\{ #1 \right\}} 

\theoremstyle{definition}
\newtheorem{defi}{Definition}[section] 

\theoremstyle{plain}
\newtheorem{prop}[defi]{Proposition} 

\theoremstyle{remark}
\newtheorem{remark}{Remark}

\numberwithin{equation}{section} 

\begin{document}
\
\vskip0.5cm

\begin{center}

  {\Large \bf
    Projective relativistic dynamics from a quantum Lorentz algebra:\\[5pt]    time-dependent deformed Hamiltonians on (anti-)de Sitter spaces}

\end{center}

\medskip

\begin{center}

  {\sc  Rutwig Campoamor-Stursberg$^{1,2}$, Oscar Carballal$^{1,2}$\\[2pt] and Francisco J.~Herranz$^3$}

\end{center}

\medskip

\noindent
$^1$ Instituto de Matem\'atica Interdisciplinar, Universidad Complutense de Madrid, E-28040 Madrid,  Spain

\noindent
$^2$ Departamento de \'Algebra, Geometr\'{\i}a y Topolog\'{\i}a,  Facultad de Ciencias
Matem\'aticas, Universidad Complutense de Madrid, Plaza de Ciencias 3, E-28040 Madrid, Spain

\noindent
{$^3$ Departamento de F\'isica, Universidad de Burgos,
  E-09001 Burgos, Spain}

\medskip

\noindent  E-mail: {\small
  \href{mailto:rutwig@ucm.es}{\texttt{rutwig@ucm.es}}, \href{mailto:oscarbal@ucm.es}{\texttt{oscarbal@ucm.es}}, \href{mailto:fjherranz@ubu.es}{\texttt{fjherranz@ubu.es}}
}

\medskip

\begin{abstract}
  \noindent
  A (non-standard) quantum deformation of the Lorentz algebra $\mathfrak{so}(1,3)$ is used to construct novel time-dependent Hamiltonian systems on the (anti-)de Sitter spacetimes in $(1+1)$-dimensions. The specific deformation  considered here contains, among others, the non-standard  quantum deformation of   $\mathfrak{sl}(2, \mathbb{R})$ and that of the `book' algebra  $\mathfrak{b}_{2}$ as quantum subalgebras. The guiding idea is to apply the theory of Poisson--Hopf deformations of Lie--Hamilton systems to the Lorentz deformation,
  giving rise to a broad family of  time-dependent deformed systems that share the same deformed  Poisson--Hopf Lorentz algebra symmetry. In particular, this approach provides Hamiltonian vector fields  expressed in terms of certain projective canonical variables, which, in turn, are related to the usual Beltrami  variables via a canonical transformation. Under this construction, it is shown that the underlying Lorentizan spaces possess  constant curvature $\kappa$ and   the quantum deformation  parameter $z$ can be identified with $\kappa$, such that  the cosmological constant is $\Lambda=-z\equiv -\kappa$. Therefore, we deal with the anti-de Sitter, de Sitter and Minkowskian spaces for $z>0$, $z<0$ and $z=0$ (the undeformed case), respectively. Finally, as applications, we study  time-dependent deformed oscillators on the (anti-)de Sitter spaces in detail, as well as  time-dependent deformed Bateman and  generalised coupled Caldirola--Kanai Hamiltonians.

\end{abstract}
\medskip
\medskip
\medskip

\medskip


\medskip
\noindent
\textbf{PACS 2010 codes}:  {02.20.Uw, 02.20.Sv, 45.10.Na, 45.20.Jj, 05.45.-a}\\

\medskip

\newpage

\tableofcontents



\section{Introduction}

The theory of quantum groups is intimately connected with integrable systems since its earliest developments in the late 1970s, particularly through the framework of the quantum inverse scattering method, chiefly represented in the pioneering works of Faddeev, Takhtajan, Sklyanin and Reshetikhin, among others (see \cite{Faddeev1995} for a historical survey). The field underwent a rapid expansion when Drinfel'd and Jimbo (in \cite{Drinfeld1985} and \cite{Jimbo1986,Jimbo1985}, respectively) independently extended the earlier constructions of Kulish and Reshetikhin \cite{Kulish1981} and of Sklyanin \cite{Sklyanin1985}. The term {\it quantum group} itself was introduced by Drinfel'd in his plenary lecture at the International Congress of Mathematicians in 1986 \cite{Drinfeld1986} (see~\cite{Chari1995,Majid1995}  for more details on quantum group theory). Since then, interest in quantum groups has grown exponentially, encompassing a wide spectrum of areas in both mathematics and physics, ranging from knot theory and non-commutative geometry to integrable systems, statistical physics, and quantum gravity (see, e.g.,  \cite{Woronowicz1987,Faddeev1988,Witten1988,Majid1988,Pasquier1988,Witten1989,LNR1992fieldtheory,MS2002,FKS2004,Amelino-Camelia:2008aez,Addazi:2021xuf}
and references therein).

Notably, quantum groups have likewise been used within the so-called {\it coalgebra method}, originally developed by Ballesteros and Ragnisco in \cite{BCR96,Ballesteros1998}, for the systematic construction of (super)integrable systems \cite{Ballesteros1999,Ballesteros2005,Ballesteros2005b,Ragnisco2007,BBHMR2009}, and which was based on a  Hopf algebra structure~\cite{Abe}. In particular, it has been shown in \cite{Ballesteros2005, Ballesteros2005b, Ragnisco2007,BBHMR2009} that deformation procedures based on quantum groups give rise to spaces whose curvature is deformed through the underlying quantum group structure.

In this context, it is worth stressing that this approach has also been applied to the analysis of `integrable' deformations of several relevant Hamiltonian systems \cite{Ballesteros2011, Ballesteros2016, Ballesteros2017, Ballesteros2024}. Remarkably, these techniques have recently been extended to a significant class of non-autonomous Hamiltonian systems with relevant applications, namely the so-called Lie--Hamilton systems \cite{Lucas2020,Ballesteros2015,CampoamorStursberg2025,CampoamorStursberg2024c}, thereby yielding the {\it Poisson--Hopf deformations of  Lie--Hamilton systems} \cite{Ballesteros2018, Ballesteros2021, Wysocki2023, CampoamorStursberg2023, CampoamorStursberg2024, CampoamorStursberg2025b}.

Lie--Hamilton systems constitute a distinguished subclass of the so-called {\it Lie systems}, that is, first-order non-autonomous systems of ordinary differential equations in normal form that admit a superposition rule, through which the general solution of the system can be expressed in terms of a finite, generic family of particular solutions and a set of significant constants \cite{Lucas2020,Carinena2011,Anderson1982,Harnad1983,Ueno1972,Turbiner1999,Oppenheim1964}. One of the advantages of Lie--Hamilton systems, in contrast with general Lie systems, lies in the fact that the coalgebra method allows the systematic derivation of time-independent constants of motion, thereby facilitating the construction of a superposition rule more readily than in the general setting \cite{Lucas2020, CampoamorStursberg2025a}.

In this regard, we emphasise that only Poisson--Hopf deformations of Lie--Hamilton systems on the plane have been  analysed, and exclusively those stemming from the non-standard quantum deformations of the book algebra $\mathfrak{b}_{2}$, the simple Lie algebra $\mathfrak{sl}(2,\mathbb{R})$, the oscillator algebra $\mathfrak{h}_{4}$, and the two-photon Lie algebra $\mathfrak{h}_{6}$ (see \cite{Ballesteros2018, Ballesteros2021, CampoamorStursberg2023, CampoamorStursberg2024, CampoamorStursberg2025b}).

In this work, we present the first known construction of a higher-dimensional (that is, non-planar) Poisson--Hopf deformation of Lie--Hamilton systems. Our deformation is based on the non-standard quantum deformation algebra $\mathcal{U}_{z}(\mathfrak{so}(1,3))$ of the Lorentz algebra $\mathfrak{so}(1,3)$, which arises as a graded contraction of the non-standard quantum deformations introduced in \cite{Herranz2000}. The procedure relies on a careful  use of the dual Poisson--Lie group $G_{z}^{*}$ associated with the bialgebra structure  on $\mathfrak{so}(1,3)$ underlying  the quantum deformation $\mathcal{U}_{z}(\mathfrak{so}(1,3))$, which may be regarded as a `dual' construction of $\mathcal{U}_{z}( \mathfrak{so}(1,3))$ by virtue of the so-called quantum duality principle (see \cite{Ballesteros2013} and references therein). We apply the resulting deformation to the Lie--Hamilton systems on $\mathrm{T}^{*} \mathbb{R}^{2}$ based on the Lorentz algebra constructed in \cite{CampoamorStursberg2025}, which were obtained via  the regular embedding $\mathfrak{so}(1,3) \subset \mathfrak{sp}(4, \mathbb{R})$ and the fundamental $4$-dimensional representation of $\mathfrak{sp}(4, \mathbb{R})$ on $\mathbb{R}^{4}$.

The Lie--Hamilton systems based on $\mathfrak{so}(1,3)$ introduced in \cite{CampoamorStursberg2025} may, as noted therein, be interpreted as Lie--Hamilton systems on $\mathrm{T}^{*}\mathbf{M}^{1+1}$, where $\mathbf{M}^{1+1}$ denotes the Minkowskian plane. These systems encompass, among others, a time-dependent Bateman Hamiltonian and a generalised coupled Caldirola--Kanai Hamiltonian. Our formulation of the corresponding Poisson--Hopf deformation naturally leads to time-dependent Hamiltonian systems on the cotangent bundle  $\mathrm{T}^{*} \mathcal{M}_{z}$  of certain open subset
$\mathcal{M}_{z} \subset \mathbb{R}^{2}$, where $z$ denotes a real deformation parameter such that the usual quantum deformation parameter $q$ is expressed as $q={\rm e}^z$. In addition, a direct relationship    is established between $\mathcal{M}_{z}$ and the  {\it $(1+1)$-dimensional relativistic Cayley--Klein spaces} of constant curvature $\kappa$ \cite{CampoamorStursberg2024c,Herranz2002}, collectively denoted by $\mathbf{S}^{2}_{[\kappa]}$, expressed in the so-called {\it Beltrami coordinates} \cite{Nurowski2025, Guo2003}. Within this framework, the deformation parameter $z$ of the Poisson--Hopf deformation is simply identified with the curvature,  $z \equiv \kappa$, and $ \mathcal{M}_{z} $ can be understood as the open and connected subset of $\mathbf{S}^{2}_{[\kappa]}$ on which Beltrami coordinates are defined. These spaces include the anti-de Sitter spacetime $(\kappa > 0)$, the de Sitter spacetime $(\kappa < 0)$, and the Minkowskian plane $(\kappa = 0)$. This result is consistent with earlier works in which curvature is  realised  as an integrable deformation parameter (see \cite{Ballesteros2005,Ballesteros2005b,Ragnisco2007,CampoamorStursberg2024c}). The approach developed here is novel with respect to previous Poisson--Hopf deformations of Lie--Hamilton systems on the plane \cite{Ballesteros2018, Ballesteros2021, CampoamorStursberg2023, CampoamorStursberg2024, CampoamorStursberg2025b}, as it represents the first instance in which such deformations give rise to substantive geometric insights.

We also develop a comprehensive study  of the Poisson geometry of the dual Poisson--Lie group $G_{z}^{*}$ associated with the non-standard  quantum  deformation of $\mathfrak{so}(1,3)$, retrieving the non-standard quantum deformations of the Lie algebras $\mathfrak{b}_{2}$ and $\mathfrak{sl}(2,\mathbb{R})$ as particular  submanifolds of $G_{z}^{*}$. Furthermore, we establish a connection between the Poisson geometry of $G_{z}^{*}$ and that of $\mathfrak{so}(1,3)^{*}$ by means of a Poisson map between a certain open subset of $\mathfrak{so}(1,3)^{*}$, equipped with its usual Kirillov--Konstant--Souriau Poisson structure, and $G_{z}^{*}$. Remarkably, the inverse of this map coincides with the expression of the twist map of $\mathcal{U}_{z} (\mathfrak{so}(1,3))$ obtained in \cite{Herranz2000}.
We appropriately compose this map with the momentum map associated with the Lie--Hamilton systems based on $\mathfrak{so}(1,3)$ to obtain a Poisson map $\mathrm{T}^{*}\mathcal{M}_z \to G_{z}^{*}$. The pull-back of the coordinate functions of $G_{z}^{*}$ associated with the generators of $\mathfrak{so}(1,3)$ produces, in a natural and constructive manner, a Poisson--Hopf deformation of the Lie--Hamilton systems based on $\mathfrak{so}(1,3)$. This procedure gives rise to a variety of interesting deformed systems, including systems solvable by quadratures, time-dependent deformed oscillators on  the (anti-)de Sitter spacetimes, and time-dependent deformed Bateman and Caldirola--Kanai Hamiltonians.  We also stress that, as the Poisson--Hopf deformation can be expressed in  Beltrami coordinates, some of the deformed systems may be interpreted as {\it projective relativistic dynamical systems} (see \cite{Albouy2008,Albouy2013} for details on projective dynamics).

This work is structured as follows. In Section~\ref{section:preliminaries}, we review the fundamental notions concerning Poisson manifolds, Lie--Hamilton systems and Poisson--Hopf deformations of Lie--Hamilton systems. The construction of   Lie--Hamilton systems on $\mathrm{T}^{*} \mathbb{R}^{2}$ based on the Lorentz algebra and some of their relevant applications is revisited in Section~\ref{section:LorentzLH}.  Section~\ref{section:nonstandardLorentz} is entirely devoted to the non-standard  quantum  deformation of $\mathfrak{so}(1,3)$. Here, the dual Poisson--Lie group $G_{z}^{*}$ associated with this deformation is obtained explicitly, providing exact expressions for the corresponding multiplication law, Poisson structure and coproduct. A geometrical interpretation of the twist map of $\mathcal{U}_{z}(\mathfrak{so}(1,3))$ and the retrieval of the non-standard quantum deformations of $\mathfrak{b}_{2}$ and $\mathfrak{sl}(2, \mathbb{R})$ as Poisson--Dirac submanifolds of $G_{z}^{*}$ are also included. Section~\ref{section:curvature} addresses how the construction of the two-dimensional Cayley--Klein spaces in Beltrami coordinates follows naturally from the preceding considerations.   The corresponding Poisson--Hopf deformation is studied in detail in Section~\ref{section:dynamics}, along with the construction of  time-independent constants of motion for all deformed Lorentz systems, making use of deformed prolongations of the generic time-dependent deformed Lorentz vector field. Next, in Section~\ref{section:applications},  several applications are explicitly presented yielding remarkable deformed systems and discussing their relation to previously studied Poisson--Hopf deformations. In particular, in the case of $\mathfrak{b}_{2}$, the resulting systems are solvable by quadratures, and our construction retrieves the previously studied deformation by restricting to a suitable submanifold. For the non-standard  quantum  deformation of $\mathfrak{sl}(2,\mathbb{R})$, we obtain time-dependent deformed oscillators on the (anti-)de Sitter spacetimes. Moreover, the complete general deformed Lorentz system is also examined, and its first-order approximation in the deformation parameter $z$ is discussed. Finally, Section~\ref{section:conclusion} summarises the main results of this work and considers possible extensions for future research.

\newpage


\section{Preliminaries}
\label{section:preliminaries}

Let us first establish the fundamental concepts and definitions used throughout the paper (more details can be found e.g.~in \cite{ Ballesteros2018,Ballesteros2021,Wysocki2023,Carinena2011,Lucas2020,Va_94,Lichnerowicz1997,Crainic2021,PW83,CGM07,CLS13}). Unless stated otherwise, all structures are assumed to be smooth. Manifolds are real, Hausdorff, paracompact, separable, and finite-dimensional.

A {\it Poisson manifold} is a pair $(M, \Pi)$, where $\Pi \in \Gamma (\wedge^{2} \mathrm{T}M  )$ is a bivector field on a manifold $M$, referred to as the {\it Poisson bivector}, such that $[\Pi, \Pi] = 0$, where $[\cdot, \cdot]$ denotes the Schouten--Nijenhuis bracket \cite{Va_94,Lichnerowicz1997,Crainic2021}. The Poisson bivector $\Pi$ induces a Poisson algebra structure on $C^{\infty}(M)$ via the Poisson bracket $ \set{\cdot, \cdot}_{\Pi}: C^{\infty} (M) \times C^{\infty}(M) \to C^{\infty}(M)$ defined by
\begin{equation}
  \set{f, f'}_{\Pi} := \Pi( \dd f, \dd f'), \qquad f, f' \in C^{\infty}(M).
  \label{eq:pre:pb}
\end{equation}
A {\it Casimir function} of the Poisson manifold $(M, \Pi)$ is a function $\mathscr{C} \in C^{\infty}(M)$ such that $\{\mathscr{C}, f\}_{\Pi} = 0$ for all $f \in C^{\infty}(M)$. The Poisson bivector $\Pi$ further determines a vector bundle morphism
\[ \sharp_{\Pi}: \mathrm{T}^{*}M \ni \eta_{x} \longmapsto \iota_{\eta_{x}} \Pi_{x} \in   \mathrm{T} M. \] A {\it symplectic manifold} is a Poisson manifold $(M, \Pi)$ for which $\sharp_{\Pi}$ is an isomorphism. In this situation, there exists a symplectic form $\omega \in \Omega^{2}(M)$ such that   $-(\flat_{\omega})^{-1} = \sharp_{\Pi}$, where $\flat_{\omega}$ is the vector bundle morphism $\flat_{\omega}: \mathrm{T} M \ni v_{x} \mapsto \iota_{v_{x}} \omega_{x} \in \mathrm{T}^{*}M$. Accordingly,  the symplectic manifold is denoted by $(M, \omega)$  and the corresponding Poisson bracket  \eqref{eq:pre:pb} on $C^{\infty}(M)$ by $\set{\cdot, \cdot}_{\omega}$.

Let $(M, \omega)$ be a symplectic manifold. The {\it Hamiltonian vector field} associated with a function $f \in C^{\infty}(M)$ is the vector field $\mathbf{X}_{f}$ on $M$ determined by the condition $\iota_{\mathbf{X}_{f}} \omega = \dd f$. The set of Hamiltonian vector fields on $M$ forms a Lie subalgebra $\mathrm{Ham}(M, \omega)$ of $\mathfrak{X}(M)$. Provided that $M$ is connected, we have an exact sequence of Lie algebras
\[ \begin{tikzcd}
    0  \arrow[r,hook]& \mathbb{R}  \arrow[r, hook]& \big(C^{\infty}(M), \set{\cdot, \cdot}_{\omega} \big) \arrow[r,"\psi"]& \big(\mathrm{Ham}(M, \omega), [\cdot, \cdot] \big)  \arrow[r,twoheadrightarrow]& 0,
  \end{tikzcd} \]
where $\psi$ maps each $f \in C^{\infty}(M)$ onto the Hamiltonian vector field $\mathbf{X}_{-f}$.

As we are mainly concerned with physical applications in the Hamiltonian setting, we will mostly work on the cotangent bundle $\pi_{Q}: \mathrm{T}^{*} Q \to Q$ of a manifold $Q$, endowed with its canonical symplectic form $\omega = - \dd \theta$. Here $\theta \in \Omega^{1}(\mathrm{T}^{*}Q)$ is the so-called tautological one-form, defined by $\theta(\alpha_{x}) := \mathrm{T}^{*}_{\alpha_{x}} \pi_{Q}(\alpha_{x})$ for every $\alpha_{x} \in \mathrm{T}^{*}Q$.

A {\it time-dependent vector field} on a manifold $M$ is a map $\mathbf{X}: \mathbb{R} \times M \ni (t, x) \mapsto \mathbf{X}(t,x) \in \mathrm{T}M$ such that, for each $t \in \mathbb{R}$, the map $\mathbf{X}(t, \cdot): M \to \mathrm{T}M$ is a vector field on $M$ \cite{Carinena2011,Lucas2020}.   Time-dependent vector fields are usually referred to as $t$-dependent vector fields, where $t \in \mathbb{R}$ is interpreted as time. The {\it associated system} of a $t$-dependent vector field $\mathbf{X}$  on $M$ is the system of first-order ordinary differential equations (ODEs) given by
\begin{equation}
  \dv{x}{t} = \mathbf{X}(t, x), \qquad (t,x) \in \mathbb{R} \times M.
  \label{eq:preliminaries:system}
\end{equation}
There is a one-to-one correspondence between first-order systems of ODEs on $M$ in normal form, such as \eqref{eq:preliminaries:system}, and $t$-dependent vector fields on $M$. Hence, we use $\mathbf{X}$ to refer both to a $t$-dependent vector field $\mathbf{X}$ and to its associated system.

A {\it Lie system} on a manifold $M$ is a $t$-dependent vector field $\mathbf{X}$ on $M$ taking values in a finite-dimensional Lie algebra of vector fields $V \subset \mathfrak{X}(M)$, a so-called {\it Vessiot--Guldberg Lie algebra} (VG Lie algebra) of $\mathbf{X}$ \cite{Carinena2011,Lucas2020,PW83}; that is, $\mathbf{X}_{t} \in V$ for all $t \in \mathbb{R}$. If $V$ is isomorphic to an (abstract) Lie algebra $\mathfrak{g}$, we say that $\mathbf{X}$ is a {\it $\mathfrak{g}$-Lie system}.  The main feature of Lie systems is the existence of a superposition rule \cite{Carinena2011,Lucas2020,PW83,CGM07}. A {\it superposition rule} for a system $\mathbf{X}$ on $M$, depending on $k \geq 1$ particular solutions, is a $t$-independent map $\Phi: M^{k} \times M \to M$ such that the general solution $x(t)$ of $\mathbf{X}$ can be cast into the form $x(t) = \Phi \big(x_{(1)}(t), \ldots, x_{(k)}(t); p \big)$, where $x_{(1)}(t), \ldots, x_{(k)}(t)$ is a generic family of particular solutions of $\mathbf{X}$ and $p \in M$ is a point related to the initial conditions. The fundamental Lie--Scheffers Theorem (see \cite[Section 3.8]{Lucas2020}) states that a system $\mathbf{X}$ admits a superposition rule if and only if it is a Lie system.

A {\it Lie--Hamilton system} (LH system) on a symplectic manifold $(M, \omega)$ is a Lie system $\mathbf{X}$ on $M$ possessing a VG Lie algebra $V$ formed by Hamiltonian vector fields relative to $\omega$ \cite{Lucas2020,CLS13}. In this context, $\psi^{-1}(V) \subset C^{\infty}(M)$ spans a finite-dimensional Lie algebra $\mathcal{H}_{\omega} \subset C^{\infty}(M)$ relative  to the Poisson bracket $\set{\cdot, \cdot}_{\omega}$, and is referred to as a {\it Lie--Hamilton algebra} (LH algebra) of $\mathbf{X}$ (see \cite[Chapter 4]{Lucas2020}).

Let $\mathbf{X}$ be a LH system on a symplectic manifold $(M, \omega)$ and $\mathcal{H}_{\omega}$ be a LH algebra of $\mathbf{X}$ with basis $\set{h_{1}, \ldots, h_{r}}$. A {\it Poisson--Hopf deformation} (PH deformation) of $\mathbf{X}$, depending on a deformation parameter $z \in \mathbb{R}$, is a family of $z$-parameterised functions $\{h_{z, 1}, \ldots, h_{z,r}\}$ satisfying the following conditions (see \cite{Ballesteros2018,Ballesteros2021,Wysocki2023}):
\begin{itemize}
  \item[(i)] For every $z \in \mathbb{R}$ there exists an open subset $U_{z} \subset M$ such that $h_{z, \alpha} \in C^{\infty}(U_{z})$ for every $ 1\leq \alpha \leq r$,
  \item[(ii)] $\lim_{z \to 0} h_{z, \alpha} = h_{\alpha}$ for every $1 \leq \alpha \leq r$, and
  \item[(iii)] their commutation relations satisfy $\set{h_{z, \alpha}, h_{z, \beta}}_{\omega} =  F_{z, \alpha \beta}(h_{z,1}, \ldots, h_{z,r})$ for certain  functions $F_{z, \alpha \beta} \in C^{\infty}(\mathbb{R}^{r})$ and all $1 \leq \alpha, \beta \leq r$.
\end{itemize}


\section{Lie--Hamilton systems from the Lorentz algebra}
\label{section:LorentzLH}

The {\it Lorentz algebra} $\mathfrak{so}(1,3)$ is the Lie algebra of the  orthocromous  Lorentz group $\mathrm{SO}(1,3)$, which corresponds to the connected component at the identity of the group of linear isometries of $\mathbb{R}^{4}$ preserving the Minkowski metric $\mathrm{diag}(-, +, +, +)$. As a real Lie algebra, $\mathfrak{so}(1,3)$ is isomorphic to $\mathfrak{sl}(2, \mathbb{C})$, and hence corresponds to a real form of the compact Lie algebra $\mathfrak{so}(4, \mathbb{C})$ generated by outer involutive automorphisms.  For our purposes, it is worth remembering that  $\mathfrak{so}(1,3)$   is isomorphic to the Lie algebra of isometries of the $(2+1)$-dimensional de Sitter spacetime~\cite{Bacry1968}, which is a Lorentzian space with a negative  constant sectional curvature $\kappa<0$ (that is, with a positive cosmological constant). In addition, the Lorentz algebra is also isomorphic to the Lie algebra of conformal transformations in the two-dimensional Euclidean space~\cite{Herranz2002}. Next, we will review both constructions.

Let us first consider $\mathfrak{so}(1,3)$ over a basis $\set{J, P_{0}, P_{i}, K_{i}: 1 \leq i \leq 2}$ with commutation relations given by
\begin{equation}
  \begin{aligned}
     & [J, P_{i}] = - \epsilon_{ij} P_{j}, \qquad &  & [J, K_{i}] = - \epsilon_{ij}K_{j}, \qquad &  & [J, P_{0}] = 0, \qquad         &  & [P_{1}, P_{2}] = -J,    \\[2pt]
     & [K_{1}, K_{2}] = J, \qquad                 &  & [P_{i}, K_{j}] = \delta_{ij}P_{0}, \qquad &  & [P_{0}, P_{i}] = K_{i}, \qquad &  & [P_{0}, K_{i}] = P_{i},
  \end{aligned}
  \label{eq.LorentzLH:comm}
\end{equation}
being $\epsilon_{ij}$ the skew-symmetric tensor defined by $\epsilon_{12} = 1$ and where $1 \leq i, j \leq 2$. The generators $J$, $P_{0}$, $P_{i}$ and $K_{i}$ $(1 \leq i \leq 2)$ can be interpreted, in this order,  as infinitesimal generators of spatial rotations, time translations, space translations and boosts in the $(2+1)$-dimensional de Sitter spacetime, such that we have set the speed of light $c=1$ and the curvature $\kappa=-1$  in (\ref{eq.LorentzLH:comm}). This justifies referring to $\set{J, P_{0}, P_{i}, K_{i}: 1 \leq i \leq 2}$ as a {\it kinematical basis} of $\mathfrak{so}(1,3)$ (see \cite{Bacry1968}).  The two quadratic Casimir invariants of $\mathfrak{so}(1,3)$ read
\begin{equation}
  \mathscr{C}_{2} :=  J^{2} + P_{1}^{2} + P_{2}^{2} - P_{0}^{2} - K_{1}^{2} - K_{2}^{2}, \qquad \mathscr{C}_{2}' := - JP_{0} - P_{1}K_{2} + P_{2}K_{1},
  \label{eq.LorentzLH:CasimirKin}
\end{equation}
where $\mathscr{C}_{2}$ is provided by the Killing--Cartan form of $\mathfrak{so}(1,3)$, while $\mathscr{C}_{2}'$ is related to the Pauli--Lubanski vector.

Secondly, we stress that $\mathfrak{so}(1,3)$ can also be regarded as the conformal algebra of the Euclidean plane. Indeed, the change of basis
\begin{equation}
  \begin{aligned}
     & H:= J - K_{2}, \qquad          &  & D:= K_1, \qquad   &  & C_{1}:= J + K_{2}, \\[2pt]
     & C_{2} := P_{1} - P_{0}, \qquad &  & K:= P_{2}, \qquad &  & P:= P_{0} + P_{1},
  \end{aligned}
  \label{eq:LorentzLH:conformalbasis}
\end{equation}
yields a so-called {\it conformal basis} $\set{H, D, C_{1}, C_{2}, K, P}$ of $\mathfrak{so}(1,3)$. In this context, the subalgebra $\langle K, H, P \rangle$ corresponds to the algebra of infinitesimal isometries of the Euclidean plane, where $K$ generates rotations, while $H$ and $P$ generate space translations. The similitude algebra (also known as the Weyl--Euclidean algebra) of the Euclidean plane is then $\langle K, H, P, D \rangle$, where $D$ is the generator of dilations. The full conformal algebra $\mathfrak{so}(1,3)$ emerges upon adding $C_{1}$ and $C_{2}$, the generators of specific conformal transformations, to the Weyl--Euclidean algebra~\cite{Herranz2002,Herranz2000}.
The  non-zero  commutation relations of $\mathfrak{so}(1,3)$ over the conformal basis \eqref{eq:LorentzLH:conformalbasis} are given by
\begin{equation}
  \begin{aligned}
     & [K, H] = -P, \qquad           &                           & [K, P] = H, \qquad         &                         & [D, H] = -H , \qquad       &                  & [D, P] = -P ,       \\[2pt]
     & [K, C_{1}] = -C_{2}  , \qquad &                           & [K, C_{2}] = C_{1}, \qquad &                         & [D, C_{1}] = C_{1}, \qquad &                  & [D,C_{2}] = C_{2} , \\[2pt]
     & [H, C_{1}] = 2D, \qquad
     &                               & [H, C_{2}] = -2K , \qquad &                            & [P, C_{1}] = 2K, \qquad &                            & [P, C_{2}] = 2D,
  \end{aligned}
  \label{eq.LorentzLH:confbasis}
\end{equation}
while the Casimir invariants \eqref{eq.LorentzLH:CasimirKin} in this basis turn out to be
\begin{equation}
  \mathscr{C}_{2} = K^{2} - D^{2} + HC_{1} +C_{2} P, \qquad \mathscr{C}_{2}' = KD + \frac{1}{2}(HC_{2} - C_{1}P ).
  \label{eq:LorentzLH:CasimirKin2}
\end{equation}

These descriptions of the Lorentz Lie algebra can be directly translated into a Poisson--Lie framework by expressing the commutation relations (\ref{eq.LorentzLH:comm}) and (\ref{eq.LorentzLH:confbasis}) as Poisson brackets. We can then  analyse these structures from a LH  perspective. It turns out that
on $\mathrm{T}^{*}\mathbb{R}$ there is no LH algebra isomorphic to $\mathfrak{so}(1,3)$ \cite{Ballesteros2015}.  On the other hand, the following result, recently obtained in \cite{CampoamorStursberg2025}, demonstrates that this is not the case on $\mathrm{T}^{*}\mathbb{R}^{2}$. Let us consider $\mathrm{T}^{*} \mathbb{R}^{2}$ endowed with global bundle coordinates $(\xi_{1}, \xi_{2}, p_{\xi_{1}}, p_{\xi_{2}})$, with respect to which its canonical symplectic form reads $\omega = \dd \xi_{1} \wedge \dd p_{\xi_{1}} + \dd \xi_{2} \wedge \dd p_{\xi_{2}}$.

\begin{prop}[\cite{CampoamorStursberg2025}] \label{prop:LorentzLH:morphism} There exists an injective morphism of Lie algebras $$\phi: \mathfrak{so}(1,3) \to C^{\infty}\big(\mathrm{T}^{*}\mathbb{R}^{2}\big),$$  defined by
  \begin{equation*}
    \begin{aligned}
       & h_{1} := \phi(H)=  -\xi_{1}^{2} + \xi_{2}^{2}, \qquad   &  & h_{2} :=\phi(D)= \frac{1}{2}(\xi_{1} p_{\xi_{1}} + \xi_{2} p_{\xi_{2}}), \qquad  &  & h_{3}:= \phi(C_{1})=- \frac{1}{4}\big(p_{\xi_{1}}^{2} - p_{\xi_{2}}^{2} \big), \\[2pt]
       & h_{4}:= \phi(C_{2}) =  \frac{1}{2}p_{\xi_{1}} p_{\xi_{2}}, \qquad &  & h_{5} := \phi(K)= \frac{1}{2}(\xi_{1}p_{\xi_{2}} - \xi_{2} p_{\xi_{1}}) , \qquad &  & h_{6}:= \phi(P)= 2 \xi_{1} \xi_{2}.
    \end{aligned}
  \end{equation*}
  Therefore, the Hamiltonian functions $h_{\alpha} \in C^{\infty}(\mathrm{T}^{*}\mathbb{R}^{2})$ $(1 \leq \alpha \leq 6)$, with  non-zero  commutation relations  given by
  \begin{equation}
    \begin{aligned}
             & \set{h_{1}, h_{2}}_{\omega} = h_{1}, \qquad &    & \set{h_{1}, h_{3}}_{\omega} = 2 h_{2}, \qquad &                                                & \set{h_{1}, h_{4}}_{\omega} = -2 h_{5}, \qquad   &    & \set{h_{1}, h_{5}}_{\omega} = h_{6},     \\[2pt] 
   & \set{h_{2}, h_{3}}_{\omega} = h_{3}, \qquad &&\set{h_{2}, h_{4}}_{\omega} = h_{4},       \qquad &    & \set{h_{2}, h_{6}}_{\omega} = -h_{6}, \qquad &       & \set{h_{3}, h_{5}}_{\omega} = h_{4},         \\[2pt] 
   & \set{h_{3},h_{6}}_{\omega} = -2 h_{5}, \qquad       &       & \set{h_{4}, h_{5}}_{\omega} = -h_{3}, \qquad &                                               & \set{h_{4}, h_{6}}_{\omega} = -2 h_{2}, \qquad &       & \set{h_{5}, h_{6}}_{\omega} = h_{1},
    \end{aligned}
    \label{eq:LorentzLH:LHalgebra}
  \end{equation}
  span a LH algebra $\mathcal{H}_{\omega}$ isomorphic to $\mathfrak{so}(1,3)$.
\end{prop}

\begin{remark}
  Relative to the canonical coordinates $(q_{1}, q_{2}, p_{1}, p_{2})$ of $\mathrm{T}^{*}\mathbb{R}^{2}$ considered in \cite{CampoamorStursberg2025}, we have applied the canonical transformation defined by
  \[ \xi_{1} = \frac{1}{2}(q_{1}-q_{2}), \qquad \xi_{2} = \frac{1}{2}(q_{1} + q_{2}), \qquad p_{\xi_{1}} = p_{1}- p_{2}, \qquad p_{\xi_{2}} = p_{1} + p_{2}. \]
  The motivation for this change of coordinates will become apparent in Sections~\ref{section:curvature} and \ref{section:dynamics}.
\end{remark}
 
As a consequence of Proposition~\ref{prop:LorentzLH:morphism}, the Hamilton equations of the $t$-dependent Hamiltonian given by
\begin{equation}
  \mathbb{R} \times \mathrm{T}^{*} \mathbb{R}^{2} \ni (t, \bm{\xi}, \bm{p}_{\xi}) \longmapsto h(t, \bm{\xi}, \bm{p}_{\xi}) := \sum_{\alpha = 1}^{6} b_{\alpha}(t) h_{\alpha}(\bm{\xi}, \bm{p}_{\xi}) \in \mathbb{R},
  \label{eq.LorentzLH:tdepHam}
\end{equation}
where $b_{\alpha}(t) \in C^{\infty}(\mathbb{R})$ is an arbitrary $t$-dependent function $(1 \leq \alpha \leq 6)$,  $\bm{\xi} := (\xi_{1}, \xi_{2})$ and $\bm{p}_{\xi}:= (p_{\xi_{1}}, p_{\xi_{2}})$, describe the integral curves of a LH system $\mathbf{X}$ on $\mathrm{T}^{*}\mathbb{R}^{2}$. More specifically, if $\mathbf{X}_{\alpha} \in \mathfrak{X}\big(\mathrm{T}^{*}\mathbb{R}^{2}\big)$ is the Hamiltonian vector field associated with $h_{\alpha}$, determined by the condition $\iota_{\mathbf{X}_{\alpha}} \omega = \dd h_{\alpha}$  $(1 \leq \alpha \leq 6)$, it follows that  $\mathbf{X}$ can be expressed as
\begin{equation}
  \mathbf{X} = \sum_{\alpha = 1}^{6} b_{\alpha}(t) \mathbf{X}_{\alpha},
  \label{eq:LorentzLH:tdepHamVF}
\end{equation}
where the vector fields
\begin{equation*}
  \begin{split}
     & \mathbf{X}_{1} = 2 \left( \xi_{1} \pdv{p_{\xi_{1}}} - \xi_{2} \pdv{p_{\xi_{2}}} \right),                                                                      \\[2pt]
     & \mathbf{X}_{2} = \frac{1}{2} \left( \xi_{1} \pdv{\xi_{1}} + \xi_{2} \pdv{\xi_{2}}   - p_{\xi_{1}} \pdv{p_{\xi_{1}}} - p_{\xi_{2}} \pdv{p_{\xi_{2}}} \right),  \\[2pt]
     & \mathbf{X}_{3} = -\frac{1}{2} \left(   p_{\xi_{1}} \pdv{\xi_{1}} - p_{\xi_{2}} \pdv{\xi_{2}} \right),                                                         \\[2pt]
     & \mathbf{X}_{4}  = \frac{1}{2} \left( p_{\xi_{2}} \pdv{\xi_{1}} + p_{\xi_{1}} \pdv{\xi_{2}} \right),                                                           \\[2pt]
     & \mathbf{X}_{5} =- \frac{1}{2} \left(   \xi_{2} \pdv{\xi_{1}} - \xi_{1} \pdv{\xi_{2}} + p_{\xi_{2}} \pdv{p_{\xi_{1}}} - p_{\xi_{1}} \pdv{p_{\xi_{2}}} \right), \\[2pt]
     & \mathbf{X}_{6} =  -2 \left( \xi_{2} \pdv{p_{\xi_{1}}} + \xi_{1} \pdv{p_{\xi_{2}}} \right),
  \end{split}
\end{equation*}
span a VG Lie algebra of $\mathbf{X}$ isomorphic to $\mathfrak{so}(1,3)$ satisfying the  following   non-zero commutation relations:
\begin{equation*}
  \begin{aligned}
     & [\mathbf{X}_{1}, \mathbf{X}_{2}] =- \mathbf{X}_{1}, \qquad  &     & [\mathbf{X}_{1}, \mathbf{X}_{3}] = -2 \mathbf{X}_{2}, \qquad &      & [\mathbf{X}_{1}, \mathbf{X}_{4}] = 2 \mathbf{X}_{5}, \qquad &                                                      & [\mathbf{X}_{1},                                                                                                                                                                                                                                                                                                                             \mathbf{X}_{5}] = - \mathbf{X}_{6}, \\[2pt]
     & [\mathbf{X}_{2}, \mathbf{X}_{3}] = - \mathbf{X}_{3}, \qquad &       & [\mathbf{X}_{2}, \mathbf{X}_{4}] = - \mathbf{X}_{4},     &   & [\mathbf{X}_{2}, \mathbf{X}_{6}] = \mathbf{X}_{6}, \qquad  &                                                              & [\mathbf{X}_{3}, \mathbf{X}_{5}]  = - \mathbf{X}_{4},                                                                                                                                                                   \\[2pt] 
 & [\mathbf{X}_{3}, \mathbf{X}_{6}] = 2 \mathbf{X}_{5}, \qquad  &   & [\mathbf{X}_{4}, \mathbf{X}_{5}] =  \mathbf{X}_{3}, \qquad &     & [\mathbf{X}_{4}, \mathbf{X}_{6}] =2 \mathbf{X}_{2}, \qquad &                                                             & [\mathbf{X}_{5}, \mathbf{X}_{6}] = - \mathbf{X}_{1}.
  \end{aligned}
\end{equation*}
The first-order system of ODEs associated with \eqref{eq:LorentzLH:tdepHamVF} is given by
\begin{equation*}
  \begin{pmatrix}
    \dot{\xi}_{1}     \\[2pt]
    \dot{\xi}_{2}     \\[2pt]
    \dot{p}_{\xi_{1}} \\[2pt]
    \dot{p}_{\xi_{2}}
  \end{pmatrix} = \frac{1}{2} \begin{pmatrix}
    b_{2}(t)   & -b_{5}(t)  & -b_{3}(t) & b_{4}(t)   \\[2pt]
    b_{5}(t)   & b_{2}(t)   & b_{4}(t)  & b_{3}(t)   \\[2pt]
    4b_{1}(t)  & -4b_{6}(t) & -b_{2}(t) & - b_{5}(t) \\[2pt]
    -4b_{6}(t) & -4b_{1}(t) & b_{5}(t)  & - b_{2}(t)
  \end{pmatrix} \begin{pmatrix}
    \xi_{1}     \\[2pt]
    \xi_{2}     \\[2pt]
    p_{\xi_{1}} \\[2pt]
    p_{\xi_{2}}
  \end{pmatrix},
\end{equation*}
which, as expected, are just the Hamilton equations derived from  the
$t$-dependent Hamiltonian  (\ref{eq.LorentzLH:tdepHam}).


\subsection{Applications of Lie--Hamilton Lorentz systems}
\label{subsection:LorentzLHapplications}

In \cite{CampoamorStursberg2025}, the starting point to study physical applications of the $t$-dependent Hamiltonian \eqref{eq.LorentzLH:tdepHam} was to consider a minimal set of generators of the $\mathfrak{so}(1,3)$-LH algebra \eqref{eq:LorentzLH:LHalgebra}; namely, $h_{1} = - \xi_{1}^{2} + \xi_{2}^{2}$, $h_{3} = -\frac{1}{4}\big( p_{\xi_{1}}^{2} - p_{\xi_{2}}^{2} \big)$ and $h_{5}= \frac{1}{2}(\xi_{1} p_{\xi_{2}} - \xi_{2} p_{\xi_{1}})$. From this, we set $b_{2}(t) = b_{4}(t) = b_{6}(t) \equiv 0$ in \eqref{eq:LorentzLH:tdepHam_gen}, yielding the following $t$-dependent  coupled Hamiltonian:
\begin{equation}
  h = -\frac{b_{3}(t)}{4}\,\big(p_{\xi_{1}}^{2} - p_{\xi_{2}}^{2}\big) - b_{1}(t) \big(\xi_{1}^{2} - \xi_{2}^{2}\big) + \frac{b_{5}(t)}{2}\,(\xi_{1} p_{\xi_{2}} - \xi_{2} p_{\xi_{1}}).
  \label{eq:LorentzLH:tdepHam_gen}
\end{equation}
The key observation is that \eqref{eq:LorentzLH:tdepHam_gen} can be interpreted as a $t$-dependent oscillator Hamiltonian on the {\it $(1+1)$-dimensional Minkowskian spacetime} $\mathbf{M}^{1+1}$, which consists of $\mathbb{R}^{2}$ equipped with the Lorentzian metric given by
\begin{equation}
  g:= \dd \xi_{1} \otimes \dd \xi_{1} - \dd \xi_{2} \otimes \dd \xi_{2}.
  \label{eq:LorentzLH:Minkowskian_metric}
\end{equation}
More precisely, $-2 h_{3} = \frac{1}{2}\big(p_{\xi_{1}}^{2} - p_{\xi_{2}}^{2}\big)$ is the kinetic energy derived from the free Lagrangian induced by the metric \eqref{eq:LorentzLH:Minkowskian_metric}, while $- \frac{1}{2} h_{1} = \frac{1}{2}\big(\xi_{1}^{2} - \xi_{2}^{2}\big)$ is the potential of the isotropic oscillator on $\mathbf{M}^{1+1}$ and $2h_{5} = \xi_{1} p_{\xi_{2}}- \xi_{2}p_{\xi_{1}}$ is the angular momentum.

\begin{remark}
  On $\mathbf{M}^{1+1}$ let us consider the so-called \textit{light-cone} coordinates
  \begin{equation}
    \xi_{+}:= \xi_{1} + \xi_{2}, \qquad \xi_{-} := \xi_{1} - \xi_{2},
    \label{eq:LorentzLH:lightcone}
  \end{equation}
  in which the Lorentzian metric \eqref{eq:LorentzLH:Minkowskian_metric} of $\mathbf{M}^{1+1}$ takes the form
  \begin{equation}
    g =\frac{1}{2}(\dd \xi_{+} \otimes \dd \xi_{-} + \dd \xi_{-} \otimes \dd \xi_{+}).
    \nonumber
  \end{equation}
  If $p_{\xi_\pm}$ denotes the conjugate momentum of $\xi_{\pm}$, a canonical transformation on $\mathrm{T}^{*} \mathbf{M}^{1+1}$ is given by \eqref{eq:LorentzLH:lightcone} together with
  \begin{equation*}
    p_{\xi_+} = \frac{1}{2}(p_{\xi_{1}} + p_{\xi_{2}}), \qquad p_{\xi_-} = \frac{1}{2}(p_{\xi_{1}}- p_{\xi_{2}}).
  \end{equation*}
  The kinetic energy and the potential of the isotropic oscillator on $\mathbf{M}^{1+1}$ now read as $-2h_{3} = 2 p_{\xi_+}p_{\xi_-}$ and $- \frac{1}{2}h_{1} = \frac{1}{2} \xi_{+} \xi_{-}$, respectively, while the expression of angular momentum is $2h_{5} =  \xi_{-}p_{\xi_+} -\xi_{+}p_{\xi_-}$. Regarding the $\mathfrak{so}(1,3)$-LH system (\ref{eq:LorentzLH:tdepHam_gen}) on $\mathrm{T}^{*}\mathbf{M}^{1+1}$, this shows that, when working with light-cone coordinates, the $t$-dependent Hamiltonian becomes
  \begin{equation}
    h = - {b_{3}(t)} p_{\xi_+}p_{\xi_-}  - b_{1}(t)  \xi_{+} \xi_{-}+ \frac{b_{5}(t)}{2}\,(   \xi_{-}p_{\xi_+} -\xi_{+}p_{\xi_-}).
    \nonumber
  \end{equation}
\end{remark}

Among others, some specific applications of \eqref{eq:LorentzLH:tdepHam_gen} obtained in \cite{CampoamorStursberg2025} were the following:
\begin{itemize}
  \item {\it Coupled systems}. Consider $\mathrm{T}^{*} \mathbf{M}^{1+1}$ as the direct product $\mathrm{T}^{*} \mathbf{M}^{1+1} = \mathrm{T}^{*}\mathbb{R} \times \mathrm{T}^{*}\mathbb{R}$, with projections onto its factors $\mathrm{pr}_{i}: \mathrm{T}^{*}\mathbf{M}^{1+1} \to \mathrm{T}^{*}\mathbb{R}$  given by $\mathrm{pr}_{i}(\bm{\xi}, \bm{p}_{\xi}) := (\xi_{i}, p_{\xi_{i}})$ $(1 \leq i \leq 2)$. Then, the $t$-dependent Hamiltonian \eqref{eq:LorentzLH:tdepHam_gen} on $\mathrm{T}^{*} \mathbf{M}^{1+1}$ can be expressed as a coupling of two systems on $\mathrm{T}^{*} \mathbb{R}$ through a $t$-dependent angular momentum term in the form
 \begin{equation*}
h = \mathrm{pr}_{1}^{*} (h') - \mathrm{pr}_{2}^{*}(h') + \frac{b_{5}(t)}{2}\, (\xi_{1}p_{\xi_{2}}- \xi_{2}p_{\xi_{1}}),
 \end{equation*}
 where $h'$ is the $t$-dependent Hamiltonian on $\mathrm{T}^{*}\mathbb{R}$ defined, in   canonical coordinates $(\xi,p_\xi)$ of $\mathrm{T}^{*}\mathbb{R}$, by
\begin{equation}
  h'= -\frac{b_{3}(t)}{4}\,p_\xi^{2} - b_{1}(t) \xi^{2}.
 \label{eq:LorentzLH:1D}
  \end{equation}

  \item {\it A $t$-dependent Bateman Hamiltonian}. Consider positive functions $m(t), k(t), \gamma(t) \in C^{\infty}(\mathbb{R})$ such that $4 m(t) k(t) > \gamma^{2}(t)$ for all $t \in \mathbb{R}$. With the choices
  \begin{equation}
 \begin{split}
 b_{1}(t)  & = - \frac{1}{2}m(t) \Omega^{2}(t), \qquad b_{3}(t) = - \frac{2}{m(t)}, \qquad b_{5}(t) = \frac{\gamma(t)}{m(t)}, \\[2pt]
  \Omega(t) & := \sqrt{\frac{1}{m(t)} \left(k(t) - \frac{\gamma^{2}(t)}{4m(t)} \right)},
 \end{split}
 \label{bBateman}
\end{equation}
 the Hamiltonian  obtained from \eqref{eq:LorentzLH:tdepHam_gen} becomes
 \begin{equation}
 \begin{split}
 h^{\mathrm{B}} & = \left( \frac{1}{2 m(t)} \, p_{\xi_{1}}^{2} + \frac{1}{2} m(t) \Omega^{2}(t) \xi_{1}^{2} \right) -  \left( \frac{1}{2 m(t)} \, p_{\xi_{2}}^{2} + \frac{1}{2} m(t) \Omega^{2}(t) \xi_{2}^{2} \right) \\[2pt]
  & \qquad + \frac{\gamma(t)}{2m(t)}\, (\xi_{1}p_{\xi_{2}} - \xi_{2}p_{\xi_{1}}).
 \end{split}
 \label{eq:LorentzLH:Bateman}
 \end{equation}
 The Hamiltonian $h^{\mathrm{B}}$ was referred to as a the {\it $t$-dependent Bateman Hamiltonian} in \cite{CampoamorStursberg2025}, as  it is the Hamiltonian analogue of the generalisation proposed in \cite{CampoamorStursberg2017} for the widely studied Bateman Lagrangian \cite{Bateman1931,Blasone1996,Deguchi2019}. In this context, the functions  $m(t)$, $k(t)$, $\gamma(t)$ and $\Omega(t)$ are interpreted as a $t$-dependent mass,  spring constant,   damping constant   and  frequency, respectively.

  \item  {\it A  generalised coupled Caldirola--Kanai Hamiltonian}.  As before, let $m(t), k(t), \gamma(t) \in C^{\infty}(\mathbb{R})$ be positive functions, and define
 \begin{equation}
 \lambda(t):= \frac{\gamma(t)}{m(t)}, \qquad \Omega(t):= \sqrt{\frac{k(t)}{m(t)}}, \qquad t \in \mathbb{R}.
 \label{eq:LorentzLH:CK}
 \end{equation}
  With the choices
 \begin{equation}
 b_{1}(t) = - \frac{1}{2}m(t) \Omega^{2}(t)\, \mathrm{e}^{2\! \int_{0}^{t} \lambda(s) \, \dd s}, \qquad b_{3}(t) = -\frac{2}{m(t)}\, \mathrm{e}^{-2\! \int_{0}^{t} \lambda(s) \, \dd s},
 \label{eq:LorentzLH:CKb}
 \end{equation}
 the Hamiltonian   obtained from  \eqref{eq:LorentzLH:tdepHam_gen} reads
 \begin{equation}
 \begin{split}
h^{2\mathrm{CK}} & = \left( \frac{1}{2m(t)} \,\mathrm{e}^{- 2\! \int_{0}^{t} \lambda(s)  \dd s}\, p_{\xi_{1}}^{2} + \frac{1}{2} m(t) \Omega^{2}(t) \,\mathrm{e}^{2\! \int_{0}^{t} \lambda(s)  \dd s}\, \xi_{1}^{2} \right)                                                                         \\[2pt]
& \quad  - \left( \frac{1}{2m(t)} \,\mathrm{e}^{- 2\! \int_{0}^{t} \lambda(s) \dd s}\, p_{\xi_{2}}^{2} + \frac{1}{2} m(t) \Omega^{2}(t) \,\mathrm{e}^{2\! \int_{0}^{t} \lambda(s)  \dd s}\, \xi_{2}^{2} \right) +  \frac{b_{5}(t)}{2}\,(\xi_{1}p_{\xi_{2}} - \xi_{2}p_{\xi_{1}}).
\end{split}
 \label{eq:LorentzLH:2CK}
 \end{equation}
 In \cite{CampoamorStursberg2025},  this was named a {\it generalised coupled Caldirola--Kanai Hamiltonian}, since the corresponding Hamiltonian \eqref{eq:LorentzLH:1D} on $\mathrm{T}^{*}\mathbb{R}$,
\begin{equation*}
h' = \frac{1}{2m(t)} \, \mathrm{e}^{- 2\!  \int_{0}^{t} \lambda(s)  \dd s}\,p_\xi^{2} + \frac{1}{2} m(t) \Omega^{2}(t) \,\mathrm{e}^{2\!  \int_{0}^{t} \lambda(s)  \dd s}\, \xi^{2},
 \end{equation*}
  is the well-known Caldirola--Kanai Hamiltonian \cite{Caldirola1941,Kanai1948,Maamache2000}, which describes the motion of a particle of $t$-dependent mass $m(t)$ attached to a spring with a $t$-dependent Hooke `constant' $k(t) = m(t) \Omega^{2}(t)$ and subjected to a frictional force with a $t$-dependent friction `constant' $\gamma(t) = m(t) \lambda(t)$, being  $\Omega(t)$  the frequency.
\end{itemize}


\section{A non-standard quantum Lorentz algebra}
\label{section:nonstandardLorentz}

In this section we address the particular non-standard quantum deformation of $\mathfrak{so}(1,3)$, originally investigated in \cite{Herranz2000} from an algebraic perspective, and examine it from a geometric viewpoint.   We point out that all quantum deformations of the  Lorentz algebra have been classified in~\cite{Zakrzewski1994lorentz}, which in a kinematical basis~\cite{Bacry1968} can be found in~\cite{BGH2022}. The specific deformation of the Lorentz algebra     considered throughout this paper stems from the so-called non-standard (triangular or Jordanian) quantum deformation of $\mathfrak{sl}(2,\mathbb{R})$~\cite{Ohn1992,BHOS1995,Ballesteros1996,Shariati}.  The main point is that this deformation preserves the non-standard quantum $\mathfrak{sl}(2,\mathbb{R})$ algebra as a (deformed) Hopf subalgebra, which allows to extend previously known results (see~\cite{Ballesteros2018}) to  higher dimensions.
 
We consider the generators $\set{H, D, C_{1}, C_{2}, K, P}$ of $\mathfrak{so}(1,3)$  as global coordinates on $\mathfrak{so}(1,3)^{*}$, as usual. With respect to  these coordinates, the Kirillov--Konstant--Souriau (KKS) Poisson bracket $\set{\cdot, \cdot}$~\cite{Va_94,KKSref} on $C^{\infty}(\mathfrak{so}(1,3)^{*})$
reads  (see (\ref{eq.LorentzLH:confbasis}))
\begin{equation}
  \begin{aligned}
     & \set{K, H} = -P, \qquad      &      & \set{K, P} = H, \qquad      &    & \set{D, H} = -H , \qquad    &  & \set{D, P} = -P ,  
     \\[2pt]
     & \set{K, C_{1}} = -C_{2}  , \qquad &    & \set{K, C_{2}} = C_{1}, \qquad &    & \set{D, C_{1}} = C_{1}, \qquad &                      & \set{D,C_{2}} = C_{2} , \\[2pt]
     & \set{H, C_{1}} = 2D, \qquad      &     & \set{H, C_{2}} = -2K , \qquad &    & \set{P, C_{1}} = 2K, \qquad &                                & \set{P, C_{2}} = 2D.
  \end{aligned}
  \label{eq:nonstandardLorentz:cr}
\end{equation}
Henceforth, we will consider  $C^{\infty}(\mathfrak{so}(1,3)^{*})$ endowed  with its usual (well-behaved) Poisson coalgebra structure (see \cite{BBHMR2009,Bonneau1994,Bidegain1996,CampoamorStursberg2025a} and references therein).
For our purposes, it suffices to observe that the coproduct $\Delta: C^{\infty}(\mathfrak{so}(1,3)^{*}) \to C^{\infty}(\mathfrak{so}(1,3)^{*} \times \mathfrak{so}(1,3)^{*})$, defined as the pull-back of the vector addition of $\mathfrak{so}(1,3)^{*}$,
\begin{equation}
  \Delta(f) (\theta, \theta') = f(\theta+ \theta'), \qquad f \in C^{\infty}(\mathfrak{so}(1,3)^{*}), \qquad \theta, \theta' \in \mathfrak{so}(1,3)^{*},
  \label{eq:nonstandardLorentz:prim_coprod}
\end{equation}
is a morphism of Poisson algebras, where  $\mathfrak{so}(1,3)^{*} \times \mathfrak{so}(1,3)^{*}$ is endowed with the product Poisson structure.
It is worth noting that the coordinate functions on $\mathfrak{so}(1,3)^{*}$ are primitive elements for the coproduct $\Delta$, which verifies the coassociativity condition~\cite{Chari1995,Abe,Majid1995}:
\begin{equation}
  ({\rm Id} \otimes \Delta) \circ \Delta =(\Delta \otimes {\rm Id}) \circ \Delta,
  \label{coass}
\end{equation}
where ${\rm Id}$ denotes the identity morphism $\mathrm{Id} : C^{\infty}(\mathfrak{so}(1,3)^{*}) \to C^{\infty}(\mathfrak{so}(1,3)^{*})$.
Furthermore, the Casimir invariants of $\mathfrak{so}(1,3)$, given in \eqref{eq:LorentzLH:CasimirKin2} and now regarded as functions on $\mathfrak{so}(1,3)^{*}$, are Casimir functions of the KKS Poisson structure \eqref{eq:nonstandardLorentz:cr}.

We now derive the dual Poisson--Lie group associated with the non-standard quantum deformation of the Lorentz algebra $\mathfrak{so}(1,3)$ studied in \cite{Herranz2000}. Starting from  the classical $r$-matrix underlying the deformation, we determine the induced Lie bialgebra structure and construct the corresponding dual Poisson--Lie group. This viewpoint will allow us to describe explicitly the deformed Poisson brackets, coproducts and Casimir functions.

First of all, let us consider the Lie subalgebra $\langle D, H \rangle$ of $\mathfrak{so}(1, 3)$, given by $[D, H] = - H$, isomorphic to the so-called `book' algebra $\mathfrak{b}_{2}$. Recall that $\mathfrak{b}_{2}$ can also be identified with the Borel subalgebra of $\mathfrak{sl}(2, \mathbb{R})$. The well-known classical $r$-matrix for $\mathfrak{b}_{2}$, which is a solution of the classical Yang--Baxter equation, reads \cite{Drinfeld1986,Ogievetsky1994}
\begin{equation}
  r:= z D \wedge H,
  \label{eq:nonstandardLorentz:r}
\end{equation}
where $z$ is considered as a real (non-zero) quantum deformation parameter such that $q = \mathrm{e}^{z}$.  This classical $r$-matrix has been the initial point to construct non-standard quantum deformations of several Lie algebras such as $\mathfrak{sl}(2, \mathbb{R})$, the oscillator algebra $\mathfrak{h}_{4}$, the (centrally extended) Schr{\"o}dinger algebra $\widehat{\mathcal{S}}(1)$ in $(1+1)$ dimensions, graded contractions of $\mathfrak{so}(2,2)$, as well as some relevant conformal algebras (see \cite{Herranz2000,Ohn1992,BHOS1995,Ballesteros1996,Shariati,Negro1997,Ballesteros2000,Herranz1997,Herranz2000a} and references therein).

In particular, the classical $r$-matrix \eqref{eq:nonstandardLorentz:r} endows $\mathfrak{so}(1,3)$ with a Lie bialgebra structure through the cocommutator
\begin{equation}
  \delta: \mathfrak{so}(1,3) \ni X \longmapsto [X \otimes 1 + 1 \otimes X, r] \in  \mathfrak{so}(1,3) \otimes \mathfrak{so}(1,3)
  \label{coco}
\end{equation}
which, expressed in terms of the conformal basis \eqref{eq:LorentzLH:conformalbasis},  takes the form
\begin{equation}
  \begin{aligned}
     & \delta(H) = 0, \qquad     &  & \delta(D) = -z  D \wedge H, \qquad   &  & \delta(P) = z  P \wedge H,     \\[2pt]
     & \delta(K) = -z  D \wedge P, \qquad &  & \delta(C_{1}) = -z  C_{1} \wedge H,  \qquad &  & \delta(C_{2}) = -z C_{2} \wedge H + 2z D \wedge K.
  \end{aligned}
  \label{cocobook}
\end{equation}
Thus, the transpose map $^{t}\delta: \mathfrak{so}(1,3)^{*} \otimes \mathfrak{so}(1,3)^{*} \to \mathfrak{so}(1,3)^{*}$ of $\delta$ defines a Lie algebra structure on the dual vector space $\mathfrak{so}(1,3)^{*}$ of $\mathfrak{so}(1,3)$ on which  the transpose map $^{t} \nu$ of the structural tensor of $\nu$ of $\mathfrak{so}(1,3)$, induces a Lie bialgebra structure. With respect to the basis $\set{e^{1}, \ldots, e^{6}}$ of $\mathfrak{so}(1,3)^{*}$ dual to the basis $\set{H, D, C_{1}, C_{2}, K, P}$ of $\mathfrak{so}(1,3)$, the Lie bracket $^{t}\delta$ has   non-zero commutation relations given by
\begin{equation}
  \begin{aligned}
     & \big[e^{1}, e^{2}\big] = z  e^{2},\qquad  &  & \big [e^{1}, e^{3}\big] = z e^{3}, \qquad &  & \big[e^{1}, e^{4}\big] = z e^{4},    \\[2pt]
     & \big[e^{1}, e^{6}\big] = -z e^{6}, \qquad &  & \big[e^{2}, e^{6}\big] = - ze^{5}, \qquad &  & \big[e^{2}, e^{5}\big] = 2 z  e^{4}.
  \end{aligned}
  \label{eq:nonstandardLorentz:dualLorentzalg}
\end{equation}
A routine inspection shows that $(\mathfrak{so}(1,3)^{*},\, ^{t}\delta)$ is a solvable Lie algebra. Moreover, the basis $\set{e^{1}, \ldots, e^{6}}$ of $\mathfrak{so}(1,3)^{*}$ leads to an ascending chain $ \mathfrak{a}_{1} \subset \cdots \subset \mathfrak{a}_{6} = \mathfrak{so}(1,3)^{*}$ with   $\mathfrak{a}_{i} := \big\langle e^{7-i}, \ldots, e^{6} \big\rangle$  $(1 \leq i \leq 6)$. These are Lie subalgebras of $(\mathfrak{so}(1,3)^{*}, \,^{t}\delta)$ such that  $\mathfrak{a}_{i}$ is an ideal of $\mathfrak{a}_{i+1}$ $(1 \leq i \leq 5)$.  Hence, if
$G_{z}^{*}$ is the 1-connected Lie  group whose Lie algebra is isomorphic to $(\mathfrak{so}(1,3)^{*}, \,^{t}\delta)$, the map
\begin{equation}
  \begin{aligned}
    \mathbb{R}^{6}    & \longrightarrow G_{z}^{*}     \\
    (H, D, C_{1}, C_{2}, K, P) & \longmapsto   \mathrm{exp} \big(P e^{6}\big) \mathrm{exp} \big(K e^{5}\big) \mathrm{exp} \big(C_{2} e^{4}\big) \mathrm{exp} \big(C_{1} e^{3}\big) \mathrm{exp} \big(D e^{2}\big)\mathrm{exp} \big(H e^{1}\big)
  \end{aligned}
  \label{eq:nonstandardLorentz:diffeo}
\end{equation}
is a diffeomorphism \cite[Theorem 3.18.11]{Varadarajan1984}, where $\mathrm{exp}: \mathfrak{so}(1,3)^{*} \to G_{z}^{*}$ is the exponential map associated with the Lie algebra $(\mathfrak{so}(1,3)^{*},\,^{t}\delta)$. This shows that the generators of $\mathfrak{so}(1,3)$ can be considered as global coordinates on $G_{z}^{*}$ in a natural way.
Now, according to the Drinfel'd theorem (see \cite[Theorem 2.2]{Etingof2010}) there exists a unique Poisson--Lie group structure $\Pi_{z}$ on $G_{z}^{*}$ such that $(\mathfrak{so}(1,3)^{*}, \,^{t}\nu)$ is its tangent Lie bialgebra. The latter entails that $\Pi_{z} \in \Gamma ( \wedge^{2} \mathrm{T} G_{z}^{*}  )$ is a multiplicative Poisson bivector on $G_{z}^{*}$ such that $^{t} \nu = \mathrm{T}_{e} \Pi_{z}^{R}$, where $\Pi_{z}^{R}: G_{z}^{*} \to \mathfrak{so}(1,3)^{*} \otimes \mathfrak{so}(1,3)^{*}$ denotes the right translate of $\Pi_{z}$ to the neutral element $e\in G_{z}^{*}$ (see \cite{Etingof2010,Lu1990,KosmannSchwarzbach2004} for more details).

The multiplication law of $G_{z}^{^*}$, $\mu: G_{z}^{*} \times G_{z}^{*} \ni (a, a') \mapsto aa' \in G_{z}^{*}$, takes the explicit form
\begin{multline}
  \mu(a, a')  = \left(H + H', D \mathrm{e}^{-z H'} + D', C_{1} \mathrm{e}^{-zH'} + C_{1}' ,\right. \\
  \left.  C_{2} \mathrm{e}^{-z H'} + C_{2}' + 2z D \mathrm{e}^{-zH'}K' - z^{2} D^{2} \mathrm{e}^{-2z H'}P', \right. \\
  \left.
  K + K' -z D \mathrm{e}^{-zH'}P', P \mathrm{e}^{zH'} + P'\right),
  \label{eq:nonstandardLorentz:mult}
\end{multline}
where $a = (H, D, C_{1}, C_{2}, K, P)$ and $a' = (H', D', C_{1}', C_{2}', K', P')$ are arbitrary elements of $G_{z}^{*}$.
Within this context, $(G_{z}^{*}, \Pi_{z})$ is referred to as the {\it dual Poisson--Lie group}  associated with the Lie bialgebra $(\mathfrak{so}(1,3), \delta)$ (cf.~\cite{Etingof2010,Lu1990,KosmannSchwarzbach2004}). The Poisson structure $\Pi_{z}$ is determined by the following non-zero fundamental brackets:
\begin{equation}
  \begin{aligned}
     & \set{K, H}_{\Pi_{z}} = - \mathrm{e}^{-z H} P, \qquad &  & \set{K,P}_{\Pi_{z}} = \frac{\mathrm{e}^{zH} -1}{z},\qquad &  & \set{D, H}_{\Pi_{z}} = - \frac{1 - \mathrm{e}^{-zH}}{z},   \\[2pt]
     & \set{K, C_{1}}_{\Pi_{z}} = - C_{2}, \qquad       &  & \set{K, C_{2}}_{\Pi_{z}} = C_{1} - z D^{2},  \qquad       &  & \set{D, P}_{\Pi_{z}} = -P,     & \\[2pt]
     & \set{H, C_{1}}_{\Pi_{z}} = 2 D, \qquad    &  & \set{H, C_{2}}_{\Pi_{z}} = -2K \mathrm{e}^{-zH},  \qquad  &  & \set{D, C_{1}}_{\Pi_{z}} = C_{1} - z D^{2},      \\[2pt]
     & \set{P, C_{1}}_{\Pi_{z}} = 2K + 2z DP,\qquad         &  & \set{P, C_{2}}_{\Pi_{z}} = 2D, \qquad     &  & \set{D, C_{2}}_{\Pi_{z}} = C_{2},      \\[2pt]
     & \set{C_{1}, C_{2}}_{\Pi_{z}} = 2zD C_{2}.     &  &       &  &
  \end{aligned}
  \label{eq:nonstandardLorentz:defcr}
\end{equation}
These commutation relations  indeed reproduce those of the Poisson analogue of the quantum deformation $\mathcal{U}_{z}(\mathfrak{so}(1,3))$ of the universal enveloping algebra $\mathcal{U}(\mathfrak{so}(1,3))$ of $\mathfrak{so}(1,3)$ obtained in \cite{Herranz2000}.
\begin{remark}
  From the  commutation relations  \eqref{eq:nonstandardLorentz:dualLorentzalg}, it readily follows that  the Lie algebra $(\mathfrak{so}(1,3)^{*}, \,^{t}\delta)$ possesses a trivial centre. Hence, the associated adjoint representation $\mathrm{ad}: \mathfrak{so}(1,3)^{*} \to \mathfrak{gl}(6, \mathbb{R})$ is a faithful representation. Through matrix exponentiation, one thereby obtains a faithful representation of the Lie group $G_{z}^{*}$ on $\mathbb{R}^{6}$, which, together with the diffeomorphism  \eqref{eq:nonstandardLorentz:diffeo}, shows that $G_{z}^{*}$ is a matrix Lie group. From this, one easily obtains the expression \eqref{eq:nonstandardLorentz:mult} of the multiplication law of $G_{z}^{*}$ expressed within the global coordinates $(H, D, C_{1}, C_{2}, K, P)$ of $G_{z}^{*}$. A straightforward computation shows that the Poisson bivector $\Pi_{z}$, defined by the brackets \eqref{eq:nonstandardLorentz:defcr}, is indeed multiplicative.
\end{remark}
The Poisson structure $\Pi_{z}$ on $G_{z}^{*}$ admits two independent Casimir functions given by
\begin{equation}
  \begin{split}
     & \mathscr{C}_{z,2} := K^{2} - \mathrm{e}^{zH} D^{2} + \left(\frac{\mathrm{e}^{zH} -1}{z} \right)C_{1} + PC_{2},                \\[2pt]
     & \mathscr{C}_{z,2}' := KD + \frac{1}{2} \left( \left( \frac{\mathrm{e}^{zH} -1}{z} \right) C_{2} - P C_{1} + z PD^{2} \right).
    \label{eq:nonstandardLorentz:defCas}
  \end{split}
\end{equation}
We also consider $C^{\infty}(G_{z}^{*})$ equipped with its usual (well-behaved) Poisson algebra structure. As in the case of $\mathfrak{so}(1,3)^{*}$, the multiplication law $\mu: G_{z}^{*} \times G_{z}^{*} \to G_{z}^{*}$ of  $G_{z}^{*}$ induces a coproduct $\Delta_{z}: C^{\infty}(G_{z}^{*}) \to  C^{\infty}(G_{z}^{*} \times G_{z}^{*})$ through its pull-back, namely
\begin{equation*}
  \Delta_{z}(f)(a, a')  := f(\mu(a,a')), \qquad f\in C^{\infty}(G_{z}^{*}), \qquad a, a' \in G_{z}^{*}.
\end{equation*}
The expressions of the coproduct $\Delta_{z}$ for the coordinate functions of $G_{z}^{*}$ stem directly from  \eqref{eq:nonstandardLorentz:mult} and read as follows:
\begin{equation}
  \begin{split}
    \Delta_{z}(H)     & = H \otimes 1 + 1 \otimes H,      \\[2pt]
    \Delta_{z}(D)     & = D \otimes \mathrm{e}^{-z H} + 1 \otimes D,      \\[2pt]
    \Delta_{z}(C_{1}) & = C_{1} \otimes \mathrm{e}^{-zH} + 1 \otimes C_{1},    \\[2pt]
    \Delta_{z}(C_{2}) & = C_{2} \otimes \mathrm{e}^{-zH} + 1 \otimes C_{2} + 2z D \otimes \mathrm{e}^{-zH }K - z^{2}D^{2} \otimes \mathrm{e}^{-2zH}P, \\[2pt]
    \Delta_{z}(K)     & = K \otimes 1 + 1 \otimes K - z D \otimes \mathrm{e}^{-zH}P,    \\[2pt]
    \Delta_{z}(P)     & = P \otimes \mathrm{e}^{zH} + 1 \otimes P,
  \end{split}
  \label{eq:nonstandardLorentz:defcoprod}
\end{equation}
which satisfies the coassociativity condition (\ref{coass}).
Note that, since the multiplication law $\mu: G_{z}^{*} \times G_{z}^{*} \to G_{z}^{*}$ is a Poisson map, the coproduct $\Delta_{z}$ is a morphism of Poisson algebras, where $G_{z}^{*}\times G_{z}^{*}$ is equipped with the product Poisson structure. In addition, observe that  the
coordinate functions   $(D,H)\subset (D,P,H)\subset (D,K,P,H)\subset G_{z}^{*}$ generate a sequence of PH subalgebra deformations, such that $(D,K,P,H)$ determines a PH  deformation of the Weyl--Euclidean algebra, while $(D,H,C_1)$ give rise  to a deformation of $\mathfrak{sl}(2, \mathbb{R})$.

As expected, when the undeformed limit $z \to 0$ is applied  to the  commutation relations \eqref{eq:nonstandardLorentz:defcr} and the coproduct \eqref{eq:nonstandardLorentz:defcoprod}, we recover the commutation relations  \eqref{eq:nonstandardLorentz:cr} and the primitive coproduct \eqref{eq:nonstandardLorentz:prim_coprod} of the Poisson coalgebra structure of $C^{\infty}(\mathfrak{so}(1,3)^{*})$. This follows from the fact that the expressions for the Poisson bracket and the coproduct of $G_{z}^{*}$ coincide, in a Poisson setting, with those of the quantum deformation $\mathcal{U}_{z}(\mathfrak{so}(1,3))$ of $\mathcal{U}(\mathfrak{so}(1,3))$  constructed in \cite{Herranz2000}. Note also that the Casimir functions \eqref{eq:nonstandardLorentz:defCas} of $G_{z}^{*}$ reduce, when $z \to 0$, to the Poisson counterpart of the Casimir functions \eqref{eq:LorentzLH:CasimirKin2} of the Lorentz Lie algebra.


\subsection{On a twist map and its Poisson geometry}

Let us now show how the Poisson structures of $G_{z}^{*}$ and $\mathfrak{so}(1,3)^{*}$ can be (locally) related.
Consider the open subset $U_{z}:= \set{z H + 1 > 0} \subset \mathfrak{so}(1,3)^{*}$, and let $\Pi_{\mathfrak{so}(1,3)^{*}} \vert_{U_{z}}$ be the restriction of the Poisson bivector $\Pi_{\mathfrak{so}(1,3)^{*}}$ associated with the KKS Poisson structure on $\mathfrak{so}(1,3)^{*}$ to $U_{z}$, so that $\big(U_{z}, \Pi_{\mathfrak{so}(1,3)^{*}} \vert_{U_{z}} \big)$ is a Poisson submanifold of $\big(\mathfrak{so}(1,3)^{*}, \Pi_{\mathfrak{so}(1,3)^{*}} \big)$.  The diffeomorphism $\varphi_{z}:   U_{z} \to G_{z}^{*}$ given by
\begin{equation}
  \varphi_{z}(H, D, C_{1}, C_{2}, K, P) := \left( \frac{\log(zH + 1)}{z}, D, C_{1} + z D^{2}, C_{2}, K, P \right)
  \label{eq:nonstandardLorentz:twistmap}
\end{equation}
is a Poisson map; that is, $(\varphi_{z})_{*} \Pi_{\mathfrak{so}(1,3)^{*}} \vert_{U_{z}} = \Pi_{z}$. Its inverse, $\varphi_{z}^{-1}: G_{z}^{*} \to U_{z}$, is also a Poisson map, and it reads
\[ \varphi_{z}^{-1}(H, D, C_{1}, C_{2}, K, P) = \left( \frac{\mathrm{e}^{zH}-1}{z}, D, C_{1} - zD^{2}, C_{2}, K, P\right). \]
Notably, the expression of $\varphi_{z}^{-1}$ coincides with the twist map of $\mathcal{U}_{z}(\mathfrak{so}(1,3))$  obtained in \cite{Herranz2000}.
In particular, if we define
\begin{equation}
  \begin{aligned}
     & \widetilde{H} :=  \frac{\mathrm{e}^{zH}-1}{z},  \qquad &  & \widetilde{D} := D,  \qquad &  & \widetilde{C}_{1} := C_{1} - zD^{2}, \\
     & \widetilde{C}_{2} :=  C_{2}, \qquad      &  & \widetilde{K} := K, \qquad  &  & \widetilde{P} := P,
  \end{aligned}
  \label{eq:nonstandardLorentz:inv_twist}
\end{equation}
we have that $\big(\widetilde{H}, \widetilde{D}, \widetilde{C}_{1}, \widetilde{C}_{2}, \widetilde{K}, \widetilde{P} \big)$ are global coordinates on $G_{z}^{*}$ fulfilling the undeformed commutation relations \eqref{eq:nonstandardLorentz:cr} relative to the Poisson bracket $\set{\cdot, \cdot}_{\Pi_{z}}$  on $C^{\infty}(G_{z}^{*})$. Hence, the map $\varphi_{z}$ relates the deformed  commutation relations  \eqref{eq:nonstandardLorentz:defcr} with the undeformed commutation relations  \eqref{eq:nonstandardLorentz:cr}.

We stress that, although the functions \eqref{eq:nonstandardLorentz:inv_twist} satisfy the undeformed commutation relations \eqref{eq:nonstandardLorentz:cr}, they nevertheless remain as non-primitive elements for the coproduct $\Delta_{z}$. More precisely,
\begin{equation*}
  \begin{split}
     & \Delta_{z}\big(\widetilde{H}\big) = \widetilde{H} \otimes 1 + 1 \otimes \widetilde{H} + z \widetilde{H} \otimes \widetilde{H},      \\
& \Delta_{z}\big(\widetilde{D}\big) = \widetilde{D} \otimes \frac{1}{1 + z \widetilde{H}} + 1 \otimes \widetilde{D},  \\
     & \Delta_{z}\big(\widetilde{C}_{1}\big)  = \widetilde{C}_{1} \otimes \frac{1}{1 + z \widetilde{H}} + 1 \otimes \widetilde{C}_{1} - 2z \widetilde{D} \otimes \frac{\widetilde{D}}{1 + z \widetilde{H}} + z^{2} \widetilde{D}^{2} \otimes \frac{\widetilde{H}}{\big(1 +z \widetilde{H}\big)^{2}}, \\
 &  \Delta_{z}\big(\widetilde{C}_{2}\big) = \widetilde{C}_{2} \otimes \frac{1}{1 + z \widetilde{H}} + 1 \otimes \widetilde{C}_{2} + 2 z \widetilde{D} \otimes \frac{\widetilde{K}}{1 + z \widetilde{H} }  - z^{2} \widetilde{D}^{2} \otimes \frac{\widetilde{P}}{\big(1 + z \widetilde{H}\big)^{2}},    \\
 & \Delta_{z}\big(\widetilde{K}\big) = \widetilde{K} \otimes 1 + 1 \otimes \widetilde{K} - z \widetilde{D} \otimes \frac{\widetilde{P}}{1 + z \widetilde{H}},                           \\
& \Delta_{z}\big(\widetilde{P}\big) = \widetilde{P} \otimes 1 + 1 \otimes \widetilde{P} + z \widetilde{P} \otimes \widetilde{H}.
  \end{split}
\end{equation*}


\subsection{On the non-standard quantum deformation of $\mathfrak{sl}(2, \mathbb{R})$ and the book algebra}
\label{subsection42}

Let us now  describe how the non-standard quantum deformation of $\mathfrak{sl}(2, \mathbb{R})$ ~\cite{Ohn1992,BHOS1995,Ballesteros1996,Shariati} and that of the book algebra $\mathfrak{b}_{2}$ \cite{Drinfeld1986,Ogievetsky1994} can be recovered from our construction as  particular types of submanifolds of the dual Poisson--Lie group $(G_{z}^{*}, \Pi_{z})$.

To begin with,  consider the  $\mathfrak{sl}(2, \mathbb{R})$ sub-bialgebra of $(\mathfrak{so}(1,3), \delta)$ spanned by $H$, $D$ and  $C_{1}$, with underlying classical $r$-matrix (\ref{eq:nonstandardLorentz:r}), and commutation relations expressed in (\ref{eq.LorentzLH:confbasis}),
\begin{equation}
  [D, H] = -H, \qquad [D, C_{1}] = C_{1}, \qquad [H, C_{1}] = 2D,
  \label{eq:nonstandardLorentz:sl2_cr}
\end{equation}
and cocommutator given by (\ref{cocobook}),
\begin{equation}
  \delta (H) = 0, \qquad \delta(D) = -z D \wedge H, \qquad \delta(C_{1}) = -z C_{1} \wedge H.
  \label{eq:nonstandardLorentz:sl2_sub}
\end{equation}
Let   $S_{z}^{*}$ be the Lie subgroup of $G_{z}^{*}$ corresponding to the Lie subalgebra $\big\langle e^{1}, e^{2}, e^{3} \big\rangle$ of the  Lie algebra $(\mathfrak{so}(1,3)^{*},\, ^{t} \delta)$ shown in \eqref{eq:nonstandardLorentz:dualLorentzalg}. Thus, $S_{z}^{*}$ can be identified with the submanifold $S_{z}^{*} \equiv \set{C_{2} = K = P = 0}$ of $G_{z}^{*}$, with global coordinates $(H, D, C_{1})$. For each $a \in S_{z}^{*}$, let us denote by  $(\mathrm{T}_{a}S_{z}^{*})^{\perp \Pi_{z}}$ the $\Pi_{z}$-ortogonal of the tangent space $\mathrm{T}_{a}S_{z}^{*}$, namely
\[ (\mathrm{T}_{a}S_{z}^{*})^{\perp \Pi_{z}} := \sharp_{\Pi_{z,a}}\big((\mathrm{T}_{a}S_{z}^{*})^{\circ}\big), \]
where $(\mathrm{T}_{a}S_{z}^{*})^{\circ}$ is the annihilator of $\mathrm{T}_{a}S_{z}^{*
    } \subset \mathrm{T}_{a}G_{z}^{*}$ and $\sharp_{\Pi_{z}}: \mathrm{T}^{*}G_{z}^{*} \to \mathrm{T}G_{z}^{*}$ is the morphism of vector bundles determined by $\Pi_{z}$. We have that
\begin{equation}
  \mathrm{T}_{a}S_{z}^{*} \cap (\mathrm{T}_{a}S_{z}^{*})^{\perp \Pi_{z}} = \set{0}.
  \label{eq:nonstandardLorentz:PD_S}
\end{equation}
Therefore, we can define a bivector field $\Pi_{S_{z}^{*}} \in \Gamma(\wedge^{2} \mathrm{T} S_{z}^{*})$ by
\begin{equation}
  \Pi_{S_{z}^{*},a} (\zeta, \eta):= \Pi_{z,a} \big(\tilde{\zeta}, \tilde{\eta} \big), \qquad a \in S_{z}^{*}, \qquad \zeta, \eta \in \mathrm{T}_{a}^{*}S_{z}^{*},
  \label{eq:nonstandardLorentz:bivector_S}
\end{equation}
where $\tilde{\zeta}, \tilde{\eta} \in \big((\mathrm{T}_{a}S_{z}^{*})^{\perp \Pi_{z}} \big)^{\circ}$ are extensions of $\zeta$ and $\eta$, respectively, and  whose existence follows from \eqref{eq:nonstandardLorentz:PD_S}. A routine inspection shows that $\Pi_{S_{z}^{*}}$ is smooth and, consequently, $(S_{z}^{*}, \Pi_{S_{z}^{*}})$ is a Poisson manifold. In fact, $\Pi_{S_{z}^{*}}$ is determined by the following fundamental Poisson brackets coming from \eqref{eq:nonstandardLorentz:defcr}:
\begin{equation}
  \set{D, H}_{\Pi_{S_{z}^{*}}} = - \frac{1 - \mathrm{e}^{-zH}}{z}, \qquad \set{D, C_{1}}_{\Pi_{S_{z}^{*}}} = C_{1} - z D^{2}, \qquad \set{H, C_{1}}_{\Pi_{S_{z}^{*}}} = 2D.
  \label{eq:nonstandardLorentz:bookgroup}
\end{equation}
These are, precisely, the commutation relations of the non-standard quantum deformation of $\mathfrak{sl}(2,\mathbb{R})$~\cite{Ohn1992,BHOS1995,Ballesteros1996,Shariati}. Moreover, from the definition \eqref{eq:nonstandardLorentz:bivector_S} of $\Pi_{S_{z}^{*}}$ and the fact that $S_{z}^{*}$ is a Lie subgroup of $G_{z}^{*}$, it follows that $(S_{z}^{*}, \Pi_{S_{z}^{*}})$ is the 1-connected dual Poisson--Lie group associated with the $\mathfrak{sl}(2, \mathbb{R})$ sub-bialgebra of $(\mathfrak{so}(1,3), \delta)$ defined by \eqref{eq:nonstandardLorentz:sl2_cr} and  \eqref{eq:nonstandardLorentz:sl2_sub}.

Then, since $S_{z}^{*}$ is a Lie subgroup of $G_{z}^{*}$, the definition of $\Pi_{S_{z}}^{*}$ immediately implies that the coproduct $\Delta_{S_{z}^{*}}: C^{\infty}(S_{z}^{*}) \to C^{\infty}(S_{z}^{*} \times S_{z}^{*})$ of the (well-behaved) Poisson coalgebra structure of $C^{\infty}(S_{z}^{*})$  can be directly obtained from the coproduct \eqref{eq:nonstandardLorentz:defcoprod} of $C^{\infty}(G_{z}^{*})$ as
\begin{equation}
  \begin{split}
    \Delta_{S_{z}^{*}}(H)     & = H \otimes 1 + 1 \otimes H, \qquad \Delta_{S_{z}^{*}}(D) = D \otimes \mathrm{e}^{-zH} + 1 \otimes D , \\[2pt]
    \Delta_{S_{z}^{*}}(C_{1}) & = C_{1} \otimes \mathrm{e}^{-zH} + 1 \otimes C_{1} .
  \end{split}
  \label{eq:nonstandardLorentz:defcoprod:sl2}
\end{equation}
In this context,  the well-known Casimir function of the non-standard PH deformation of $\mathfrak{sl}(2, \mathbb{R})$ is given by
\begin{equation}
  \mathscr{V}_{z,2} := - \mathrm{e}^{zH} D^{2} + \left( \frac{\mathrm{e}^{zH} -1}{z} \right) C_{1},
  \label{eq:nonstandardLorentz:defCas:sl2}
\end{equation}
which is just the restriction of the Casimir function $\mathscr{C}_{z,2}$ of $G_{z}^{*}$, defined in \eqref{eq:nonstandardLorentz:defCas}, to the submanifold $S_{z}^{*} \equiv \set{C_{2} = K = P = 0} \subset G_{z}^{*}$. Note that the restriction of $\mathscr{C}_{z,2}'$ to $S_{z}^{*}$ vanishes identically.

Analogously, the non-standard quantum book algebra can be derived from the $\mathfrak{b}_{2}$ sub-bialgebra of $(\mathfrak{so}(1,3), \delta)$ determined by $[D, H] = - H$, together with $\delta(H) = 0$ and $\delta(D) = -z D \wedge H$.  The corresponding 1-connected  dual Poisson--Lie group is the Lie subgroup $B_{z}^{*}\subset S_{z}^{*} \subset G_{z}^{*}$ associated with the Lie subalgebra $\big\langle e^{1}, e^{2} \big\rangle$ of $(\mathfrak{so}(1,3)^{*}, ^{t} \delta)$. This subgroup is identified with the submanifold $B_{z}^{*} \equiv \set{C_{1} = C_{2} = K = P = 0}$ of $G_{z}^{*}$, with global coordinates $(H, D)$.  It should be noted  that all non-coboundary Poisson--Lie group structures on $B_{z}^{*}$ have been classified in \cite{Ballesteros2012}. As in the previous case, the condition
\[ \mathrm{T}_{a} B_{z}^{*} \cap (\mathrm{T}_{a}B_{z}^{*})^{\perp \Pi_{z}} = \set{0}, \qquad a \in B_{z}^{*}\]
is satisfied, and  the bivector field $\Pi_{B_{z}^{*}} \in \Gamma(\wedge^{2} \mathrm{T} B_{z}^{*})$ defined by
\[ \Pi_{B_{z}^{*},a}(\zeta, \eta):= \Pi_{z,a}\big(\tilde{\zeta}, \tilde{\eta} \big), \qquad a \in B_{z}^{*}, \qquad \zeta, \eta \in \mathrm{T}_{a}^{*}B_{z}^{*}, \]
where $\tilde{\zeta}, \tilde{\eta} \in \big((\mathrm{T}_{a}B_{z}^{*})^{\perp \Pi_{z}} \big)^{\circ}$ are extensions of $\zeta$ and $\eta$, respectively,
is smooth. The fundamental Poisson bracket determining $\Pi_{B_{z}^{*}}$, inherited from \eqref{eq:nonstandardLorentz:defcr}, reads
\begin{equation}
  \set{D, H}_{\Pi_{B_{z}^{*}}} = - \frac{1 - \mathrm{e}^{-zH}}{z},
  \label{eq:nonstandardLorentz:book_defcr}
\end{equation}
while the coproduct $\Delta_{B_{z}^{*}}: C^{\infty}(B_{z}^{*}) \to C^{\infty}(B_{z}^{*} \times B_{z}^{*})$ of the corresponding   Poisson coalgebra structure is
\begin{equation}
  \Delta_{B_{z}^{*}}(H) = H \otimes 1 + 1 \otimes H, \qquad \Delta_{B_{z}^{*}}(D) = D \otimes \mathrm{e}^{-zH} + 1 \otimes D.
  \nonumber
\end{equation}
This construction reproduces the non-standard PH  deformation of $\mathfrak{b}_{2}$ \cite{Drinfeld1986,Ogievetsky1994}.

\begin{remark}
  A straightforward inspection shows that  $(S_{z}^{*}, \Pi_{S_{z}^{*}})$ and $(B_{z}^{*}, \Pi_{B_{z}^{*}})$ are not immersed Poisson submanifolds of $(G_{z}^{*}, \Pi_{z})$, but rather  Poisson--Dirac submanifolds of $(G_{z}^{*}, \Pi_{z})$ (see \cite[Proposition 8.30]{Crainic2021}).
\end{remark}

\begin{remark}
  The non-standard (triangular or Jordanian) quantum deformation of $\mathfrak{sl}(2,\mathbb{R})$, corresponding to the classical $r$-matrix (\ref{eq:nonstandardLorentz:r}), is completely different from the
  well-known Drinfel'd--Jimbo    deformation~\cite{Drinfeld1986} from both the mathematical and physical viewpoints. In the conformal basis, the latter is provided by the standard (or quasi-triangular) classical $r$-matrix $r=z H\wedge C_1$, which is a solution of the modified classical Yang--Baxter equation. From (\ref{coco}) and (\ref{eq:nonstandardLorentz:sl2_cr}), it follows that the Lie bialgebra structure is determined by the cocommutator given by
  $$
    \delta(D) =0, \qquad \delta(H) = 2z  H \wedge D, \qquad  \delta(C_1) = 2z  C_1 \wedge D ,
  $$
  to be compared with (\ref{eq:nonstandardLorentz:sl2_sub}). Therefore, the primitive generator is now the dilation $D$, instead of the translation $H$. Notice that there is no   Drinfel'd--Jimbo  deformation for the
  book algebra.
\end{remark}


\section{(1+1)-dimensional (anti-)de Sitter spacetimes from a twist map}
\label{section:curvature}

In this section, we show how the Poisson map $\varphi_{z}: U_{z} \to G_{z}^{*}$, defined  in \eqref{eq:nonstandardLorentz:twistmap} and associated with the `twist' map $\varphi_{z}^{-1}: G_{z}^{*} \to U_{z}$, naturally leads, when combined with the morphism of Lie algebras $\phi: \mathfrak{so}(1,3) \to C^{\infty}(\mathrm{T}^{*} \mathbb{R}^{2})$ given in Proposition~\ref{prop:LorentzLH:morphism}, to the so-called $(1+1)$-dimensional relativistic {\it Cayley--Klein spacetimes} \cite{Herranz2002,CampoamorStursberg2024c} expressed in {\it Beltrami coordinates} \cite{Nurowski2025, Guo2003}.

The morphism of Lie algebras $\phi: \mathfrak{so}(1,3) \to C^{\infty}(\mathrm{T}^{*} \mathbb{R}^{2})$ defined in Proposition~\ref{prop:LorentzLH:morphism} gives rise to an {\it associated momentum map} $\mathbf{J}_{\phi}: \mathrm{T}^{*} \mathbb{R}^{2} \to \mathfrak{so}(1,3)^{*}$, defined by
\[  \langle \mathbf{J}_{\phi}(\bm{\xi}, \bm{p}_{\xi}), v \rangle := [\phi(v)](\bm{\xi}, \bm{p}_{\xi}), \qquad v \in \mathfrak{so}(1,3),  \]
where $\langle \cdot, \cdot \rangle$ is the canonical pairing between $\mathfrak{so}(1,3)^{*}$ and $\mathfrak{so}(1,3)$. The remarkable fact is that $\mathbf{J}_{\phi}$ is a Poisson map, with $\mathrm{T}^{*}\mathbb{R}^{2}$ equipped with the Poisson structure induced by its canonical symplectic form and $\mathfrak{so}(1,3)^{*}$ endowed with its KKS Poisson structure (see \cite{CampoamorStursberg2025a,CampoamorStursberg2026} and references therein).
In terms of the basis $\set{e^{1}, \ldots, e^{6}}$ of $\mathfrak{so}(1,3)^{*}$ dual to the basis $\set{H, D, C_{1}, C_{2}, K, P}$ of $\mathfrak{so}(1,3)$, we have that
\[
  \mathbf{J}_{\phi}(\bm{\xi}, \bm{p}_{\xi}) = \sum_{\alpha = 1}^{6}   h_{\alpha} (\bm{\xi}, \bm{p}_{\xi})e^{\alpha}, \qquad (\bm{\xi}, \bm{p}_{\xi}) \in \mathrm{T}^{*} \mathbb{R}^{2},
\]
  where $h_{\alpha}$ $(1 \leq \alpha \leq 6)$ are the Hamiltonian functions given in Proposition~\ref{prop:LorentzLH:morphism}. 
The preimage of the open subset $U_{z} = \set{z H + 1 > 0} \subset \mathfrak{so}(1,3)^{*}$ under $\mathbf{J}_{\phi}$ is the open subset
\[
  \mathbf{J}_{\phi}^{-1}(U_{z}) = \big\{(\bm{\xi}, \bm{p}_{\xi}) \in \mathrm{T}^{*} \mathbb{R}^{2}: 1 -z\big(\xi_{1}^{2} - \xi_{2}^{2}\big) > 0 \big\} \subset \mathrm{T}^{*} \mathbb{R}^{2}.
\]
Now let $\mathcal{M}_{z}$ be the open subset
\begin{equation}
  \mathcal{M}_{z}:=\big \{\bm{\xi} \in \mathbb{R}^{2}: 1 - z \big(\xi_{1}^{2} - \xi_{2}^{2}\big) > 0\big\} \subset \mathbb{R}^{2}.
  \label{eq:twist:domain}
\end{equation}
It readily follows that $\mathbf{J}_{\phi}^{-1}(U_{z})$ can be identified with the cotangent bundle of $\mathcal{M}_{z}$, that is, $\mathrm{T}^{*} \mathcal{M}_{z} = \mathbf{J}_{\phi}^{-1}(U_{z})$. Since both the restriction of the momentum map $\mathbf{J}_{\phi}$ to $\mathrm{T}^{*} \mathcal{M}_{z}$, $\mathbf{J}_{\phi} \vert_{\mathrm{T}^{*} \mathcal{M}_{z}}: \mathrm{T}^{*} \mathcal{M}_{z} \to U_{z}$, and the map $\varphi_{z}: U_{z} \to G_{z}^{*}$ given in \eqref{eq:nonstandardLorentz:twistmap} are Poisson maps, their composition yields a Poisson map
\begin{equation}
  \Psi_{z} := \varphi_{z} \circ \mathbf{J}_{\phi} \vert_{\mathrm{T}^{*} \mathcal{M}_{z}}  : \mathrm{T}^{*} \mathcal{M}_{z} \to G_{z}^{*}.
  \label{eq:twist:def_map}
\end{equation}
From the previous analysis in Section~\ref{section:LorentzLH}, we recall  that $-2 \mathbf{J}_{\phi}^{*}(C_{1}) = -2h_{3} = \frac{1}{2}(p_{\xi_{1}}^{2} - p_{\xi_{2}}^{2})$ is the kinetic energy on the Minkowskian plane $\mathbf{M}^{1+1}$.  From  Proposition~\ref{prop:LorentzLH:morphism} and the twist relations (\ref{eq:nonstandardLorentz:inv_twist}), we obtain that
\begin{equation}
  \begin{split}
    - 2 \Psi_{z}^{*} (C_{1}) & = -2\big(h_3+ z h_2^2\big)    \\
  & = \frac{1}{2} \left[ \big(1 - z\xi_{1}^{2} \big)p_{\xi_{1}}^{2} - 2z \xi_{1}\xi_{2}p_{\xi_{1}}p_{\xi_{2}} - \big(1+z \xi_{2}^{2} \big)p_{\xi_{2}}^{2} \right]\equiv T_z,
  \end{split}
  \label{eq:twist:def_energy}
\end{equation}
which corresponds to the kinetic energy $T_z$ associated with the metric $g_{z}$ on $\mathcal{M}_{z}$ given by
\begin{equation}
  g_{z} := \frac{(1 + z \xi_{2}^{2}) \dd \xi_{1} \otimes \dd \xi_{1} - z \xi_{1} \xi_{2} (\dd \xi_{1} \otimes \dd \xi_{2} + \dd \xi_{2} \otimes \dd \xi_{1}) - (1 - z \xi_{1}^{2}) \dd \xi_{2} \otimes \dd \xi_{2}}{1 - z(\xi_{1}^{2} - \xi_{2}^{2})} .
  \label{eq:twist:def_metric}
\end{equation}
This is (regardless of the value of $z$) a Lorentzian metric of constant (Gaussian) curvature equal to the deformation parameter $z$ on $\mathcal{M}_{z}$. As expected, when $z \to 0$ we recover the Minkowskian metric \eqref{eq:LorentzLH:Minkowskian_metric} on $\mathbf{M}^{1+1} = \mathcal{M}_{0}$,
\[ \lim_{z \to 0} g_{z} = \dd \xi_{1} \otimes \dd \xi_{1} - \dd \xi_{2} \otimes \dd \xi_{2}. \]

All these results can be expressed in terms of the well-known (projective)  canonical Beltrami variables, $ \big(\tilde{\bm{ \xi}}, \bm{p}_{\tilde\xi} \big) \in \mathrm{T}^{*} \mathbb{R}^{2}$ (see, e.g.,  \cite{Nurowski2025, Guo2003}  and references therein), through the canonical transformation defined by
\begin{equation}
  \begin{aligned}
     & {\xi_1} := \frac{\tilde\xi_1}{\sqrt{1+z\big( \tilde\xi_1^2-  \tilde\xi_2^2\big)}} ,  \qquad &  & p_{\xi_1}  := \sqrt{1+z\big( \tilde\xi_1^2-  \tilde\xi_2^2\big)}  \left(\big(1+z \tilde\xi_1^2 \big)p_{\tilde \xi_1}+ z\tilde\xi_1 \tilde\xi_2p_{\tilde \xi_2}  \right)   , \\
     & {\xi_2} := \frac{\tilde\xi_2}{\sqrt{1+z\big( \tilde\xi_1^2-  \tilde\xi_2^2\big)}}, \qquad   &  & p_{\xi_2}  := \sqrt{1+z\big( \tilde\xi_1^2-  \tilde\xi_2^2\big)}  \left(\big(1-z \tilde\xi_2^2 \big)p_{\tilde \xi_2}- z\tilde\xi_1 \tilde\xi_2p_{\tilde \xi_1}  \right)  .
  \end{aligned}
  \label{cantransform}
\end{equation}
In these new coordinates, the open subset (\ref{eq:twist:domain}) becomes
\begin{equation}
  \mathcal{M}_{z}:=\big \{\tilde{\bm{\xi}} \in \mathbb{R}^{2}: 1 +z \big(\tilde\xi_{1}^{2} - \tilde\xi_{2}^{2}\big) > 0\big\} \subset \mathbb{R}^{2}.
  \label{eq:twist:domain2}
\end{equation}
By applying the canonical transformation to the  kinetic energy $T_z$ (\ref{eq:twist:def_energy}) and to the metric  $g_{z}$ (\ref{eq:twist:def_metric}), we obtain their corresponding expressions  written in terms of  the canonical Beltrami variables:
\be
T_z=\frac{1}{2}\left(1 +z \big(\tilde\xi_{1}^{2} - \tilde\xi_{2}^{2}\big) \right) \left[ \big(1 + z\tilde\xi_{1}^{2} \big)p_{\tilde\xi_{1}}^{2} +2z \tilde\xi_{1}\tilde\xi_{2}p_{\tilde\xi_{1}}p_{\tilde\xi_{2}} - \big(1-z \tilde\xi_{2}^{2} \big)p_{\tilde\xi_{2}}^{2} \right],
\nonumber
\ee
\be
g_{z} = \frac{\big(1 - z \tilde\xi_{2}^{2}\big) \dd \tilde\xi_{1} \otimes \dd\tilde \xi_{1} + z \tilde\xi_{1} \tilde\xi_{2} (\dd \tilde\xi_{1} \otimes \dd \tilde\xi_{2} + \dd \tilde\xi_{2} \otimes \dd \tilde\xi_{1}) - \big(1 + z \tilde\xi_{1}^{2}\big) \dd \tilde\xi_{2} \otimes \dd \tilde\xi_{2}}{\big(1+ z\big(\tilde\xi_{1}^{2} -\tilde \xi_{2}^{2}\big)\big)^2}.
\label{eq:twist:beltrami_metric}
\ee
What is noteworthy here is that these latter relations  correspond precisely to the  kinetic energy $T_\kappa$ and the metric  $g_{\mathrm{CK},\kappa}$ of the  {\it $(1+1)$-dimensional relativistic Cayley--Klein spacetimes   $\mathbf{S}^{2}_{[\kappa]}$}  \cite{Herranz2002,CampoamorStursberg2024c}, expressed in  the Beltrami variables $\big(\tilde{\bm{ \xi}}, \bm{p}_{\tilde\xi} \big)$, provided that the deformation parameter $z$ is identified with the constant curvature $\kappa$ of  $\mathcal{M}_{z}$. These are just the $(1+1)$-dimensional {\it anti-de Sitter} $\mathbf{AdS}^{1+1}$ $(\kappa > 0)$, the {\it de Sitter} $\mathbf{dS}^{1+1}$ $(\kappa < 0)$ and the {\it Minkowskian} $\mathbf{M}^{1+1}$ $(\kappa = 0)$ spacetimes. Therefore, we have established the following  equivalences:
\be
 \mathcal{M}_{z}\equiv \mathcal{M}_{[\kappa]} ,\qquad T_z \big({\bm{ \xi}}, \bm{p}_{\xi}    \big)\equiv T_\kappa\big(\tilde{\bm{ \xi}}, \bm{p}_{\tilde\xi}    \big) ,  \qquad g_{z} \big({\bm{ \xi}}   \big)\equiv  g_{\mathrm{CK},\kappa} \big(\tilde{\bm{ \xi}}   \big) ,\qquad z\equiv \kappa.
\nonumber
\ee


\subsection{Beltrami coordinates on the (anti-)de Sitter spacetimes}
\label{subsecBeltrami}

Let us briefly recall the construction of the {\it Beltrami coordinates}  \cite{Nurowski2025, Guo2003} for the $(1+1)$-dimensional relativistic spacetimes of constant curvature: the Minkowski   $\mathbf{M}^{1+1}$,    anti-de Sitter $\mathbf{AdS}^{1+1}$  and de Sitter $\mathbf{dS}^{1+1}$ spacetimes.

First of all, let $\mathrm{SO}_{\kappa}(3)$ be the Lie group of isometries of the quadratic form $\mathbb{I}_{\kappa}$ on $\mathbb{R}^{3}$ given by
\begin{equation*}
  \mathbb{I}_{\kappa} := \mathrm{diag}(1, \kappa, - \kappa),
\end{equation*}
and consider   the isotropy group of the point $(1,0,0) \in \mathbb{R}^{3}$, which  is isomorphic to $\mathrm{SO}(1,1)\subset \mathrm{SO}_{\kappa}(3)$ regardless of the value of $\kappa$. In contrast, $\mathrm{SO}_{\kappa}(3)$ is isomorphic to the simple Lie group $\mathrm{SO}(2,1)$ when $\kappa \neq 0$, whereas for $\kappa = 0$ it reduces to the inhomogeneous Lie group $\mathrm{ISO}(1,1)$.
The corresponding homogeneous space
\begin{equation*}
  \mathbf{S}^{2}_{[\kappa]} := \mathrm{SO}_{\kappa}(3) / \mathrm{SO}(1,1)
\end{equation*}
is a so-called \textit{two-dimensional Cayley--Klein relativistic spacetime} \cite{Herranz2002,CampoamorStursberg2024c}. The metric $g_{\mathrm{CK}, \kappa}$ of $\mathbf{S}^{2}_{[\kappa]}$ is the projection to $\mathbf{S}^{2}_{[\kappa]}$ of the metric on
$\mathrm{SO}_{\kappa}(3)$ obtained by right translation of the metric defined by $\mathbb{I}_{\kappa}$ on the Lie algebra $\mathrm{T}_{e}\mathrm{SO}_{\kappa}(3) \simeq \mathfrak{so}_{\kappa}(3)$. Hence, $\mathbf{S}_{[\kappa]}^{2}$ is a symmetric space of   constant (Gaussian) curvature $\kappa = z$ relative to this metric. From a physical perspective, the curvature $\kappa$ is related to the cosmological constant $\Lambda$ through $\kappa = - \Lambda$. These spaces admit a natural kinematical interpretation: when $\kappa > 0$ one obtains the {\it anti-de Sitter spacetime} $\mathbf{AdS}^{1+1}$; when $\kappa < 0$, the {\it de Sitter spacetime} $\mathbf{dS}^{1+1}$; while the flat case $\kappa = 0$ retrieves the {\it Minkowskian plane} $\mathbf{M}^{1+1}$.

The natural action of $\mathrm{SO}_{\kappa}(3)$ on $\mathbb{R}^{3}$ as linear isometries of $\mathbb{I}_{\kappa}$ provides an ambient model for the three specific spaces included in $\mathbf{S}^{2}_{[\kappa]}$. Namely, $\mathbf{S}^{2}_{[\kappa]}$ is diffeomorphic to the connected component of the submanifold
\begin{equation}
  \mathbb{I}_{\kappa}^{-1}(1) =\big \{(x_{0}, x_{1}, x_{2}) \in \mathbb{R}^{3}: x_{0}^{2} + \kappa \big(x_{1}^{2} - x_{2}^{2}\big) = 1\big\}
  \label{eq:appendix:ambient}
\end{equation}
containing the point $(1,0,0)$, which will correspond to the origin in $\mathbf{S}^{2}_{[\kappa]}$. The coordinates $(x_{0}, x_{1}, x_{2})$ on $\mathbb{R}^{3}$ satisfying the constraint \eqref{eq:appendix:ambient} are called {\it ambient} or {\it Weierstrass coordinates}. In terms of these, the metric $g_{\mathrm{CK},\kappa}$ of $\mathbf{S}^{2}_{[\kappa]}$ comes from the flat ambient metric $ \dd x_{0} \otimes \dd x_{0} + \kappa \dd x_{1} \otimes \dd x_{1} - \kappa \dd x_{2} \otimes \dd x_{2}$ on $\mathbb{R}^{3}$ divided by the curvature $\kappa$ and restricted to $i: \mathbf{S}^{2}_{[\kappa]} \equiv \mathbb{I}_{\kappa}^{-1}(1) \hookrightarrow \mathbb{R}^{3}$, that is,
\begin{equation*}
  g_{\mathrm{CK},\kappa} = i^{*} \left( \frac{1}{\kappa} (\dd x_{0} \otimes \dd x_{0} + \kappa \dd x_{1} \otimes \dd x_{1} - \kappa \dd x_{2} \otimes \dd x_{2}) \right).
\end{equation*}


\begin{figure}[t!]
  \centering
  \includegraphics[width=\textwidth]{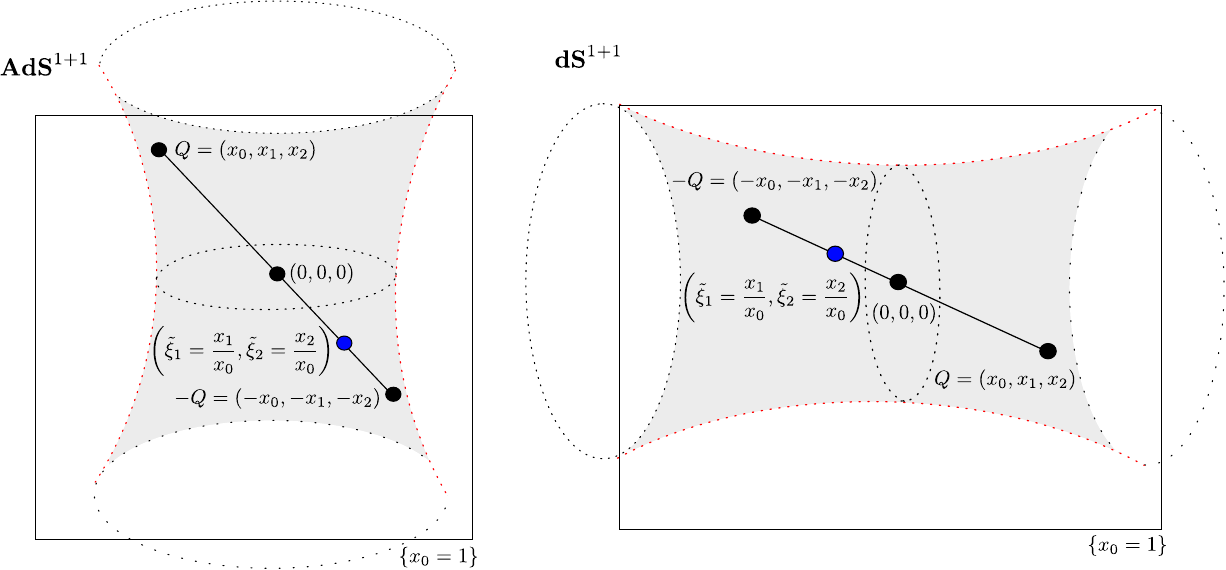}
  \small
  \caption{\small Beltrami coordinates $\big(\tilde\xi_{1}, \tilde\xi_{2} \big)$  on $\mathbf{AdS}^{1+1} = \mathbf{S}_{[+]}^{2}$ and on $\mathbf{dS}^{1+1} = \mathbf{S}^{2}_{[-]}$, defined in \eqref{eq:appendix:Beltrami_coord} as the coordinates of the corresponding point on the affine plane $\set{x_{0} = 1}$. Red dots indicate the points of $\mathbf{AdS}^{1+1} \cap \set{x_{0} = 0}$ and $\mathbf{dS}^{1+1} \cap \set{x_{0} = 0}$.}
  \label{fig:appendix:Beltrami}
\end{figure}


Let us now describe the so-called {\it Beltrami coordinates} $\big(\tilde\xi_{1}, \tilde\xi_{2}\big)$ on the spaces $\mathbf{S}^{2}_{[\kappa]}$. Consider $\pi_{0}$  the central projection of $\mathbf{S}^{2}_{[\kappa]}$ onto the affine plane $\set{x_{0} = 1} \subset \mathbb{R}^{3}$ with centre at  the origin $(0,0,0)\in \mathbb{R}^{3}$. Of course, $\pi_{0}$ is only defined on the open subset $\mathbf{S}^{2}_{[\kappa]} \cap \mathbb{R}^{3}_{x_{0} \neq 0} \subset \mathbf{S}^{2}_{[\kappa]}$. More precisely, given a point $Q = (x_{0}, x_{1}, x_{2}) \in \mathbf{S}^{2}_{[\kappa]} \cap \mathbb{R}^{3}_{x_{0} \neq 0}$, one has that $\pi_{0}(Q) = \big(\tilde\xi_{1}, \tilde\xi_{2}\big)$, where $\big(1,\tilde \xi_{1},\tilde \xi_{2} \big)$ is the intersection point of the affine line $\set{(0,0,0) + \beta (x_{0}, x_{1}, x_{2}): \beta \in \mathbb{R}}$ with the affine plane $\set{x_{0}= 1}$. Thus,
\begin{equation}
  \tilde\xi_{1} = \frac{x_{1}}{x_{0}}, \qquad \tilde\xi_{2} = \frac{x_{2}}{x_{0}},
  \label{eq:appendix:Beltrami_coord}
\end{equation}
where
\begin{equation*}
  x_{0} =  \frac{\mathrm{sgn}(x_{0})}{\sqrt{1 + \kappa \big(\tilde\xi_{1}^{2} - \tilde\xi_{2}^{2} \big)}}, \qquad 1 + \kappa \big(\tilde\xi_{1}^{2}- \tilde\xi_{2}^{2} \big) >0.
\end{equation*}
Thus, the image of $\pi_{0}$ is the open subset $\mathcal{M}_{z}\equiv\mathcal{M}_{[\kappa]} \subset \mathbb{R}^{2}$ given in \eqref{eq:twist:domain2}.
For $\kappa = 0$, the projection $\pi_{0}$ is just the identity map and $\big(\tilde\xi_{1}, \tilde\xi_{2}\big) = (x_{1}, x_{2})$, with $\mathcal{M}_{[\kappa]}= \mathbb{R}^{2}$. On the other hand,  $\pi_{0}$ is not injective whenever $\kappa \neq 0$, since  the antipodal point $-Q = (-x_{0},-x_{1},-x_{2}) \in \mathbf{S}^{2}_{[\kappa]} \cap \mathbb{R}^{3}_{x_{0} \neq 0}$  of $Q$ is such  that $\pi_{0}(Q) = \pi_{0}(-Q)$. Nevertheless, $\pi_{0}$ becomes injective when restricted to the connected components $\mathbf{S}^{2}_{[\kappa]} \cap \mathbb{R}^{3}_{x_{0} > 0}$ and $\mathbf{S}^{2}_{[\kappa]} \cap \mathbb{R}^{3}_{x_{0}<0}$ of $\mathbf{S}^{2}_{[\kappa]} \cap \mathbb{R}^{3}_{x_{0} \neq 0}$. Indeed,
\begin{equation*}
  \pi_{0} \vert_{\mathbf{S}^{2}_{[\kappa]} \cap \mathbb{R}^{3}_{x_{0}>0}} \to \mathcal{M}_{[\kappa]}, \qquad \pi_{0} \vert_{\mathbf{S}^{2}_{[\kappa]} \cap \mathbb{R}^{3}_{x_{0}<0}} \to \mathcal{M}_{[\kappa]}
\end{equation*}
are diffeomorphisms, implying that $\big(\mathbf{S}^{2}_{[\kappa]} \cap \mathbb{R}^{3}_{x_{0} > 0}, \big(\tilde\xi_{1},\tilde \xi_{2}\big) \big)$ and  $\big(\mathbf{S}^{2}_{[\kappa]} \cap \mathbb{R}^{3}_{x_{0} < 0}, \big(\tilde\xi_{1}, \tilde\xi_{2} \big) \big)$ are local coordinate systems on $\mathbf{S}^{2}_{[\kappa]}$. The coordinates $\big(\tilde\xi_{1}, \tilde\xi_{2} \big) \in \mathcal{M}_{[\kappa]}$ given by \eqref{eq:appendix:Beltrami_coord} are called \textit{Beltrami coordinates} (see Figure~\ref{fig:appendix:Beltrami}). The expression of the metric $g_{\mathrm{CK}, \kappa}$ of $\mathbf{S}^{2}_{[\kappa]}$ in these coordinates is given in \eqref{eq:twist:beltrami_metric}.

\begin{remark}
  When $\kappa \neq 0$, the Beltrami coordinates \eqref{eq:appendix:Beltrami_coord} only cover the points of the open subset $\mathbf{S}^{2}_{[\kappa]} \cap \mathbb{R}^{3}_{x_{0} \neq 0}$. The previous construction can be easily generalised in order to find an atlas of the whole $\mathbf{S}^{2}_{[\kappa]}$ formed by Beltrami-type coordinates.
\end{remark}


\section{Projective relativistic dynamics}
\label{section:dynamics}

In this section, we derive a PH deformation of the $\mathfrak{so}(1,3)$-LH system \eqref{eq:LorentzLH:tdepHamVF}, based on the non-standard quantum Lorentz algebra \eqref{eq:nonstandardLorentz:defcr}, by means of the Poisson map $\Psi_{z}: \mathrm{T}^{*} \mathcal{M}_z\to G_{z}^{*}$ introduced in \eqref{eq:twist:def_map}.  The associated deformation parameter is the constant curvature, $\kappa \equiv z$,  of the   metric \eqref{eq:twist:def_metric} on $\mathcal{M}_z$, obtained from the Hamiltonian function $-2 \Psi_{z}^{*}(C_{1})$ given in \eqref{eq:twist:def_energy}.
The resulting deformed system will be expressed in the canonical variables $(\bm{\xi}, \bm{p}_{\xi})$ of $\mathrm{T}^{*} \mathcal{M}_z$, although it should be noted that it is always possible to express all the results presented hereafter  in terms of the (projective)  canonical Beltrami variables  $ \big(\tilde{\bm{ \xi}}, \bm{p}_{\tilde\xi} \big)$, described in Section~\ref{subsecBeltrami}, by applying the  canonical transformation defined in (\ref{cantransform}).
In this regard, it is worth emphasising that the fact that the role of the curvature as an `integrable' deformation parameter has already been explored in previous works \cite{Ballesteros2005,Ballesteros2005b,Ragnisco2007,CampoamorStursberg2024c}.  Furthermore, since the kinetic energy \eqref{eq:twist:def_energy} is a polynomial in the canonical variables $(\bm{\xi}, \bm{p}_{\xi})$, several  of the systems obtained herein may be interpreted as particular instances   of {\it projective relativistic dynamics} (see \cite{Albouy2008,Albouy2013} for details on projective dynamics). In addition, we also construct $t$-independent constants of motion for the deformed Lorentz system.
 
The following result summarises our construction of the PH deformation of the $\mathfrak{so}(1,3)$-LH system \eqref{eq:LorentzLH:tdepHamVF}, based on the non-standard quantum deformation of the Lorentz algebra \eqref{eq:nonstandardLorentz:defcr}.

\begin{prop}
  \label{propALL}
  The Hamiltonian functions $h_{z,\alpha} \in C^{\infty}(\mathrm{T}^{*}\mathcal{M}_z)$  given by
  \begin{equation}
    \begin{split}
       & h_{z,1} := \Psi_{z}^{*}(H)=  \frac{\log \big(1 -z \big(\xi_{1}^{2} - \xi_{2}^{2} \big) \big)}{z},   \\[2pt]
       & h_{z,2} := \Psi_{z}^{*}(D) = \frac{1}{2}(\xi_{1} p_{\xi_{1}} + \xi_{2} p_{\xi_{2}}),     \\[2pt]
       & h_{z,3} := \Psi_{z}^{*}(C_{1})=  - \frac{1}{4} \left(\big(1 -z \xi_{1}^{2} \big)p_{\xi_{1}}^{2} - 2 z \xi_{1} \xi_{2} p_{\xi_{1}}p_{\xi_{2}} - \big(1 + z \xi_{2}^{2} \big) p_{\xi_{2}}^{2} \right), \\[2pt]
       & h_{z,4}:= \Psi_{z}^{*}(C_{2}) = \frac{1}{2}p_{\xi_{1}} p_{\xi_{2}},     \\[2pt]
       & h_{z,5} := \Psi_{z}^{*}(K) = \frac{1}{2}(\xi_{1}p_{\xi_{2}} - \xi_{2} p_{\xi_{1}}),     \\[2pt]
       & h_{z,6} := \Psi_{z}^{*}(P) = 2 \xi_{1} \xi_{2},
    \end{split}
    \label{eq:applications:def_ham}
  \end{equation}
  where $\Psi_{z}: \mathrm{T}^{*} \mathcal{M}_z \to G_{z}^{*}$ is the Poisson map given in \eqref{eq:twist:def_map}, satisfy the following  non-zero  commutation relations:
  \begin{equation}
    \begin{aligned}
       & \set{h_{z,1}, h_{z,2}}_{\omega} = \frac{1- \mathrm{e}^{- z h_{z,1}}}{z}, \qquad &  & \set{h_{z,1}, h_{z,3}}_{\omega} = 2 h_{z,2},      \\[2pt]
       & \set{h_{z,1}, h_{z,4}}_{\omega} = - 2 h_{z,5} \,\mathrm{e}^{-z h_{z,1}}, \qquad &  & \set{h_{z,1}, h_{z,5}}_{\omega} =  h_{z,6} \,\mathrm{e}^{-z h_{z,1}}, \\[2pt]
 & \set{h_{z,2}, h_{z,3}}_{\omega} = h_{z,3} -z h_{z,2}^{2}, \qquad  & & \set{h_{z,2}, h_{z,4}}_{\omega} = h_{z,4},\\[2pt]
 & \set{h_{z,2}, h_{z,6}}_{\omega} = - h_{z,6}, \qquad &  & \set{h_{z,3}, h_{z,4}}_{\omega} =  2 z h_{z,2} h_{z,4},\\[2pt]
& \set{h_{z,3}, h_{z,5}}_{\omega} = h_{z,4}, \qquad &  & \set{h_{z,3}, h_{z,6}}_{\omega} = - 2 h_{z,5} - 2z h_{z,2} h_{z,6},   \\[2pt]
&\set{h_{z,4}, h_{z,5}}_{\omega} = - h_{z,3} + z h_{z,2}^{2}, \qquad &&\set{h_{z,4}, h_{z,6}}_{\omega} = - 2 h_{z,2},\\[2pt]
& \set{h_{z,5}, h_{z,6}}_{\omega} = \frac{ \mathrm{e}^{z h_{z,1}} -1}{z},
    \end{aligned}
    \label{eq:PHdef:defcr_ham}
  \end{equation}
 where $\omega = \dd \xi_{1} \wedge \dd p_{\xi_{1}} + \dd \xi_{2} \wedge \dd p_{\xi_{2}}$.
  Hence, they define a PH deformation of the LH system \eqref{eq:LorentzLH:tdepHamVF} on $\mathrm{T}^{*} \mathbf{M}^{1+1}$.
\end{prop}

The undeformed counterparts of the Hamiltonian functions $h_{z, \alpha}$ given in \eqref{eq:applications:def_ham} are, precisely, the Hamiltonian functions on $\mathrm{T}^{*}\mathbf{M}^{1+1} \equiv \mathrm{T}^{*} \mathbb{R}^{2}$ given in Proposition~\ref{prop:LorentzLH:morphism}; that is, $h_{\alpha} = \lim_{z \to 0} h_{z,\alpha}$ for every $1 \leq \alpha \leq 6$.
In this way, the $t$-dependent Hamiltonian on $\mathrm{T}^{*} \mathcal{M}_{z}$ defined by
\begin{equation}
  \mathbb{R} \times \mathrm{T}^{*} \mathcal{M}_{ z} \ni (t, \bm{\xi}, \bm{p}_{\xi}) \longmapsto h_{z}(t, \bm{\xi}, \bm{p}_{\xi}) := \sum_{\alpha = 1}^{6} b_{\alpha}(t) h_{z, \alpha}(\bm{\xi}, \bm{p}_{\xi}) \in \mathbb{R}
  \label{eq:applications:tdep_def}
\end{equation}
is a deformation of the $t$-dependent Hamiltonian \eqref{eq.LorentzLH:tdepHam} on $\mathrm{T}^{*} \mathbf{M}^{1+1}$.

 \begin{remark}
  \label{remdeformedCas}

  Under the Poisson map $\Psi_{z}$ \eqref{eq:twist:def_map}, the deformed Casimir invariants \eqref{eq:nonstandardLorentz:defCas} turn out to be
  \begin{equation}
    \begin{split}
       & \Psi_{z}^{*} (\mathscr{C}_{z,2} )=  h_5^{2} - h_2^{2} \,\mathrm{e}^{z h_{z,1}} + \left(  \frac{ \mathrm{e}^{z h_{z,1}} -1}{z}\right)h_3+h_4h_6\equiv 0,        \\[2pt]
       & \Psi_{z}^{*} ( \mathscr{C}_{z,2}' )= h_2h_5 + \frac{1}{2}\left[\left(  \frac{ \mathrm{e}^{z h_{z,1}} -1}{z}\right) h_4- h_3 h_6 + z h_2^2 h_6 \right]\equiv 0.
      \nonumber
    \end{split}
  \end{equation}
\end{remark}
 
The Hamiltonian vector fields $\mathbf{X}_{z, \alpha}$ on $\mathrm{T}^{*} \mathcal{M}_z$ associated with the Hamiltonian functions $h_{z, \alpha}$, determined by the condition $\iota_{\mathbf{X}_{z, \alpha}}\omega = \dd h_{z, \alpha}$, read
\begin{equation*}
  \begin{split}
     & \mathbf{X}_{z,1} = \frac{2}{1 -z \big(\xi_{1}^{2} - \xi_{2}^{2}\big)} \left( \xi_{1} \pdv{p_{\xi_{1}}} - \xi_{2} \pdv{p_{\xi_{2}}} \right),  \\[2pt]
     & \mathbf{X}_{z,2} = \frac{1}{2} \left( \xi_{1} \pdv{\xi_{1}} + \xi_{2} \pdv{\xi_{2}}   - p_{\xi_{1}} \pdv{p_{\xi_{1}}} - p_{\xi_{2}} \pdv{p_{\xi_{2}}} \right),   \\[2pt]
& \begin{split}
 \mathbf{X}_{z,3}= \frac{1}{2} & \left[- \left( \big(1 -z \xi_{1}^{2}\big) p_{\xi_{1}} -z \xi_{1} \xi_{2} p_{\xi_{2}} \right) \pdv{\xi_{1}}  + \left( \big(1 +z \xi_{2}^{2}\big) p_{\xi_{2}} +z \xi_{1} \xi_{2} p_{\xi_{1}} \right) \pdv{\xi_{2}} \right. \\
&   \left. - z (\xi_{1} p_{\xi_{1}} + \xi_{2} p_{\xi_{2}} ) \left(p_{\xi_{1}} \pdv{p_{\xi_{1}}} + p_{\xi_{2}} \pdv{p_{\xi_{2}}} \right) \right],
 \end{split} \\[2pt]
& \mathbf{X}_{z,4} = \frac{1}{2} \left( p_{\xi_{2}} \pdv{\xi_{1}} + p_{\xi_{1}} \pdv{\xi_{2}} \right),     \\[2pt]
     & \mathbf{X}_{z, 5} = -\frac{1}{2} \left(  \xi_{2} \pdv{\xi_{1}} - \xi_{1} \pdv{\xi_{2}} +p_{\xi_{2}} \pdv{p_{\xi_{1}}} - p_{\xi_{1}} \pdv{p_{\xi_{2}}} \right),  \\[2pt]
     & \mathbf{X}_{z,6} = -2 \left( \xi_{2} \pdv{p_{\xi_{1}}} + \xi_{1} \pdv{p_{\xi_{2}}} \right).
  \end{split}
\end{equation*}
These vector fields do not longer span a VG Lie algebra for $z \neq 0$, but rather a generalised distribution on $\mathrm{T}^{*}\mathcal{M}_z$, which is integrable in the Stefan--Sussman sense (see~\cite{Va_94,Pa57,CIMM2015,Lavau2018,Lewis2023}  and references therein), given by the following non-zero commutation relations:
\begin{equation*}
  \begin{aligned}
     & [\mathbf{X}_{z,1}, \mathbf{X}_{z,2}] = - \mathrm{e}^{-z h_{z, 1}} \mathbf{X}_{z, 1}, \quad     &  & [\mathbf{X}_{z, 1}, \mathbf{X}_{z, 3}] = - 2 \mathbf{X}_{z, 2},   \\[2pt]
     & [\mathbf{X}_{z, 1}, \mathbf{X}_{z, 4}] = -2 \mathrm{e}^{-z h_{z,1}} (z h_{z, 5} \mathbf{X}_{z, 1}- \mathbf{X}_{z, 5}),\quad &  & [\mathbf{X}_{z, 1}, \mathbf{X}_{z, 5}] = \mathrm{e}^{-z h_{z, 1}} (z h_{z, 6} \mathbf{X}_{z, 1} - \mathbf{X}_{z, 6}),   \\[2pt]
     & [\mathbf{X}_{z, 2}, \mathbf{X}_{z, 3}] = - \mathbf{X}_{z, 3} + 2 z h_{z, 2} \mathbf{X}_{z, 2}, \quad  &  & [\mathbf{X}_{z, 2}, \mathbf{X}_{z, 4}] = - \mathbf{X}_{z, 4},   \\[2pt]
     & [\mathbf{X}_{z, 2}, \mathbf{X}_{z, 6}] = \mathbf{X}_{z, 6}, \quad &  & [\mathbf{X}_{z, 3}, \mathbf{X}_{z, 4}] = -2 z(h_{z, 4} \mathbf{X}_{z, 2} + h_{z, 2} \mathbf{X}_{z, 4}),  \\[2pt]
     & [\mathbf{X}_{z, 3}, \mathbf{X}_{z, 5}] = - \mathbf{X}_{z, 4}, \quad   &  & [\mathbf{X}_{z, 3}, \mathbf{X}_{z, 6}] = 2\big (\mathbf{X}_{z, 5} +z(h_{z,6} \mathbf{X}_{z,2} + h_{z,2} \mathbf{X}_{z,6}) \big), \\[2pt]
     & [\mathbf{X}_{z,4}, \mathbf{X}_{z,5}] = \mathbf{X}_{z,3} - 2z h_{z, 2} \mathbf{X}_{z, 2}, \qquad  &  & [\mathbf{X}_{z, 4}, \mathbf{X}_{z,6}] = 2 \mathbf{X}_{z, 2},   \\[2pt]
     & [\mathbf{X}_{z, 5}, \mathbf{X}_{z, 6}] = - \mathrm{e}^{z h_{z, 1}} \mathbf{X}_{z, 1}.
  \end{aligned}
\end{equation*}
The Hamilton equations of the $t$-dependent Hamiltonian \eqref{eq:applications:def_ham} on $\mathrm{T}^{*} \mathcal{M}_z$ describe the integral curves of the $t$-dependent vector field $\mathbf{X}_{z}$ on $\mathrm{T}^{*} \mathcal{M}_z$ given by
\begin{equation}
  \mathbf{X}_{z} = \sum_{\alpha = 1}^{6} b_{\alpha}(t) \mathbf{X}_{z, \alpha}.
  \label{eq:applications:defthamVF}
\end{equation}
More precisely, its associated first-order system of ODEs reads
\begin{equation}
  \begin{split}
    \dv{\xi_{1}}{t}     & = \frac{1}{2} \left[ b_{2}(t) \xi_{1} - b_{3}(t) \left( \big(1-z \xi_{1}^{2}\big) p_{\xi_{1}} -z \xi_{1}\xi_{2}p_{\xi_{2}} \right) + b_{4}(t) p_{\xi_{2}} - b_{5}(t) \xi_{2} \right],   \\[2pt]
    \dv{\xi_{2}}{t}     & = \frac{1}{2} \left[ b_{2}(t) \xi_{2} + b_{3}(t) \left( \big( 1 +z \xi_{2}^{2}\big) p_{\xi_{2}} +z \xi_{1} \xi_{2} p_{\xi_{1}} \right) +  b_{4}(t) p_{\xi_{1}} + b_{5}(t) \xi_{1} \right],   \\[2pt]
    \dv{p_{\xi_{1}}}{t} & = \frac{1}{2} \left[  \frac{4 b_{1}(t) \xi_{1}}{1 -z \big(\xi_{1}^{2} - \xi_{2}^{2}\big)} - b_{2}(t) p_{\xi_{1}} -z b_{3}(t)  p_{\xi_{1}} ( \xi_{1} p_{\xi_{1}} + \xi_{2} p_{\xi_{2}}) - b_{5}(t) p_{\xi_{2}} - 4 b_{6}(t) \xi_{2} \right], \\[2pt]
    \dv{p_{\xi_{2}}}{t} & = \frac{1}{2} \left[  -\frac{4 b_{1}(t) \xi_{2}}{1 -z \big(\xi_{1}^{2} - \xi_{2}^{2}\big)} - b_{2}(t) p_{\xi_{2}} -z b_{3}(t)  p_{\xi_{2}} ( \xi_{1} p_{\xi_{1}} + \xi_{2} p_{\xi_{2}}) +b_{5}(t) p_{\xi_{1}} - 4 b_{6}(t) \xi_{1} \right],
  \end{split}
  \label{eq:applicactions:defHameq}
\end{equation}
which are also the Hamilton equations derived from the $t$-dependent Hamiltonian (\ref{eq:applications:tdep_def}).


\subsection{Deformed prolongations and time-independent constants of motion}
\label{subsection:contants}

Since $\Psi_{z}: \mathrm{T}^{*} \mathcal{M}_z\to G_{z}^{*}$ is a Poisson map, it follows that, for every Casimir function $\mathscr{C} \in C^{\infty}(G_{z}^{*})$, the pull-back $\Psi_{z}^{*}(\mathscr{C})$ is a constant of motion of $\mathbf{X}_{z}$. For the Casimir functions $\mathscr{C}_{z,2}$ and $\mathscr{C}_{z,2}'$, given in \eqref{eq:nonstandardLorentz:defCas}, we have that $\Psi_{z}^{*}(\mathscr{C}_{z,2}) = \Psi_{z}^{*}(\mathscr{C}_{z,2}') = 0$, as noted in Remark~\ref{remdeformedCas}.
In addition, the coproduct (\ref{eq:nonstandardLorentz:defcoprod}) $\Delta_{z}: C^{\infty}(G_{z}^{*}) \to C^{\infty}(G_{z}^{*} \times G_{z}^{*})$  of $C^{\infty}(G_{z}^{*})$  is, by construction,  a morphism of Poisson algebras, which implies that the composition
\[ (\Psi_{z} \times \Psi_{z})^{*} \circ \Delta_{z} : C^{\infty}(G_{z}^{*}) \to C^{\infty}(\mathrm{T}^{*} \mathcal{M}_z \times \mathrm{T}^{*} \mathcal{M}_z) \]
is likewise a morphism of Poisson algebras. Remarkably, this construction yields the so-called $2^{\textup{nd}}$-deformed prolongation of $\mathbf{X}_{z}$ (\ref{eq:applications:defthamVF}), denoted by $\mathbf{X}_{z}^{[2]} $  (see \cite{Ballesteros2021} for further details).  To deduce specific expressions for $\mathbf{X}_{z}^{[2]} $, let us consider the product manifold $\mathrm{T}^{*}\mathcal{M}_z \times \mathrm{T}^{*} \mathcal{M}_z$ endowed with canonical variables $\big(\bm{\xi}^{(1)}, \bm{p}_{\xi}^{(1)}, \bm{\xi}^{(2)}, \bm{p}_{\xi}^{(2)}\big)$, where $\big(\bm{\xi}^{(i)}, \bm{p}_{\xi}^{(i)}\big)$ denote the canonical variables on the $i^{\textup{th}}$-copy of $\mathrm{T}^{*} \mathcal{M}_z$, such that    the corresponding symplectic form  is given by
\be
\omega^{[2]}=\dd \xi_{1}^{(1)} \wedge \dd  p_{\xi_{1}}^{(1)}  + \dd \xi_{2}^{(1)} \wedge \dd  p_{\xi_{2}}^{(1)} +\dd \xi_{1}^{(2)} \wedge \dd  p_{\xi_{1}}^{(2)}  + \dd \xi_{2}^{(2)} \wedge \dd  p_{\xi_{2}}^{(2)} .
\label{symplectic2}
\ee
Therefore, the  kinetic energy $T_z$ on $\mathcal{M}_{z}$,   previously presented in (\ref{eq:twist:def_energy}) and associated with the metric  $g_{z}$ (\ref{eq:twist:def_metric}), now appears as $T_z= - 2 \Psi_{z}^{*}(C_{1}) =- 2 h_{z,3} $ in (\ref{eq:applications:def_ham}). This    can be  expressed directly in terms of  the variables  of  the $i^{\textup{th}}$-copy   $\big(\bm{\xi}^{(i)}, \bm{p}_{\xi}^{(i)}\big)$ $(1 \leq i \leq 2)$ in the form
\begin{equation}
  T_{z} ^{(i)} =   \frac{1}{2}  \left[ \left(1 - z \left( \xi_{1}^{(i)} \right)^{2} \right) \left( p_{\xi_{1}}^{(i)} \right)^{2} - 2 z  \xi_{1}^{(i)}  \xi_{2}^{(i)} p_{\xi_{1}}^{(i)} p_{\xi_{2}}^{(i)} - \left(1 + z \left( \xi_{2}^{(i)} \right)^{2} \right) \left( p_{\xi_{2}}^{(i)} \right)^{2}   \right] .
  \label{eq:applications:kinetic}
\end{equation}

For convenience, we introduce the following shorthand notations, which will be used hereafter:
\begin{equation}
  \norm{\bm{\xi}^{(i)}}^{2} := \left(\xi_{1}^{(i)}\right)^{2} - \left(\xi_{2}^{(i)}\right)^{2} , \qquad \bm{\xi}^{(i)} \! \bm{\cdot}   \bm{p}_{\xi}^{(i)} := \xi_{1}^{(i)} p_{\xi_{1}}^{(i)} + \xi_{2}^{(i)} p_{\xi_{2}}^{(i)}, \qquad L^{(i)} := \xi_{1}^{(i)} p_{\xi_{2}}^{(i)} - \xi_{2}^{(i)} p_{\xi_{1}}^{(i)}.
  \label{eq:applications:notation}
\end{equation}
Note that $\norm{\cdot}^{2}$ is the norm associated with  the Clifford algebra $\mathrm{Cl}_{1,1}(\mathbb{R})$, and that $L^{(i)}$ is the angular momentum on the $i^{\textup{th}}$-copy of $\mathrm{T}^{*} \mathcal{M}_z$.
Then, the $2^{\textup{nd}}$-order deformed prolongation of $\mathbf{X}_{z}$ is the $t$-dependent vector field $\mathbf{X}_z^{[2]}$ on the product manifold $\mathrm{T}^{*} \mathcal{M}_z\times \mathrm{T}^{*} \mathcal{M}_z$ defined by
\be
\mathbf{X}_{z}^{[2]} := \sum_{\alpha = 1}^{6} b_{\alpha}(t) \mathbf{X}_{z, \alpha}^{[2]},
\label{Xz2}
\ee
where $\mathbf{X}_{z,\alpha}^{[2]}$ is the Hamiltonian vector field associated with the Hamiltonian function $h_{z, \alpha}^{[2]}$.  Therefore,  the corresponding  $t$-dependent Hamiltonian on $\mathrm{T}^{*} \mathcal{M}_z\times \mathrm{T}^{*} \mathcal{M}_z$  is defined by
\be
h_{z}^{[2]}  := \sum_{\alpha = 1}^{6} b_{\alpha}(t) h_{z, \alpha}^{[2]}  .
\label{hz2}
\ee

The explicit expressions of the $2^{\textup{nd}}$-deformed prolongation $\mathbf{X}_{z}^{[2]} $ are displayed in Table~\ref{table:def_1} in terms of the functions $h_{z, \alpha}^{[2]}$ $(1\le \alpha\le 6)$,   which satisfy the same formal commutation relations (\ref{eq:PHdef:defcr_ham}) with respect to the symplectic form (\ref{symplectic2}).


\begin{table}[t!]
  \centering
  { \small
    \caption{\small   The  deformed Hamiltonian functions $h_{z, \alpha}^{[2]}$  associated with the   vector fields   $\mathbf{X}_{z,\alpha}^{[2]}$  that provide the   $2^{\textup{nd}}$-deformed prolongation  of $\mathbf{X}_{z}$ (\ref{eq:applications:defthamVF}) with $t$-dependent vector field $\mathbf{X}_{z}^{[2]} $ (\ref{Xz2}) and  Hamiltonian $h_{z}^{[2]}  $ (\ref{hz2}).  We   use   the shorthand notations   \eqref{eq:applications:notation} and $T_z^{(i)}$ is the kinetic energy \eqref{eq:applications:kinetic} on the $i^{\textup{th}}$-copy of $\mathrm{T}^{*} \mathcal{M}_z$.   }
    \begin{tabular}{l}
      \hline

      \hline
      \\[-4pt]
$\displaystyle h_{z,1}^{[2]} = (\Psi_{z} \times \Psi_{z})^{*}(\Delta_{z}(H)) =    \frac{\log\left[ \left(1 -z \norm{\bm{\xi}^{(1)}}^{2} \right) \left(1 -z \norm{\bm{\xi}^{(2)}}^{2} \right) \right] }{z} $                                                                                                                                                                                                                                                                   \\[12pt]
 $  \displaystyle h_{z,2}^{[2]} = (\Psi_{z} \times \Psi_{z})^{*}(\Delta_{z}(D))= \frac{1}{2} \left( \frac{\bm{\xi}^{(1)}  \! \bm{\cdot} \bm{p}_{\xi}^{(1)}}{1 -z \norm{\bm{\xi}^{(2)}}^{2}} + \bm{\xi}^{(2)}  \! \bm{\cdot} \bm{p}_{\xi}^{(2)} \right)   $                                                                                                                                                                                                                      \\[16pt]
$  \displaystyle h_{z,3}^{[2]} = (\Psi_{z} \times \Psi_{z})^{*}(\Delta_{z}(C_{1}))= - \frac{1}{2} \left( \frac{T_{z}^{(1)}}{1-z \norm{\bm{\xi}^{(2)}}^{2}} + T_{z}^{(2)} \right)$                                                                                                                                                                                                                                                                                                             \\[16pt]
 $  \displaystyle h_{z, 4}^{[2]}  = (\Psi_{z} \times \Psi_{z})^{*}(\Delta_{z}(C_{2}))= \frac{1}{2} \left[ p_{\xi_{1}}^{(2)} p_{\xi_{2}}^{(2)} - \frac{z^{2} \xi_{1}^{(2)} \xi_{2}^{(2)} \left( \bm{\xi}^{(1)}  \! \bm{\cdot} \bm{p}_{\xi}^{(1)} \right)^{2}}{\left(1-z \norm{\bm{\xi}^{(2)}}^{2} \right)^{2}} + \frac{p_{\xi_{1}}^{(1)} p_{\xi_{2}}^{(1)} +zL^{(2)}\! \left( \bm{\xi}^{(1)}  \! \bm{\cdot} \bm{p}_{\xi}^{(1)} \right)}{1 -z \norm{\bm{\xi}^{(2)}}^{2}} \right]$ \\[20pt]
$  \displaystyle h_{z,5}^{[2]} = (\Psi_{z} \times \Psi_{z})^{*}(\Delta_{z}(K))= \frac{1}{2} \left[ L^{(1)}+L^{(2)} - \frac{2 z\xi_{1}^{(2)} \xi_{2}^{(2)}\left(\bm{\xi}^{(1)}  \! \bm{\cdot} \bm{p}_{\xi}^{(1)} \right)}{1 -z\norm{\bm{\xi}^{(2)}}^{2}}\right]$                                                                                                                                                                                                                \\[18pt]
$  \displaystyle h_{z,6}^{[2]}  = (\Psi_{z} \times \Psi_{z})^{*}(\Delta_{z}(P))= 2 \left[ \xi_{1}^{(1)} \xi_{2}^{(1)} \left( 1 -z \norm{\bm{\xi}^{(2)}}^{2} \right) + \xi_{1}^{(2)} \xi_{2}^{(2)} \right]$                                                                                                                                                                                                                                                                     \\[16pt]
 \hline

 \hline
    \end{tabular}
    \label{table:def_1}
  }
\end{table}


In contrast to the  system $\mathbf{X}_{z}$ on $\mathrm{T}^{*} \mathcal{M}_z$, the Casimir functions $\mathscr{C}_{z,2}$ and $\mathscr{C}_{z,2}'$ on $G_{z}^{*}$, given in  \eqref{eq:nonstandardLorentz:defCas}, now give rise to two non-zero $t$-independent constants of motion of $\mathbf{X}_{z}^{[2]}$, namely
\begin{equation}
  F_{z,2}^{[2]} := (\Psi_{z} \times \Psi_{z})^{*}(\Delta_{z}(\mathscr{C}_{z,2})), \qquad F_{z,2}'^{[2]}:= (\Psi_{z} \times \Psi_{z})^{*}(\Delta_{z}(\mathscr{C}_{z,2}')),
  \label{eq:applications:defcts2}
\end{equation}
which are explicitly shown in Table~\ref{table:def_2}.  This, in turn, means that both constants Poisson-commute with all  Hamiltonian functions $h_{z,\alpha}^{[2]} $ given in  Table~\ref{table:def_1},  that is,
\be
\left\{F_{z,2}^{[2]} ,h_{z,\alpha}^{[2]}  \right\}_{\omega^{[2]}}=\left\{F_{z,2}'^{[2]} ,h_{z,\alpha}^{[2]}  \right\}_{\omega^{[2]}}=0,\qquad 1\le \alpha\le 6,
\nonumber
\ee
as well as with the  $t$-dependent Hamiltonian $ h_{z}^{[2]}$ (\ref{hz2})
for any value of the $t$-dependent functions $b_{\alpha}(t)$.   Furthermore, in this case, $F_{z,2}^{[2]} $ and $F_{z,2}'^{[2]} $ also commute with each other:
\be
\left\{ F_{z,2}^{[2]} ,F_{z,2}'^{[2]} \right\}_{\omega^{[2]}}= 0.
\nonumber
\ee


\begin{table}[t!]

  \small
  \caption{\small      $t$-independent deformed constants of motion on $\mathrm{T}^{*} \mathcal{M}_z \times \mathrm{T}^{*} \mathcal{M}_z$, as defined in \eqref{eq:applications:defcts2}, of the $2^{\textup{nd}}$-deformed prolongation of the deformed system \eqref{eq:applications:defthamVF} with $t$-dependent vector field $\mathbf{X}_{z}^{[2]} $ (\ref{Xz2}) and  Hamiltonian $h_{z}^{[2]}  $ (\ref{hz2}). We again use   the shorthand notations   \eqref{eq:applications:notation} and    $T_z^{(i)}$ denotes the kinetic energy \eqref{eq:applications:kinetic}. }
  \begin{center}
    \begin{tabular}{l}
      \hline

      \hline                                                                                                                                                                                                                                                                                                                                                          \\[-2pt]
 {\small    $\begin{array}{l}
\displaystyle 4F_{z, 2}^{[2]} =      \displaystyle{\left( L^{(1)} + L^{(2)} - \frac{2 z\, \xi_{1}^{(2)} \xi_{2}^{(2)} \left(\bm{\xi}^{(1)}\! \bm{\cdot} \bm{p}_{\xi}^{(1)} \right)}{1 -z   \norm{\bm{\xi}^{(2)}}^{2}} \right)^{2}    }                                                    \\[16pt]
 \qquad \qquad\quad    \displaystyle{  -\left(1 -z \norm{\bm{\xi}^{(1)}}^{2}\right)\left(1 -z \norm{\bm{\xi}^{(2)}}^{2}\right) \left( \bm{\xi}^{(2)}  \! \bm{\cdot} \bm{p}_{\xi}^{(2)} +\frac{\bm{\xi}^{(1)} \! \bm{\cdot} \bm{p}_{\xi}^{(1)} }{1 -z \norm{\bm{\xi}^{(2)}}^{2}}\right)^2 } \\[16pt]
 \qquad \qquad\quad    \displaystyle{ + 2  \left(  \frac{T_{z}^{(1)} }{1-z \norm{\bm{\xi}^{(2)}}^{2}} +T_{z}^{(2)}  \right) \left( \norm{\bm{\xi}^{(1)}}^{2} + \norm{\bm{\xi}^{(2)}}^{2} -z \norm{\bm{\xi}^{(1)}}^{2} \norm{\bm{\xi}^{(2)}}^{2}  \right)   }                               \\[16pt]
 \qquad \qquad\quad      \displaystyle{+4  \left[ p_{\xi_{1}}^{(2)} p_{\xi_{2}}^{(2)}+
 \frac{p_{\xi_{1}}^{(1)} p_{\xi_{2}}^{(1)}+ z L^{(2)} \left(\bm{\xi}^{(1)} \!\bm{\cdot}\bm{p}_{\xi}^{(1)} \right)   }{1 -z  \norm{\bm{\xi}^{(2)}}^{2}} - \frac{z^{2} \xi_{1}^{(2)} \xi_{2}^{(2)} \left( \bm{\xi}^{(1)}\! \bm{\cdot} \bm{p}_{\xi}^{(1)}\right)^2  }{\left(1-z \norm{\bm{\xi}^{(2)}}^{2} \right)^{2}} \right] }                                                                                        \\[16pt]
\qquad\qquad \qquad\qquad      \displaystyle{\times \left[ \left( 1 -z \norm{\bm{\xi}^{(2)}}^{2} \right) \xi_{1}^{(1)} \xi_{2}^{(1)} + \xi_{1}^{(2)} \xi_{2}^{(2)} \right]  }
 \end{array}
  $ }  \\[3.4cm]
 {\small  $ \begin{array}{l}  \displaystyle 4F_{z,2}'^{[2]}=
            \displaystyle{ \left( \frac{\bm{\xi}^{(1)} \! \bm{\cdot} \bm{p}_{\xi}^{(1)}}{1-z\norm{\bm{\xi}^{(2)}}^{2}} + \bm{\xi}^{(2)}  \! \bm{\cdot} \bm{p}_{\xi}^{(2)} \right) \left( L^{(1)} + L^{(2)} -  \frac{2 z\,\xi_{1}^{(2)} \xi_{2}^{(2)}\left( \bm{\xi}^{(1)}  \! \bm{\cdot} \bm{p}_{\xi}^{(1)} \right)}{1 -z \norm{\bm{\xi}^{(2)}}^{2}} \right)} \\[16pt]
            \qquad \qquad\quad  \displaystyle{ + z \left( \frac{\bm{\xi}^{(1)} \!\bm{\cdot} \bm{p}_{\xi}^{(1)}}{1 -z \norm{\bm{\xi}^{(2)}}^{2}} + \bm{\xi}^{(2)} \!\bm{\cdot} \bm{p}_{\xi}^{(2)} \right)^{2} \left[ \left( 1-z \norm{\bm{\xi}^{(2)}}^{2} \right) \xi_{1}^{(1)} \xi_{2}^{(1)} + \xi_{1}^{(2)} \xi_{2}^{(2)} \right] }                          \\[16pt]
            \qquad \qquad\quad  \displaystyle{ + 2    \left(  \frac{T_{z}^{(1)} }{1-z \norm{\bm{\xi}^{(2)}}^{2}} +T_{z}^{(2)}  \right) \left[ \left( 1-z \norm{\bm{\xi}^{(2)}}^{2} \right) \xi_{1}^{(1)} \xi_{2}^{(1)} + \xi_{1}^{(2)} \xi_{2}^{(2)} \right]}                                                                                                 \\[16pt]
 \qquad \qquad\quad      \displaystyle{-  \left[ p_{\xi_{1}}^{(2)} p_{\xi_{2}}^{(2)}+
 \frac{p_{\xi_{1}}^{(1)} p_{\xi_{2}}^{(1)}+ z L^{(2)} \left(\bm{\xi}^{(1)} \!\bm{\cdot}\bm{p}_{\xi}^{(1)} \right)   }{1 -z  \norm{\bm{\xi}^{(2)}}^{2}}
 - \frac{z^{2} \xi_{1}^{(2)} \xi_{2}^{(2)} \left( \bm{\xi}^{(1)}\! \bm{\cdot} \bm{p}_{\xi}^{(1)}\right)^2  }{\left(1-z \norm{\bm{\xi}^{(2)}}^{2} \right)^{2}} \right] }                                                                                                                                                \\[16pt]
 \qquad\qquad \qquad\qquad       \displaystyle{ \times \left(\norm{\bm{\xi}^{(1)}}^{2} + \norm{\bm{\xi}^{(2)}}^{2} -z \norm{\bm{\xi}^{(1)}}^{2} \norm{\bm{\xi}^{(2)}}^{2} \right) }
 \end{array} $} \\
      \\[-4pt]
      \hline

      \hline
    \end{tabular}
  \end{center}
  \label{table:def_2}
\end{table}


\begin{remark}
  As a direct consequence  of the above results, we obtain the
  $t$-independent constants of motion for the  $t$-dependent undeformed Hamiltonian    defined
  on $\mathrm{T}^{*}\mathbf{M}^{1+1} \times \mathrm{T}^{*}\mathbf{M}^{1+1}$,  that is, $h^{[2]} \equiv  h_{0}^{[2]}$ determined by (\ref{eq.LorentzLH:tdepHam}) and with primitive coproduct,   in the form
  \begin{equation}
    \begin{split}
      \lim_{z\to 0}4F_{z, 2}^{[2]}  & =  (L_1+L_2)^2-\left( \bm{\xi}^{(1)}  \! \bm{\cdot} \bm{p}_{\xi}^{(1)} +\bm{\xi}^{(2)}  \! \bm{\cdot} \bm{p}_{\xi}^{(2)}  \right)^2
      +2 \left(\norm{\bm{\xi}^{(1)}}^{2} +\norm{\bm{\xi}^{(2)}}^{2} \right) \left( T^{(1)}+ T^{(2)} \right)                                                                                                                                                                        \\[2pt]
& \qquad +4 \left( \xi_{1}^{(1)} \xi_{2}^{(1)}  + \xi_{1}^{(2)} \xi_{2}^{(2)}  \right)\left(p_{\xi_{1}}^{(1)} p_{\xi_{2}}^{(1)} +p_{\xi_{1}}^{(2)} p_{\xi_{2}}^{(2)}  \right),    \\[4pt]
      \lim_{z\to 0}4 F_{z,2}'^{[2]} & =   \left( \bm{\xi}^{(1)}  \! \bm{\cdot} \bm{p}_{\xi}^{(1)} +\bm{\xi}^{(2)}  \! \bm{\cdot} \bm{p}_{\xi}^{(2)}  \right)(L_1+L_2)+ 2 \left( T^{(1)}+ T^{(2)} \right)\left( \xi_{1}^{(1)} \xi_{2}^{(1)}  + \xi_{1}^{(2)} \xi_{2}^{(2)}  \right) \\[2pt]
 & \qquad - \left(p_{\xi_{1}}^{(1)} p_{\xi_{2}}^{(1)} +p_{\xi_{1}}^{(2)} p_{\xi_{2}}^{(2)}  \right)\left(\norm{\bm{\xi}^{(1)}}^{2} +\norm{\bm{\xi}^{(2)}}^{2} \right) ,
    \end{split}
    \nonumber
  \end{equation}
  such that the  shorthand notations (\ref{eq:applications:notation}) are maintained and
  the kinetic energy in the $i^{\textup{th}}$-copy of the  Minkowskian plane is given by
  \be
  T^{(i)}= \lim_{z\to 0}T_z^{(i)} =\frac 12 \left( \left( p_{\xi_{1}}^{(i)} \right)^{2}  -\left( p_{\xi_{2}}^{(i)} \right)^{2}  \right) .
  \nonumber
  \ee
\end{remark}


\section{Applications of  deformed Lie--Hamilton Lorentz systems}
\label{section:applications}

From the general results concerning the PH deformation of the $\mathfrak{so}(1,3)$-LH system presented previously, we will now analyse some of its possible applications. In particular, we will construct specific deformed LH systems based on the quantum book subalgebra, deformed oscillators on the (anti-)de Sitter spacetimes, which requires the use of the
quantum $\mathfrak{sl}(2, \mathbb{R})$ subalgebra, and, finally, deformed coupled  LH systems considering the   full  deformation of  $\mathfrak{so}(1,3)$. All these results can be expressed in terms of
the Beltrami variables  $ \big(\tilde{\bm{ \xi}}, \bm{p}_{\tilde\xi} \big)$ via the canonical transformation given in (\ref{cantransform}).


\subsection{Solvability by quadratures from the quantum book algebra}
\label{subsection:quadratures}

One of the simplest  $t$-dependent deformed Hamiltonian \eqref{eq:applications:tdep_def} on $\mathrm{T}^{*} \mathcal{M}_z$ we can consider  is when $b_{1}(t), b_{2}(t) \neq 0$ and $b_{\alpha}(t) = 0$ for all $3 \leq \alpha \leq 6$ and $t \in \mathbb{R}$. This yields the $t$-dependent Hamiltonian
\begin{equation}
  h_{z} = b_{1}(t) h_{z, 1} + b_{2}(t) h_{z, 2} = b_{1}(t) \,\frac{\mathrm{log} \left(1 -z \big(\xi_{1}^{2} - \xi_{2}^{2} \big) \right)}{z} + \frac{b_{2}(t)}{2}\, \big(\xi_{1} p_{\xi_{1}} + \xi_{2} p_{\xi_{2}} \big).
  \label{eq:applications:book}
\end{equation}
We underline that this system is based on the non-standard quantum book algebra, since $h_{z, 1} = \Psi_{z}^{*}(H)$, $h_{z, 2} = \Psi_{z}^{*}(D)$ and  $(H, D)$ are global coordinates on the dual Poisson--Lie group $B_{z}^{*}$ with Poisson bracket \eqref{eq:nonstandardLorentz:book_defcr},   that is,
$$
  \set{h_{z,1}, h_{z,2}}_{\omega} = \frac{1- \mathrm{e}^{- z h_{z,1}}}{z}.
$$

In this instance, the Hamilton equations of \eqref{eq:applications:book} are much simpler than in the general case \eqref{eq:applicactions:defHameq}:
\begin{equation}
  \begin{aligned}
     & \dv{\xi_{1}}{t} = \frac{1}{2} b_{2}(t) \xi_{1}, \qquad &  & \dv{p_{\xi_{1}}}{t} = \frac{2 b_{1}(t) \xi_{1}}{1- z \big(\xi_{1}^{2} - \xi_{2}^{2} \big)} - \frac{1}{2} b_{2}(t) p_{\xi_{1}},    \\[2pt]
     & \dv{\xi_{2}}{t} = \frac{1}{2} b_{2}(t) \xi_{2}, \qquad &  & \dv{p_{\xi_{2}}}{t} = - \frac{2 b_{1}(t) \xi_{2}}{1 - z \big(\xi_{1}^{2} - \xi_{2}^{2} \big)} - \frac{1}{2} b_{2}(t) p_{\xi_{2}}.
  \end{aligned}
  \label{eq:applications:book_system}
\end{equation}
These equations are clearly separable, hence the system can be solved by quadratures. More specifically, by means of the $t$-dependent function
\begin{equation*}
  \mathbb{R} \ni t \longmapsto \sigma(t) := \int_{0}^{t} b_{2}(\tau) \: \dd \tau \in \mathbb{R},
\end{equation*}
the exact solution of \eqref{eq:applications:book_system} can be written as
\begin{equation}
  \begin{aligned}
     & \xi_{1}(t) = c_{1}\,\mathrm{e}^{\sigma(t)/2},\qquad     &  & p_{\xi_{1}}(t) = \mathrm{e}^{-\sigma(t)/2} \left( c_{3} + 2 c_{1}\int_{0}^{t} \frac{b_{1}(s) \, \mathrm{e}^{\sigma (s)}}{1 -z \, \mathrm{e}^{\sigma(s)} (c_{1}^{2} - c_{2}^{2})} \: \dd s \right),     \\[2pt]
     & \xi_{2} (t) = c_{2}\, \mathrm{e}^{\sigma (t)/2}, \qquad &  & p_{\xi_{2}}(t) = \mathrm{e}^{-\sigma(t)/2} \left( c_{4} - 2 c_{2}\int_{t_{0}}^{t} \frac{b_{1}(s) \, \mathrm{e}^{\sigma (s)}}{1 -z \, \mathrm{e}^{\sigma(s)} (c_{1}^{2} - c_{2}^{2})} \: \dd s \right),
  \end{aligned}
  \label{eq:applications:book_exact}
\end{equation}
where $c_{1}, \ldots, c_{4}$ are integration constants relative to the initial conditions at the time $t = 0$.

When $z \to 0$, the undeformed system on $\mathrm{T}^{*} \mathbf{M}^{1+1}$ is even simpler, since the interaction between $\xi_{1}$ and $\xi_{2}$ introduced by the deformation parameter $z$ in \eqref{eq:applications:book_system} disappears, leading to the equations
\begin{equation}
  \begin{aligned}
     & \dv{\xi_{1}}{t} = \frac{1}{2} b_{2}(t) \xi_{1}, \qquad &  & \dv{p_{\xi_{1}}}{t} = 2 b_{1}(t) \xi_{1} - \frac{1}{2} b_{2}(t) p_{\xi_{1}}, \\[2pt]
     & \dv{\xi_{2}}{t} = \frac{1}{2} b_{2}(t) \xi_{2}, \qquad &  & \dv{p_{\xi_{2}}}{t} = -2 b_{1}(t) \xi_{2} - \frac{1}{2}b_{2}(t) p_{\xi_{2}}.
    \label{eq:applications:book_undeformed}
  \end{aligned}
\end{equation}
In this way, the expression of the exact solution \eqref{eq:applications:book_exact} is then simplified to
\begin{equation*}
  \begin{aligned}
     & \xi_{1}(t) = c_{1}\,\mathrm{e}^{\sigma(t)/2},\qquad     &  & p_{\xi_{1}}(t) = \mathrm{e}^{-\sigma(t)/2}\left( c_{3} + 2 c_{1}\int_{0}^{t} b_{1}(s) \, \mathrm{e}^{ \sigma (s)} \: \dd s \right), \\[2pt]
     & \xi_{2} (t) = c_{2}\, \mathrm{e}^{\sigma (t)/2}, \qquad &  & p_{\xi_{2}}(t) = \mathrm{e}^{-\sigma(t)/2} \left( c_{4} - 2 c_{2}\int_{0}^{t} b_{1}(s) \, \mathrm{e}^{\sigma (s)} \: \dd s \right).
  \end{aligned}
\end{equation*}
Note also that the undeformed system \eqref{eq:applications:book_undeformed} corresponds to an uncoupled system in the sense of \cite{CampoamorStursberg2025}, since its associated Hamiltonian $h = \lim_{z \to 0} h_{z}$ is a linear combination of lower-dimensional LH systems:
\begin{equation}
  h = b_{1}(t) h_{1} + b_{2}(t) h_{2} = \left( -b_{1}(t) \xi_{1}^{2} + \frac{b_{2}(t)}{2}\, \xi_{1} p_{\xi_{1}} \right) + \left( b_{1}(t) \xi_{2}^2+ \frac{b_{2}(t)}{2}\, \xi_{2} p_{\xi_{2}} \right).
  \label{eq:applications:book_undef}
\end{equation}
This is the main reason why these type of systems were not studied in detail previously, since the motivation in \cite{CampoamorStursberg2025} was to construct higher-dimensional {\em intrinsic} LH systems. However, from the perspective of the theory of exactly solvable ODEs, interesting systems derived from PH deformations based on the quantum book algebra have been recently obtained in \cite{CampoamorStursberg2023,CampoamorStursberg2024,CampoamorStursberg2025b}. Let us now briefly sketch how  the $t$-dependent Hamiltonian \eqref{eq:applications:book} can be related to such lower-dimensional systems.

First of all, let $\mathcal{N}_{z} \subset \mathbb{R}$ be the open subset $\mathcal{N}_{z} := \set{ q \in \mathbb{R}: 1 -z q^{2} > 0} \subset \mathbb{R}$, diffeomorphic to $\mathbb{R}$ regardless of the value of $z$.   Consider the submanifold
    \begin{equation}
      i : \mathrm{T}^{*} \mathcal{N}_{z} \equiv \set{\xi_{2} = p_{\xi_{2}} = 0} \hookrightarrow \mathrm{T}^{*} \mathcal{M}_{z},
      \nonumber
    \end{equation}
 with global coordinates $(q := \xi_{1}, p := p_{\xi_{1}})$. The $t$-dependent vector field associated with \eqref{eq:applications:book},
    \begin{equation*}
      \mathbf{X}_{z} = b_{1}(t) \mathbf{X}_{z,1} + b_{2}(t)\mathbf{X}_{z,2},
    \end{equation*}
    restricts to the $t$-dependent vector field on $\mathrm{T}^{*} \mathcal{N}_{z}$ given by
    \begin{equation}
      \mathbf{X}_{z}' := b_{1}(t)  \mathbf{X}_{z, 1}' + b_{2}(t) \mathbf{X}_{z,2}',
      \label{eq:applications:book_projected}
    \end{equation}
where
    \begin{equation}
      \mathbf{X}_{z, 1}' := \mathbf{X}_{z,1} \vert_{\mathrm{T}^{*} \mathcal{N}_{z}} = \frac{2 q}{1 -z q^{2}} \pdv{p}, \qquad \mathbf{X}_{z, 2}' := \mathbf{X}_{z,2} \vert_{\mathrm{T}^{*} \mathcal{N}_{z}} = \frac{1}{2} \left( q \pdv{q} - p \pdv{p} \right).
      \label{eq:applications:book_projhamVF}
    \end{equation}
As one can expect, these are Hamiltonian vector fields on $\mathrm{T}^{*}\mathcal{N}_{z}$ associated with the Hamiltonian functions
    \begin{equation}
      h_{z,1}' :=  \frac{\log(1 -z q^2)}{z}, \qquad h_{z,2}' := \frac{1}{2} qp,
      \label{eq:applications:book_projham}
    \end{equation}
 respectively,  as $i^{*} h_{z,\alpha} = h_{z, \alpha}'$  $(1 \leq \alpha \leq 2)$. The latter Hamiltonian functions close the same deformed  commutation relations  as $h_{z, 1}$ and $h_{z,2}$, and
    \begin{equation*}
      \big[\mathbf{X}_{z, 1}',  \mathbf{X}_{z,2}' \big] = [\mathbf{X}_{z, 1}, \mathbf{X}_{z, 2}] \vert_{\mathrm{T}^{*} \mathcal{N}_{z}}.
    \end{equation*}
This implies that \eqref{eq:applications:book_projected} defines a PH deformation of the $\mathfrak{b}_{2}$-LH system on $\mathrm{T}^{*}\mathcal{N}_{z}$, from the  $t$-dependent  undeformed vector field given by
    \begin{equation*}
      \mathbf{X}' = \lim_{z \to 0}  \mathbf{X}_{z}' = \sum_{\alpha = 1}^{2} b_{\alpha}(t)  \mathbf{X}_{\alpha}' = 2b_{1}(t)q \pdv{p} + \frac{b_{2}(t)}{2} \left( q \pdv{q} -p \pdv{p} \right),
    \end{equation*}
and is associated with the $t$-dependent Hamiltonian $h_{z}' = \sum_{\alpha = 1}^{2} b_{\alpha}(t) h_{z, \alpha}'$.
Under the canonical transformation from $\big\{(q, p) \in \mathrm{T}^{*} \mathcal{N}_{z}: q \neq 0 \big\}$ onto $\mathbb{R}^{2}_{z x < 0}$ defined by
    \begin{equation*}
      x := -\frac{\mathrm{log}(1 -z q^{2})}{z} , \qquad y:= \frac{(1 -z q^{2})p}{2q},
    \end{equation*}
 the Hamiltonian functions \eqref{eq:applications:book_projham} read
    \begin{equation*}
      h_{z, 1}' = - x, \qquad h_{z, 2}' = \left( \frac{  \mathrm{e}^{z x} -1}{z} \right) y,
    \end{equation*}
while the vector fields \eqref{eq:applications:book_projhamVF} take the form
    \begin{equation*}
      \mathbf{X}_{z, 1}' =  \pdv{y}, \qquad  \mathbf{X}_{z, 2}' =  \left( \frac{  \mathrm{e}^{z x} -1}{z} \right) \pdv{x} - \mathrm{e}^{z x} y \pdv{y},
    \end{equation*}
such that
    $$
     \big[\mathbf{X}_{z, 1}',  \mathbf{X}_{z,2}' \big] = - \mathrm{e}^{- z h_{z,1}'} \mathbf{X}_{z, 1}' .
    $$
This is, in fact, the PH deformation of $\mathfrak{b}_{2}$-LH systems studied in   previous works \cite{CampoamorStursberg2023,CampoamorStursberg2024,CampoamorStursberg2025b}.


\subsection{Time-dependent deformed oscillators on  the (anti-)de Sitter spacetimes}
\label{subsection:def_Ermakov}

Although the previous system \eqref{eq:applications:book_system} is worth studying, since it can be solved by quadratures, its associated $t$-dependent Hamiltonian \eqref{eq:applications:book} is not a natural one (that is, kinetic energy plus potential), so {\it a priori} it does not have a clear physical interpretation in a context of  Lorentzian systems. For this reason, we will now analyse the $t$-dependent Hamiltonian
\begin{equation}
  \begin{split}
    h_{z} & = b_{1}(t) h_{z, 1} + b_{2}(t) h_{z, 2} + b_{3}(t) h_{z, 3}     \\[2pt]
          & = b_{1}(t)\, \frac{\log \big(1 -z \big(\xi_{1}^{2} - \xi_{2}^{2} \big) \big)}{z}+ \frac{1}{2}b_{2}(t) (\xi_{1} p_{\xi_{1}} + \xi_{2} p_{\xi_{2}})     \\[2pt]
& \qquad -  \frac{1}{4} b_{3}(t)  \left(\big(1 -z \xi_{1}^{2} \big)p_{\xi_{1}}^{2} - 2 z \xi_{1} \xi_{2} p_{\xi_{1}}p_{\xi_{2}} - \big(1 + z \xi_{2}^{2} \big) p_{\xi_{2}}^{2} \right)  ,
  \end{split}
  \label{eq:applications:nonstandardsl2_gen}
\end{equation}
emerging from \eqref{eq:applications:tdep_def} for the choices $b_{\alpha}(t) = 0$ for all $4 \leq \alpha \leq 6$ and $t \in \mathbb{R}$, and  which retrieves \eqref{eq:applications:book}  when $b_{3}(t)$ vanishes. This system is now related to the non-standard quantum $\mathfrak{sl}(2, \mathbb{R})$ algebra, since $h_{z,1} = \Psi_{z}^{*}(H)$, $h_{z, 2} = \Psi_{z}^{*}(D)$, $h_{z, 3} = \Psi_{z}^{*}(C_{1})$,  and $(H, D, C_{1})$ are global coordinates on the dual Poisson--Lie group $S_{z}^{*}$   of $G_{z}^{*}$  with commutation relations \eqref{eq:nonstandardLorentz:bookgroup}; namely (see (\ref{eq:PHdef:defcr_ham}))
\be
\set{h_{z,1}, h_{z,2}}_{\omega} = \frac{1- \mathrm{e}^{- z h_{z,1}}}{z}, \qquad   \set{h_{z,1}, h_{z,3}}_{\omega} = 2 h_{z,2}, \qquad \set{h_{z,2}, h_{z,3}}_{\omega} = h_{z,3} -z h_{z,2}^{2}.
\nonumber
\ee

\begin{figure}[t!]
  \centering
  \begin{subfigure}[l]{0.3\textwidth}
    \includegraphics[width =1.3\textwidth]{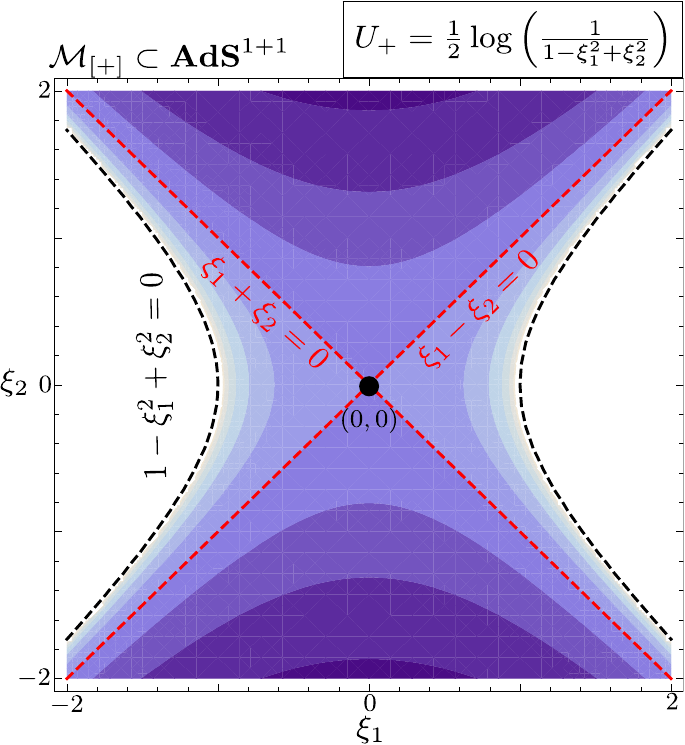}
  \end{subfigure}
  \qquad \qquad \qquad
  \begin{subfigure}[l]{0.3\textwidth}
    \includegraphics[width =1.3\textwidth]{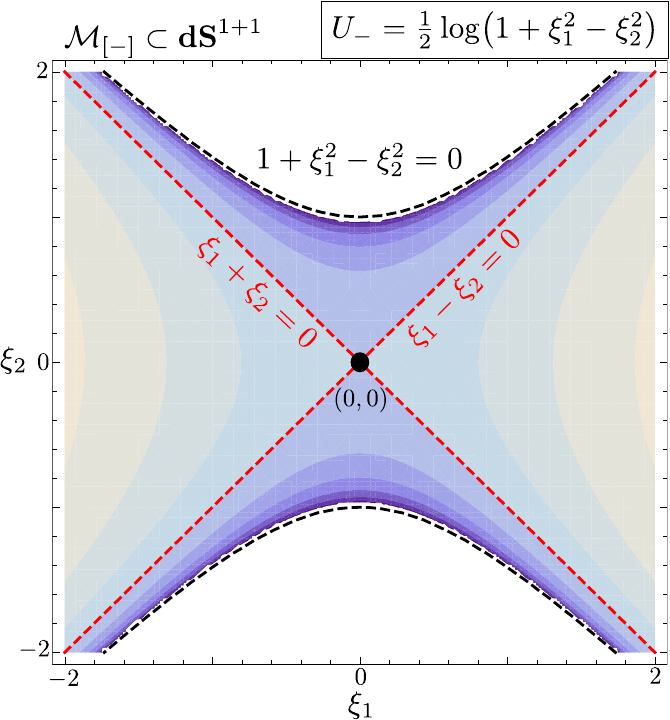}
  \end{subfigure}
  \\[10pt]
  \begin{subfigure}[l]{0.3\textwidth}
    \includegraphics[width =1.3\textwidth]{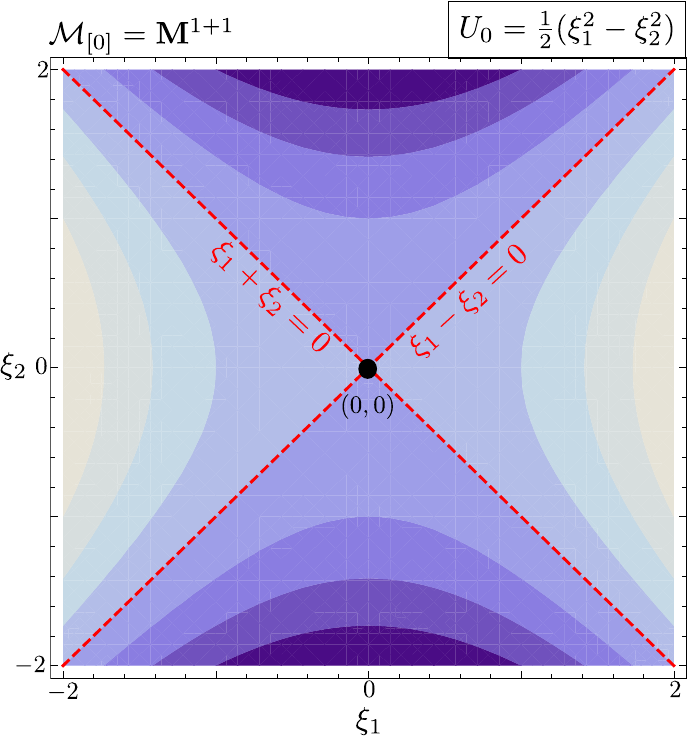}
  \end{subfigure}
  \caption{ \small  Level plots of the deformed potential $U_{z}$ \eqref{eq:applications:defPotential} on the three spaces $\mathcal{M}_{[z]}$, where $z$ is normalised  to the values $z =\{ 0, \pm 1\}$.
    Dark blue represents small values, while  higher values are `light blue'-shifted.}
  \label{fig:applications:def_potential}
\end{figure}

In order to extract a natural Hamiltonian from \eqref{eq:applications:nonstandardsl2_gen}, we first  define a potential as
\begin{equation}
  U_{z}:= - \frac{1}{2}h_{z,1}=  -\frac{1}{2z}\log(1 -z\big (\xi_{1}^{2}- \xi_{2}^{2} \big)) ,
  \label{eq:applications:defPotential}
\end{equation}
which can be expressed in terms of the isotropic harmonic oscillator potential  $U^{\mathrm{HO}}$ on the Minkowski spacetime $\mathbf{M}^{1+1}$,
\begin{equation}
  U^{\mathrm{HO}} = \frac{1}{2}  (\xi_{1}^{2} - \xi_{2}^{2}) ,
  \label{eq:applications:isotropicHO}
\end{equation}
in the form
\begin{equation}
  U_{z} = -\frac{1}{2z} \log (1 - 2 zU^{\mathrm{HO}}).
  \label{deformedU}
\end{equation}
Consequently, since the deformation parameter $z$ is the constant curvature of the metric (\ref{eq:twist:def_metric}) associated with the kinetic energy $T_z=-2   h_{z,3}$ (\ref{eq:twist:def_energy}), it follows that  $U_{z} $ (\ref{deformedU}) can be interpreted as a deformed oscillator on the
$(1+1)$-dimensional  anti-de Sitter  $\mathbf{AdS}^{1+1}$ $(z > 0)$ and the  de Sitter $\mathbf{dS}^{1+1}$ $(z < 0)$ spacetimes, such that the undeformed limit gives $ \lim_{z \to 0} U_{z}=U_{0}^{\mathrm{HO}} =U^{\mathrm{HO}}$ (\ref{eq:applications:isotropicHO}) on $\mathbf{M}^{1+1}$.
Then, to  emphasise  the role of the curvature $z$, we express $\mathcal{M}_z\equiv\mathcal{M}_{[z]}$ and for the values $z=\pm 1$ we find that the deformed potential  $U_{z} $ (\ref{eq:applications:defPotential}) is given by
\begin{equation}
  \begin{aligned}
     & \mathcal{M}_{[+]}\subset \mathbf{AdS}^{1+1}\!: &  &
    U_{+}=  \frac{1}{2 } \log( \frac{1}{  1 - \xi_{1}^{2}+ \xi_{2}^{2} } ) , \\[2pt]
     & \mathcal{M}_{[-]}\subset \mathbf{dS}^{1+1}\!:  &  &
    U_{-}=  \frac{1}{2 } \log(  {  1+    \xi_{1}^{2}- \xi_{2}^{2}  } ).
  \end{aligned}
  \nonumber
\end{equation}
In Figure~\ref{fig:applications:def_potential} the level plots of $U_{z}$ are analysed and compared depending on the value of the curvature $z$ for the three Lorentzian spaces.


Next, from this physical viewpoint, we introduce a $t$-dependent frequency $\Omega(t)$,  a $t$-dependent mass $m(t)$,    the kinetic energy $T_z= - 2h_{z,3}  $, as given in (\ref{eq:twist:def_energy}),  and the potential $U_{z}$ (\ref{eq:applications:defPotential}) in the $t$-dependent deformed Hamiltonian (\ref{eq:applications:nonstandardsl2_gen}), thus considered in the
(anti-)de Sitter spacetimes, by choosing
\be
b_1(t)=-\frac 12 m(t)\Omega^2(t),\qquad b_3(t)=- \frac{2 }{m(t)} ,
\nonumber
\ee
while keeping the coefficient $b_2(t)$ arbitrary. Hence, we arrive at a natural Hamiltonian along with a    $b_2(t)$-term:
\begin{equation}
  h_{z}  =\frac{1}{m(t)}\, T_z +m(t)\Omega^2(t) U_{z} + \frac{1}{2}b_{2}(t) (\xi_{1} p_{\xi_{1}} + \xi_{2} p_{\xi_{2}}) .
  \label{hamcoupledAdS}
\end{equation}
Note that  $m(t)$ can always be set as a constant mass.
The (flat/undeformed) Minkowskian limit of $ h_{z} $ gives rise to
\be
\lim_{z \to 0} { h_{z} }=\frac 1{2 {m(t)} }\, \big(  p_{\xi_{1}}^2-   p_{\xi_{2}}^2  \big)+\frac 12 m(t)\Omega^2(t)  \big(   {\xi_{1}}^2-    {\xi_{2}}^2  \big) + \frac{1}{2}b_{2}(t) (\xi_{1} p_{\xi_{1}} + \xi_{2} p_{\xi_{2}}) .
\label{hamcoupledM}
\ee
Therefore, a truly  genuine natural Hamiltonian is obtained by setting $b_2(t)=0$ in $ h_{z} $ (\ref{hamcoupledAdS}), which we shall consider from now on to study some of its properties. In this case,  the  Hamilton equations  turn out to be
\begin{equation}
  \begin{aligned}
     & \dot{\xi}_{1} = \frac 1{m(t)} \left( \big( 1 - z \xi_{1}^{2} \big) p_{\xi_{1}} -z \xi_{1} \xi_{2} p_{\xi_{2}} \right), \quad &  & \dot{p}_{\xi_{1}} = \frac z{m(t)} \,  p_{\xi_{1}}(\xi_{1} p_{\xi_{1}} + \xi_{2}p_{\xi_{2}})  - \frac{m(t)\Omega^{2}(t) \xi_{1}}{  1 -z\big (\xi_{1}^{2}- \xi_{2}^{2}  \big)},    \\[2pt]
     & \dot{\xi}_{2} =- \frac 1{m(t)} \left( \big(1 +z \xi_{2}^{2} \big) p_{\xi_{2}}+z \xi_{1} \xi_{2} p_{\xi_{1}} \right),  \quad  &  & \dot{p}_{\xi_{2}} =   \frac z{m(t)} \,   p_{\xi_{2}}(\xi_{1} p_{\xi_{1}} + \xi_{2}p_{\xi_{2}})  + \frac{m(t)\Omega^{2}(t) \xi_{2}}{ 1  -z\big (\xi_{1}^{2}- \xi_{2}^{2}  \big)}.
  \end{aligned}
  \label{eq:applications:pdm_def}
\end{equation}

Unlike \eqref{eq:applications:book_system}, this system of ODEs is far from being solvable by quadratures. This property still holds in the undeformed case (thereby justifying the use of Lie systems theory).    From the   two equations on the left-hand side of \eqref{eq:applications:pdm_def}, we find that
\begin{equation}
  p_{\xi_{1}} =m(t) \,\frac{ \big(1+z \xi_{2}^{2} \big) \dot{\xi}_{1} -z \xi_{1} \xi_{2} \dot{\xi}_{2}}{ 1  -z\big (\xi_{1}^{2}- \xi_{2}^{2}  \big) }, \qquad p_{\xi_{2}} =-m(t) \,  \frac{ \big(1 -z \xi_{1}^{2} \big) \dot{\xi}_{2} + z \xi_{1} \xi_{2} \dot{\xi}_{1}}{ 1  -z\big (\xi_{1}^{2}- \xi_{2}^{2}  \big)},
  \label{eq:applications:Legendre}
\end{equation}
and by substituting these expressions into the other two equations of \eqref{eq:applications:pdm_def} we arrive, after some routine computations, at the following system of second-order differential equations:
\begin{equation}
  \begin{split}
     & \ddot{\xi}_{1} +\frac{m'(t)}{m(t)}\,   \dot{\xi}_{1}+ \xi_{1} \left[\Omega^{2}(t) +z\,\frac{  \big(1 +z \xi_{2}^{2} \big) \dot{\xi}_{1}^{2} - 2 z \xi_{1} \xi_{2} \dot{\xi}_{1} \dot{\xi}_{2} - \big(1 -z \xi_{1}^{2} \big) \dot{\xi}_{2}^{2} }{1  -z\big (\xi_{1}^{2}- \xi_{2}^{2}  \big)} \right] = 0, \\[4pt]
     & \ddot{\xi}_{2} +\frac{m'(t)}{m(t)}\,   \dot{\xi}_{2}+ \xi_{2} \left[\Omega^{2}(t) +z\,\frac{  \big(1 +z \xi_{2}^{2} \big) \dot{\xi}_{1}^{2} - 2 z \xi_{1} \xi_{2} \dot{\xi}_{1} \dot{\xi}_{2} - \big(1 -z \xi_{1}^{2} \big) \dot{\xi}_{2}^{2} }{1  -z\big (\xi_{1}^{2}- \xi_{2}^{2}  \big)} \right] =0 .
  \end{split}
  \label{eq:applications:Ermakov_def}
\end{equation}
As one can expect, these are just the Euler--Lagrange equations of the $t$-dependent Lagrangian $L_{z}: \mathbb{R} \times \mathrm{T} \mathcal{M}_{[z]} \to \mathbb{R}$ associated with the Hamiltonian \eqref{hamcoupledAdS} with $b_2(t)=0$. In terms of the metric $g_{z}$  given in \eqref{eq:twist:def_metric}   and the potential $U_{z}$ (\ref{eq:applications:defPotential}), this Lagrangian takes the form
\begin{equation}
  L_{z} := \frac{1}{2} m(t) g_{z}\big (\dot{\xi}_{1}, \dot{\xi}_{2} \big) - m(t)\Omega^{2}(t) U_{z}.
  \label{eq:applications:Lagrangian}
\end{equation}
Furthermore,  the expressions \eqref{eq:applications:Legendre} determine the Legendre transform $\mathbb{F} \mathscr{L}_{z}: \mathrm{T} \mathcal{M}_{[z]} \to \mathrm{T}^{*}\mathcal{M}_{[z]}$ associated with the free Lagrangian $\mathscr{L}_{z} := \frac{1}{2} m(t)   g_{z}\big (\dot{\bm{\xi}}, \dot{\bm{\xi}}\big)$ induced by the metric $g_{z}$. Note that the $t$-dependent Hamiltonian $h_z$ and the $t$-dependent Lagrangian \eqref{eq:applications:Lagrangian} are related via the Legendre transform $\mathbb{F} \mathscr{L}_{z}$ through $\mathbb{F} \mathscr{L}_{z}^{*}(h_{z} ) =  \dot{\xi}_{1} \pdv{L_{z}}{\dot{\xi}_{1}} + \dot{\xi}_{2} \pdv{L_{z}}{\dot{\xi}_{2}} - L_{z}$.

When $z \to 0$, we obtain the undeformed counterpart of the system \eqref{eq:applications:Ermakov_def} on the Minkowskian plane,
\begin{equation}
  \ddot{\xi}_{1} +\frac{m'(t)}{m(t)}\,   \dot{\xi}_{1} + \Omega^{2}(t) \xi_{1} =0, \qquad  \ddot{\xi}_{2} +\frac{m'(t)}{m(t)}\,   \dot{\xi}_{2} + \Omega^{2}(t)\xi_{2} = 0,
  \nonumber
\end{equation}
with underlying Hamiltonian (\ref{hamcoupledM}) with $b_2(t)=0$. For a constant mass, say $m(t)\equiv 1$, it reduces to the equations of two one-dimensional harmonic oscillators:
\be
\ddot{\xi}_{1}   + \Omega^{2}(t) \xi_{1} =0, \qquad  \ddot{\xi}_{2} + \Omega^{2}(t)\xi_{2} = 0,
\nonumber
\ee
associated with the Hamiltonian of the isotropic harmonic oscillator with a $t$-dependent frequency on the Minkowskian plane given by
\begin{equation}
  h^{\mathrm{HO}} = \frac{1}{2}(p_{\xi_{1}}^{2} - p_{\xi_{2}}^{2}) + \frac{1}{2} \Omega^{2}(t)(\xi_{1}^{2} - \xi_{2}^{2}) .
  \label{eq:applications:HO}
\end{equation}
In this sense, equations  \eqref{eq:applications:Ermakov_def}  determine  a deformed second-order oscillator system with $t$-dependent frequency $\Omega(t)$ and mass $m(t)$, which is defined on  the (anti-)de Sitter spacetimes.

Let us now analyse the $t$-independent constants of motion  of the generic $t$-dependent Hamiltonian $h_{z}$ (\ref{eq:applications:nonstandardsl2_gen}), with arbitrary $t$-dependent functions $b_{\alpha}(t)$ $(1 \leq \alpha \leq 3)$. The results shown in Table~\ref{table:def_2} for $F_{z, 2}^{[2]}$ and $F_{z,2}'^{[2]}$, as well as   the trivial ones given in Remark~\ref{remdeformedCas},  also hold in this case. Nevertheless, given that  $h_{z}$ is determined by the non-standard PH deformation of $\mathfrak{sl}(2, \mathbb{R})$ presented in Section~\ref{subsection42}, which is a Lie subgroup $S_{z}^{*} \subset G_{z}^{*}$, it is possible to deduce additional  and much  simpler $t$-independent constants of motion for $h_{z}$.

Observe that, differently from \eqref{eq:applications:book},  the Hamiltonian \eqref{eq:applications:nonstandardsl2_gen} (and, in particular, also \eqref{hamcoupledAdS}) admits a constant of motion on $\mathrm{T}^{*} \mathcal{M}_{[z]}$ given by
\begin{equation}
  F_{z} := \Psi_{z}^{*}(\mathscr{V}_{z,2}) = - \mathrm{e}^{z h_{z,1}} h_{z,2}^{2} + \left( \frac{ \mathrm{e}^{z h_{z,1}} -1}{z} \right) h_{z, 3} = - \frac{1}{4} ( \xi_{1} p_{\xi_{2}} + \xi_{2} p_{\xi_{1}} )^{2},
  \label{eq:applications:constant1}
\end{equation}
where $\mathscr{V}_{z,2}$ is the Casimir function \eqref{eq:nonstandardLorentz:defCas:sl2} on the deformed $\mathfrak{sl}(2, \mathbb{R})$ group $S_{z}^{*}$. It is worth mentioning that this conserved quantity is independent of the deformation parameter $z$, and that its existence implies that the  system of ODEs \eqref{eq:applications:pdm_def} on $\mathrm{T}^{*}\mathcal{M}_{[z]}$ can be reduced by one degree of freedom following an analogue procedure to the one described in \cite[Theorem 3.1]{CampoamorStursberg2025} for Lie systems. Furthermore, we obtain a second
constant of motion on $\mathrm{T}^{*} \mathcal{M}_{[z]} \times \mathrm{T}^{*} \mathcal{M}_{[z]}$
by considering the deformed coproduct (\ref{eq:nonstandardLorentz:defcoprod:sl2}) together with the
Poisson map (\ref{eq:twist:def_map}) (see Proposition~\ref{eq:applications:def_ham}); namely:
\begin{equation}
  F_{z}^{[2]} = (\Psi_{z} \times \Psi_{z})^{*}(\Delta_{z}(\mathscr{V}_{z, 2})) = - \mathrm{e}^{z h_{z,1}^{[2]}} \, \left(h_{z, 2}^{[2]}\right)^{2} +   \left( \frac{ \mathrm{e}^{z h_{z, 1}^{[2]}} -1}{z} \right) h_{z, 3}^{[2]},
  \label{eq:applications:definvariant}
\end{equation}
with $h_{z, \alpha}^{[2]}$ $(1 \leq \alpha \leq 3)$   given in  Table~\ref{table:def_1}.
By introducing  the  kinetic energy $ T_{z} ^{(i)}$ (\ref{eq:applications:kinetic}) along with  the shorthand notations \eqref{eq:applications:notation}, the resulting expression of \eqref{eq:applications:definvariant} is significantly simplified to
\begin{equation}
  \begin{split}
    F_{z}^{[2]} & =-\frac 14\, \frac{1- z \norm{\bm{\xi}^{(1)}}^{2}  }{1- z \norm{\bm{\xi}^{(2)}}^{2}   }   \left( \bm{\xi}^{(1)}\! \bm{\cdot} \bm{p}_{\xi}^{(1)} + \bm{\xi}^{(2)}\! \bm{\cdot} \bm{p}_{\xi}^{(2)} \left( 1 -z \norm{\bm{\xi}^{(2)}}^{2}  \right) \right)^2 \\[2pt]
 & \qquad+ \frac 12 \left( \frac{T_{z}^{(1)}  }{1- z \norm{\bm{\xi}^{(2)}}^{2}} + T_{z}^{(2)}   \right)   \left( \norm{\bm{\xi}^{(1)}}^{2} + \norm{\bm{\xi}^{(2)}}^{2} -z \norm{\bm{\xi}^{(1)}}^{2} \norm{\bm{\xi}^{(2)}}^{2} \right) .
  \end{split}
  \label{eq:applications:I2k}
\end{equation}
Its undeformed counterpart $F_{0}^{[2]} = \lim_{z \to 0} F_{z}^{(2)}$ on  $\mathrm{T}^{*} \mathbf{M}^{1+1}$  reads
\begin{equation}
  F_{0}^{[2]} = -\frac{1}{4} \left[ \left( \sum_{i = 1}^{2} \xi_{1}^{(i)} p_{\xi_{1}}^{(i)} + \xi_{2}^{(i)} p_{\xi_{2}}^{(i)}\right)^{2} - \left( \sum_{i = 1}^{2} \left(\xi_{1}^{(i)} \right)^{2} - \left(\xi_{2}^{(i)}\right)^{2} \right) \left( \sum_{i = 1}^{2} \left(p_{\xi_{1}}^{(i)}\right)^{2} - \left(p_{\xi_{2}}^{(i)}\right)^{2} \right) \right].
  \label{eq:applications:I20}
\end{equation}
Therefore, if we express   $F_{z}$  (\ref{eq:applications:constant1})    in the  first-copy   $\big( \bm{\xi}^{(1)}, \bm{p}_{\xi}^{(1)} \big)$ of the product manifold $\mathrm{T}^{*} \mathcal{M}_{[z]} \times \mathrm{T}^{*} \mathcal{M}_{[z]}$ as
$$
  F_{z}^{[1]}    = - \frac{1}{4} \left( \xi_{1}^{(1)} p_{\xi_{2}}^{(1)}  + \xi_{2}^{(1)}  p_{\xi_{1}}^{(1)}  \right)^{2},
$$
we find that  the following  commutation relations  are satisfied with respect to the   Poisson bracket induced by the  symplectic form (\ref{symplectic2}):
$$
  \left\{F_{z}^{[1]} ,h_{z,\alpha}^{[2]}  \right\}_{\omega^{[2]}}=\left\{F_{z}^{[2]} ,h_{z,\alpha}^{[2]}  \right\}_{\omega^{[2]}}=0,\quad 1\le \alpha\le 3,\qquad \left\{ F_{z}^{[1]} ,F_{z}^{[2]} \right\}_{\omega^{[2]}}= 0.
$$

\begin{remark}
  Notice that when restricted to the submanifold
  $$
    \left\{\xi_{2}^{(i)} = p_{\xi_{2}}^{(i)} = 0: 1 \leq i \leq 2 \right\} \subset \mathrm{T}^{*} \mathcal{M}_{[z]} \times \mathrm{T}^{*} \mathcal{M}_{[z]},
  $$
  diffeomorphic to $\mathrm{T}^{*} \mathbb{R} \times \mathrm{T}^{*} \mathbb{R}$ and with global coordinates  $\big( \xi_{1}^{(1)}, p_{\xi_{1}}^{(1)}, \xi_{1}^{(2)}, p_{\xi_{1}}^{(2)} \big)$,
  the deformed constant of motion $F_{z}^{[2]} $ \eqref{eq:applications:I2k} takes the form
  \begin{equation}
    F_{z}^{[2]}   \leadsto  \frac{1}{4}\, \frac{\left[ \left( 1- z \left(\xi_{1}^{(1)}\right)^2 \right)\xi_{1}^{(2)}p_{\xi_{1}}^{(1)} -\left( 1- z \left(\xi_{1}^{(2)}\right)^2 \right) \xi_{1}^{(1)}p_{\xi_{1}}^{(2)} \right]^2 }{1-z \left(\xi_{1}^{(2)}\right)^2 } ,
    \nonumber
  \end{equation}
  while its undeformed counterpart $F_{0}^{[2]} $ (\ref{eq:applications:I20})   leads to
  \begin{equation}
    F_{0}^{[2]}  \leadsto  \frac{1}{4} \left(\xi_{1}^{(1)}p_{\xi_{1}}^{(2)} - \xi_{1}^{(2)} p_{\xi_{1}}^{(1)}\right)^{2},
    \nonumber
  \end{equation}
  which is just the  square of angular momentum.
\end{remark}


\subsection{Time-dependent deformed Bateman and generalised coupled Caldirola--Kanai  Hamiltonians}

The deformed systems analysed in Sections~\ref{subsection:quadratures} and \ref{subsection:def_Ermakov} are PH deformations of LH systems associated with proper subalgebras of the LH algebra $\mathcal{H}_{\omega} \simeq \mathfrak{so}(1,3)$ given in \eqref{eq:LorentzLH:LHalgebra}. Namely, \eqref{eq:applications:book} is a PH deformation of the $\mathfrak{b}_{2}$-LH system \eqref{eq:applications:book_undef}, while the Hamiltonian (\ref{hamcoupledAdS}), with $b_2(t)=0$ and $m(t)\equiv 1$,  is a PH deformation of the $\mathfrak{sl}(2, \mathbb{R})$-LH system \eqref{eq:applications:HO}. In order to obtain applications from  the PH deformation of the whole LH algebra $\mathcal{H}_{\omega} \simeq \mathfrak{so}(1,3)$, we consider the $t$-dependent Hamiltonian given by
\begin{equation}
  h_{z} := b_{1}(t) h_{z,1} + b_{3}(t) h_{z, 3} + b_{5}(t) h_{z, 5},
  \label{eq:applications:defLorentz_gen}
\end{equation}
since its undeformed counterpart
\begin{equation}
  h = \lim_{z \to 0} h_{z} = b_{1}(t) h_{1} + b_{3}(t) h_{3} + b_{5}(t) h_{5}
  \nonumber
\end{equation}
is a $\mathfrak{so}(1,3)$-LH system on $\mathrm{T}^{*} \mathbf{M}^{1+1}$ such that  the Hamiltonian functions
$h_{1} = - \xi_{1}^{2} + \xi_{2}^{2}$, $h_{3} = -\frac{1}{4} \big (p_{\xi_{1}}^{2} - p_{\xi_{2}}^{2} \big)$ and $h_{5} = \frac{1}{2} (\xi_{1} p_{\xi_{2}} - \xi_{2} p_{\xi_{1}})$ span the LH algebra $\mathcal{H}_{\omega} \simeq \mathfrak{so}(1,3)$ given in \eqref{eq:LorentzLH:LHalgebra}.

The Hamilton equations of \eqref{eq:applications:defLorentz_gen} read (see (\ref{eq:applicactions:defHameq}))
\begin{equation}
  \begin{split}
    \dot{\xi}_{1}     & = -\frac{1}{2} \left[   b_{3}(t) \left( \big(1-z \xi_{1}^{2} \big) p_{\xi_{1}} -z \xi_{1}\xi_{2}p_{\xi_{2}} \right)  + b_{5}(t) \xi_{2} \right],      \\[2pt]
    \dot{\xi}_{2}     & = \frac{1}{2} \left[  b_{3}(t) \left( \big( 1 + z \xi_{2}^{2} \big) p_{\xi_{2}} + z \xi_{1} \xi_{2} p_{\xi_{1}} \right) + b_{5}(t) \xi_{1} \right],        \\[2pt]
    \dot{p}_{\xi_{1}} & = \frac{1}{2} \left[  \frac{4 b_{1}(t) \xi_{1}}{1 - z \big(\xi_{1}^{2} - \xi_{2}^{2} \big)} - z b_{3}(t)   p_{\xi_{1}} ( \xi_{1} p_{\xi_{1}} + \xi_{2} p_{\xi_{2}}) - b_{5}(t) p_{\xi_{2}}  \right], \\[2pt]
    \dot{p}_{\xi_{2}} & =- \frac{1}{2} \left[   \frac{4 b_{1}(t) \xi_{2}}{1 -z \big(\xi_{1}^{2} - \xi_{2}^{2} \big)} + z b_{3}(t)  p_{\xi_{2}} ( \xi_{1} p_{\xi_{1}} + \xi_{2} p_{\xi_{2}}) -b_{5}(t) p_{\xi_{1}} \right].
  \end{split}
  \label{eq:applications:defLorentz_system}
\end{equation}
Like  the system \eqref{eq:applications:pdm_def}, these equations cannot be solved by quadratures whenever $b_{3}(t) b_{5}(t)  \neq 0$. Furthermore, under these conditions, the associated undeformed system of \eqref{eq:applications:defLorentz_system},
\begin{equation}
  \begin{pmatrix}
    \dot{\xi}_{1}     \\[2pt]
    \dot{\xi}_{2}     \\[2pt]
    \dot{p}_{\xi_{1}} \\[2pt]
    \dot{p}_{\xi_{2}}
  \end{pmatrix} = \frac{1}{2} \begin{pmatrix}
    0         & -b_{5}(t)  & -b_{3}(t) & 0          \\[2pt]
    b_{5}(t)  & 0          & 0         & b_{3}(t)   \\[2pt]
    4b_{1}(t) & 0          & 0         & - b_{5}(t) \\[2pt]
    0         & -4b_{1}(t) & b_{5}(t)  & 0
  \end{pmatrix} \begin{pmatrix}
    \xi_{1}     \\[2pt]
    \xi_{2}     \\[2pt]
    p_{\xi_{1}} \\[2pt]
    p_{\xi_{2}}
  \end{pmatrix},
  \label{eq:applications:undefLorentz_system}
\end{equation}
is no longer an uncoupled sum as \eqref{eq:applications:HO}, because of  the presence of the $t$-dependent angular momentum term $b_{5}(t) h_{5} = \frac{1}{2}b_{5}(t)(\xi_{1} p_{\xi_{2}}- \xi_{2} p_{\xi_{1}})$. At this point, we emphasise that the {\em deformed coupled system} \eqref{eq:applications:defLorentz_system}  can be considered as a perturbation of the undeformed system \eqref{eq:applications:undefLorentz_system} in the sense of \cite{Kummer1971}, since all the expressions in \eqref{eq:applications:defLorentz_system} involving the deformation parameter $z$ are analytic in $z$. Let us now consider the first-order approximation in $z$ of \eqref{eq:applications:defLorentz_system}, namely
\begin{equation}
  \begin{split}
    \dot{\xi}_{1}     & = - \frac{1}{2} \big(b_{3}(t) p_{\xi_{1}} + b_{5}(t) \xi_{2} \big)+ \frac{1}{2} z  b_{3}(t) \xi_{1}(\xi_{1} p_{\xi_{1}} + \xi_{2} p_{\xi_{2}}) + o[z]^{2},       \\[2pt]
    \dot{\xi}_{2}     & = \frac{1}{2} \big( b_{3}(t) p_{\xi_{2}} + b_{5}(t) \xi_{1} \big) + \frac{1}{2} z b_{3}(t) \xi_{2}(\xi_{1} p_{\xi_{1}} + \xi_{2} p_{\xi_{2}}) + o[z]^{2},       \\[2pt]
    \dot{p}_{\xi_{1}} & = \left(2 b_{1}(t) \xi_{1} - \frac{1}{2} b_{5}(t) p_{\xi_{2}} \right) + z \left( 2b_{1}(t) \xi_{1}\big(\xi_{1}^{2} -\xi_{2}^{2} \big) - \frac{1}{2} b_{3}(t) p_{\xi_{1}}(\xi_{1} p_{\xi_{1}} + \xi_{2} p_{\xi_{2}}) \right) + o[z]^{2},     \\[2pt]
    \dot{p}_{\xi_{2}} & = -\left( 2 b_{1}(t) \xi_{2} - \frac{1}{2} b_{5}(t) p_{\xi_{1}} \right) -z  \left( 2 b_{1}(t) \xi_{2} \big(\xi_{1}^{1} - \xi_{2}^{2} \big) + \frac{1}{2} b_{3}(t) p_{\xi_{2}}(\xi_{1} p_{\xi_{1}} + \xi_{2}p_{\xi_{2}}) \right) + o[z]^{2}.
  \end{split}
  \nonumber
\end{equation}
Through a lengthy yet straightforward computation, provided that $b_{3}(t) \neq 0$ for all $t \in \mathbb{R}$, we can derive a system of second-order differential equations for $\xi_{1}$ and $\xi_{2}$ of the form
\begin{equation}
  \begin{split}
    -4 b_{3} \ddot{\xi}_{1} - b_{3} \big(4 b_{1}b_{3} + b_{5}^{2} \big) \xi_{1} - 2 b_{3} \dot{b}_{5} \xi_{2} + 2 \dot{b}_{3} \big(b_{5} \xi_{2} + 2 \dot{\xi}_{1} \big) + z F_{1} + o[z]^{2} & = 0, \\[2pt]
    4 b_{3} \ddot{\xi}_{2} +b_{3} \big(4b_{1} b_{3} + b_{5}^{2} \big) \xi_{2} - 2 b_{3} \dot{b}_{5} \xi_{1} + 2 \dot{b}_{3}\big( b_{5} \xi_{1} - 2 \dot{\xi}_{2} \big) + z F_{2}+ o[z]^{2}    & = 0,
  \end{split}
  \label{eq:applications:defLorentz_2nd}
\end{equation}
where
\begin{equation}
  F_{i} = F_{i}(\xi_{1}, \dot{\xi}_{1}, \ddot{\xi}_{1},  \xi_{2}, \dot{\xi}_{2}, \ddot{\xi}_{2}, b_{1}, b_{3}, \dot{b}_{3}, b_{5}, \dot{b}_{5}), \qquad 1 \leq i \leq 2
  \nonumber
\end{equation}
are polynomial functions such that
\begin{equation}
  \pdv[3]{F_{i}}{\xi_{i}}\neq 0, \qquad \pdv[2]{F_{i}}{\dot{\xi}_{j}} \neq 0, \qquad  \pdv{F_{i}}{\ddot{\xi}_{j}}  \neq 0, \qquad \pdv[2]{F_{1}}{\xi_{2}} \neq 0, \qquad \pdv[2]{F_{2}}{\xi_{1}} \neq 0, \qquad 1 \leq i,j \leq 2.
  \label{eq:applications:Fi}
\end{equation}
We omit the explicit formulae for $F_{i}$ $(1 \leq i \leq 2)$, since they are too cumbersome. For our analysis it suffices to notice that \eqref{eq:applications:Fi} shows that the deformation parameter $z$ produces a higher-order interaction between $\xi_{1}$ and $\xi_{2}$ in \eqref{eq:applications:defLorentz_2nd}.

In the following, we make particular choices of the coefficient functions in \eqref{eq:applications:defLorentz_gen}, obtaining the corresponding deformation of the $\mathfrak{so}(1,3)$-LH systems presented in Section~\ref{subsection:LorentzLHapplications}.
\begin{itemize}

  \item {\it A  $t$-dependent  deformed Bateman Hamiltonian}. Let  $m(t), k(t), \gamma(t) \in C^{\infty}(\mathbb{R})$ be positive functions such that $4 m(t) k(t) > \gamma^{2}(t)$ for all $t \in \mathbb{R}$, and consider the Hamiltonian $h_{z}^{\mathrm{B}}$ obtained from \eqref{eq:applications:defLorentz_gen} with the same choices for $b_1(t)$,  $b_3(t)$ and  $b_5(t)$  as in (\ref{bBateman}), which, in terms of  the kinetic energy \eqref{eq:twist:def_energy} and the potential \eqref{eq:applications:defPotential}, yields
\begin{equation}
h_{z}^{\mathrm{B}} = \frac{1}{m(t)}  \, T_{z}+ m(t) \Omega^{2}(t) U_{z} + \frac{\gamma(t)}{2m(t)}(\xi_{1} p_{\xi_{2}} - \xi_{2} p_{\xi_{1}}).
\nonumber
\end{equation}
We call $h_{z}^{\mathrm{B}}$ a {\it  $t$-dependent deformed Bateman Hamiltonian}, since
\begin{equation*}
\begin{split}
\lim_{z \to 0} h_{z}^{\mathrm{B}} & =  \left( \frac{1}{2 m(t)}\, p_{\xi_{1}}^{2} + \frac{1}{2} m(t) \Omega^{2}(t) \xi_{1}^{2} \right) -  \left( \frac{1}{2 m(t)}\, p_{\xi_{2}}^{2} + \frac{1}{2} m(t) \Omega^{2}(t) \xi_{2}^{2} \right) \\
 & \qquad + \frac{\gamma(t)}{2m(t)} (\xi_{1}p_{\xi_{2}} - \xi_{2}p_{\xi_{1}})
\end{split}
\end{equation*}
retrieves the $t$-dependent Bateman Hamiltonian \eqref{eq:LorentzLH:Bateman}.

  \item {\it A deformed generalised coupled Caldirola--Kanai Hamiltonian}. Let $h^{2\mathrm{CK}}_{z}$ be the Hamiltonian obtained from \eqref{eq:applications:defLorentz_gen} by considering the same functions $\lambda(t)$ and $\Omega(t)$  defined in (\ref{eq:LorentzLH:CK})  along  with $b_1(t)$ and  $b_3(t)$ as given in (\ref{eq:LorentzLH:CKb}).   Then, $h_{z}^{2\mathrm{CK}}$ can be written in terms of the kinetic energy \eqref{eq:twist:def_energy} and the potential \eqref{eq:applications:defPotential} as
\begin{equation}
h^{2\mathrm{CK}}_{z} = \frac{1}{m(t)} \, \mathrm{e}^{-2 \int_{0}^{t} \lambda(s)  \dd s} \,T_{z} + m(t) \Omega^{2}(t)\, \mathrm{e}^{2 \int_{0}^{t} \lambda(s)  \dd s}\,U_{z} + \frac{b_{5}(t)}{2}(\xi_{1} p_{\xi_{2}} - \xi_{2} p_{\xi_{1}}).
\nonumber
\end{equation}
Therefore, $h_{z}^{2\mathrm{CK}}$ is a {\it deformed generalised coupled Caldirola--Kanai Hamiltonian}, since
\begin{equation}
\begin{aligned}
& \lim_{z \to 0} h_{z}^{2 \mathrm{CK}}  = \left( \frac{1}{2m(t)} \,\mathrm{e}^{- 2\int_{0}^{t} \lambda(s)  \dd s}\, p_{\xi_{1}}^{2} + \frac{1}{2} m(t) \Omega^{2}(t) \,\mathrm{e}^{2 \int_{0}^{t} \lambda(s) \dd s}\, \xi_{1}^{2} \right)   \\
& \qquad\qquad\qquad- \left( \frac{1}{2m(t)}\, \mathrm{e}^{- 2\int_{0}^{t} \lambda(s) \dd s}\,p_{\xi_{2}}^{2} + \frac{1}{2} m(t) \Omega^{2}(t) \,\mathrm{e}^{2 \int_{0}^{t} \lambda(s)  \dd s} \,\xi_{2}^{2} \right) +  \frac{b_{5}(t)}{2}(\xi_{1}p_{\xi_{2}} - \xi_{2}p_{\xi_{1}})
\end{aligned}
\nonumber
\end{equation}
is just the generalised coupled Caldirola--Kanai Hamiltonian shown in \eqref{eq:LorentzLH:2CK}.
\end{itemize}


\section{Concluding remarks}
\label{section:conclusion}

The starting point of this work was to consider a specific
quantum deformation of the Lorentz algebra $\mathfrak{so}(1,3)$, denoted as $\mathcal{U}_{z}(\mathfrak{so}(1,3))$ with quantum deformation real parameter $z$,  of its universal enveloping algebra $\mathcal{U}(\mathfrak{so}(1,3))$,  as  presented in \cite{Herranz2000}. This deformation has two relevant properties.
Firstly, it contains, as Hopf subalgebras, the well-known   non-standard quantum deformation of $\mathfrak{sl}(2, \mathbb{R})$, as well as the single coboundary deformation of the book algebra $\mathfrak{b}_{2}$, such that  $\mathcal{U}_{z}(\mathfrak{so}(1,3))\supset \mathcal{U}_{z}(\mathfrak{sl}(2, \mathbb{R}))\supset \mathcal{U}_{z}(\mathfrak{b}_{2})$; therefore, they all share a common classical $r$-matrix. Secondly, given that the underlying classical $r$-matrix satisfies the classical Yang--Baxter equation, the quantum deformation is of non-standard (or triangular) nature; this property, in turn, implies that, by definition,  there exists  a non-linear change of basis (a `twist' map), that allows  $\mathcal{U}_{z}(\mathfrak{so}(1,3))$ to be expressed  in terms of  undeformed commutation relations, while preserving  a deformed coproduct.

The Lorentz algebra was initially formulated in a kinematical basis as the Lie symmetry algebra of the $(2+1)$-dimensional de Sitter spacetime and, subsequently, in a conformal basis as the conformal Lie symmetry algebra of the Euclidean plane, which constitutes a much more suitable basis for the specific deformation analysed throughout the article and for its further applications presented here.

Among the main new results obtained, it is worth noting that the dual Poisson--Lie group $G_{z}^{*}$ associated with $\mathcal{U}_{z}(\mathfrak{so}(1,3))$ has been constructed together with the dual Poisson--Lie subgroups  $\mathfrak{sl}(2,\mathbb{R})$ ($S_{z}^{*}$) and $\mathfrak{b}_{2}$ ($B_{z}^{*}$) as particular submanifolds of $G_{z}^{*}$, resulting in the embedding
$G_{z}^{*} \supset S_{z}^{*} \supset B_{z}^{*}$, and their Poisson geometry has also been studied. Furthermore, the composition of the twist map, provided by the  non-standard character of the deformation, with an appropriate momentum map, has enabled us  to  deduce a  kinetic energy in `certain' projective variables, which, after a canonical transformation, have been related to the usual canonical Beltrami variables on the $(1+1)$-dimensional (anti-)de Sitter spacetimes with constant curvature equal to $z$, described,  in a unified setting, as $\mathcal{M}_z$; hence, the sign of $z$ determines the  anti-de Sitter $\mathbf{AdS}^{1+1}$ $(z>0)$ and the de Sitter $\mathbf{dS}^{1+1}$ $(z<0)$ spaces, while the undeformed limit $z\to 0$ gives rise to the Minkowskian $\mathbf{M}^{1+1}$ plane.  From this approach, a generic time-dependent deformed Hamiltonian system $h_z$ on $\mathcal{M}_z$ (with {\em six} arbitrary $t$-dependent coefficients)  has  been derived from the corresponding non-standard  PH deformation of the Lorentz algebra, and has been interpreted within the framework of projective relativistic dynamics in a natural way. Additionally, $t$-independent constants of motion for the $2^{\textup{nd}}$-order  deformed prolongations of $h_z$ have been explicitly obtained. The significance  of these results has been illustrated through several applications, which cover $t$-dependent deformed oscillators on the (anti-)de Sitter spacetimes $\mathcal{M}_z$, determined by the PH deformation of $\mathfrak{sl}(2,\mathbb{R})$ ($S_{z}^{*}$),  as well as the complete general deformed Lorentz system ($G_{z}^{*}\supset S_{z}^{*})$, for which the deformation parameter $z$
has been interpreted as an integrable deformation parameter and the first-order approximation in $z$ has also been discussed in detail.

The results of this work open up several lines of research that deserve further investigation. Among others, the following are worth mentioning:
\begin{itemize}

  \item The representation of the Hamiltonian functions, that provides the generic $\mathfrak{so}(1,3)$-LH system given in Proposition~\ref{prop:LorentzLH:morphism}, corresponds  to a zero-value of both quadratic Casimir invariants. This fact precludes the presence of non-central potentials, of the type $1/\xi^2$, known as Rosochatius--Winternitz terms, which appear in the Smorodinsky--Winternitz system~\cite{WSUF65}, formed by an $N$-dimensional isotropic Euclidean oscillator with $N$ centrifugal terms. Therefore, an analysis of other representations of $\mathfrak{so}(1,3)$, with non-zero values for the Casimirs, could be useful to introduce such terms, which in the deformed
        case may give rise to $t$-dependent additional `curved' potentials on the (anti-)de Sitter spacetimes $\mathcal{M}_z$.

  \item Another question is whether it would be possible to extend all the results presented in this article to higher dimensions, in particular, to the $(4+1)$-dimensional de Sitter algebra, $\mathfrak{so}(1,5)$, which is the conformal Lie algebra of the four-dimensional Euclidean space. The introduction of the corresponding non-standard PH deformation would lead to time-dependent systems on the $(3+1)$-dimensional  (anti-)de Sitter spaces, with the deformation parameter $z$ playing the role of their constant sectional curvature. This task appears to be feasible, given that the Hopf algebra structure of the necessary non-standard quantum deformation of $\mathfrak{so}(1,5)$,  as well as that  of $\mathfrak{so}(2,4)$, has already been obtained in~\cite{Herranz2000a} in a conformal basis, such that
        $\mathcal{U}_{z}(\mathfrak{so}(1,5))\supset\mathcal{U}_{z}(\mathfrak{so}(1,4))\supset \mathcal{U}_{z}(\mathfrak{so}(1,3))\supset \mathcal{U}_{z}(\mathfrak{sl}(2, \mathbb{R}))\supset \mathcal{U}_{z}(\mathfrak{b}_{2})$.

  \item Finally, a question that naturally arises is whether another quantum deformation of the Lorentz algebra could be useful in the context of LH systems and their PH deformations. In this regard, it is worth recalling that the classification of all possible quantum deformations of $\mathfrak{so}(1,3)$ was already obtained by Zakrzewski in~\cite{Zakrzewski1994lorentz} via their classical $r$-matrices. When  expressed in a kinematical basis, it has been shown in~\cite{BGH2022} that there are four types of (multiparametric) deformations, named A, B, C and D. Two of these types, B and D, are  non-standard; they satisfy the classical Yang--Baxter equation and, therefore,
        ensure the existence of a twist map. Types A and C, on the other hand, correspond to standard quantum deformations, as they satisfy the modified classical Yang--Baxter equation. The specific quantum Lorentz deformation considered in this article is, precisely, the type D deformation; consequently, the (biparametric) type B deformation deserves a more detailed study.

\end{itemize}

Work on these open problems is currently in progress.


\section*{Acknowledgements}

\phantomsection
\addcontentsline{toc}{section}{Acknowledgments}

This work has been partially supported by Agencia Estatal de Investigaci\'on (Spain)   under  the grant PID2023-148373NB-I00 funded by MCIN/AEI/10.13039/501100011033/FEDER, UE. R.C.-S.  acknowledges support by the research grant PID2024-156578NB-I00 funded by the Spanish MICIU/AEI/10.13039/501100011033/FEDER, EU.
O.C. acknowledges a fellowship (grant C15/23) supported by  Universidad Complutense de Madrid and Banco Santander.  F.J.H. is also supported by  the Regional Government of Castilla y Le\'on (Junta de Castilla y Le\'on, Spain) under the grant BU011P25/FEDER, UE and acknowledges the contribution of the European Cooperation in Science and Technology COST Action CA23130 (BridgeQG).


\end{document}